\documentclass[acmtog,authorversion]{acmart}
\usepackage{etex}
\usepackage{multirow}
\usepackage{soul}
\usepackage{booktabs} 
\usepackage{amsmath}
\usepackage{hyperref}
\usepackage{cleveref}
\usepackage{tabularx}
\usepackage[ruled]{algorithm2e} 
\usepackage[permil]{overpic}
\usepackage[group-separator={,}]{siunitx}
\usepackage[export]{adjustbox}
\usepackage{float}

\SetAlFnt{\small}
\SetAlCapFnt{\small}
\SetAlCapNameFnt{\small}
\SetAlCapHSkip{0pt}

\usepackage{tikz}
\usepackage[skins]{tcolorbox} 
\usepackage{shadowtext}
\shadowrgb{0,0,0}
\shadowoffset{0.5pt}

\setcitestyle{square,nosort}

\def\commentType{1}

\usepackage{subcaption}
\usepackage{rotating}

\usepackage{tabularx}

\definecolor{darkred}{RGB}{127,0,0}
\definecolor{OliveGreen}{rgb}{0,0.6,0}

\newcommand{\old}[1]{}
\newcommand{\new}[1]{#1}

\newlength\beautywidth
\newlength\zoomwidth

\newlength\trimAl
\newlength\trimAr
\newlength\trimAt
\newlength\trimAb
\newlength\trimBl
\newlength\trimBr
\newlength\trimBt
\newlength\trimBb

\usepackage[normalem]{ulem}

\ifnum\commentType=0
	\newcommand{\customComment}[3]{}
	\newcommand{\customTODO}[3]{}
\fi
\ifnum\commentType=1
	\newcommand{\customComment}[3]{\textcolor{#2}{\textsl{#1: #3}}}
	\newcommand{\customTODO}[3]{\textcolor{#2}{\textsl{#1 TODO: #3}}}
\fi
\ifnum\commentType=2
	\usepackage{pdfcomment}
	\newcommand{\customComment}[3]{\pdfcomment[icon=Comment,opacity=0.5,color=#2,author=#1]{#3}}
	\newcommand{\customTODO}[3]{\pdfcomment[icon=Note,opacity=0.5,color=#2,author=#1]{#3}}
\fi
\ifnum\commentType=3
    \usepackage{pdfcomment} 
    \usepackage{todonotes} 
    \usepackage{silence}
    \newcommand{\customComment}[3]{\todo[color=#2!40,size=\small]{\textbf{#1:} #3}}
    \newcommand{\customTODO}[3]{\todo[inline,color=#2!40,size=\small]{\textbf{#1:} #3}}
\fi

\newcommand{\putX}{740}
\newcommand{\putY}{10}
\newcommand{\putW}{.04}
\newcommand{\imgW}{.16}
\newcommand{\imgWr}{.16}
\newcommand{\plotW}{.33}

\renewcommand{\d}{\,\text{d}}    

\newcommand{\voxPhaseSlice}{P}            

\newcommand{\vox}{\mathbf{v}}
\newcommand{\volPos}{\mathbf{x}}
\newcommand{\volAng}{\mathbf{\omega}}
\newcommand{\volExit}{\mathbf{y}}
\newcommand{\volPosInt}{\hat{\mathbf{x}}}
\newcommand{\volAngInt}{\hat{\mathbf{\omega}}}
\newcommand{\voxPos}{\mathbf{v}_{o}}
\newcommand{\voxExit}{\mathbf{v}_{i}}
\newcommand{\pixPosInt}{\volPos}
\newcommand{\pixAngInt}{\volAng}
\newcommand{\transmittanceFunc}{\tau}
\newcommand{\transmittance}[2]{\transmittanceFunc({#1},{#2})}
\newcommand{\emissionFunc}{\varepsilon}
\newcommand{\emission}[1]{\emissionFunc({#1})}
\newcommand{\emissionPrefilterFunc}{G}
\newcommand{\emissionPrefilter}[2]{\emissionPrefilterFunc({#1},{#2})}
\newcommand{\absorbCoefFunc}{\sigma_a}
\newcommand{\absorbCoef}[1]{\absorbCoefFunc({#1})}
\newcommand{\scatteringCoefFunc}{\sigma_s}
\newcommand{\scatteringCoef}[1]{\scatteringCoefFunc({#1})}
\newcommand{\absScatCoefFunc}{\sigma_t}
\newcommand{\absScatCoef}[1]{\absScatCoefFunc({#1})}
\newcommand{\phaseFunc}{f_p}
\newcommand{\phase}[3]{\phaseFunc({#1},{#2},{#3})}
\newcommand{\phaseTerm}[3]{f({#1},{#2},{#3})}
\newcommand{\phasePrefilterFunc}{F}
\newcommand{\phasePrefilter}[3]{\phasePrefilterFunc({#1},{#2},{#3})}
\newcommand{\voxPhaseFunc}{\rho}
\newcommand{\voxPhase}[3]{\voxPhaseFunc({#1},{#2},{#3})}
\newcommand{\voxAlbedoFunc}{\gamma} 
\newcommand{\voxAlbedo}[2]{\voxAlbedoFunc({#1},{#2})}
\newcommand{\radianceFunc}{L}
\newcommand{\radiance}[2]{\radianceFunc({#1},{#2})}
\newcommand{\intOver}[1]{\d{#1}}
\newcommand{\pix}{I}
\newcommand{\pixFilter}[1]{H_{\pix}}
\newcommand{\sphere}{4\pi}
\newcommand{\radianceBoundaryFunc}{L_{b}}
\newcommand{\radianceBoundary}[3]{\radianceBoundaryFunc({#1},{#2},{#3})}
\newcommand{\setAllVox}{V}
\newcommand{\setVox}[1]{V_{#1}}
\newcommand{\voxSum}{\sum_{\vox\in{\setVox{\volPos\volExit}}}\hspace{-1mm}}
\newcommand{\trackedCovFunc}{\Lambda}
\newcommand{\trackedCov}[3]{\trackedCovFunc_{#1}({#2},{#3})}
\newcommand{\voxCovFunc}{\alpha}
\newcommand{\voxCov}[3]{\voxCovFunc_{#1}({#2},{#3})}
\newcommand{\voxCoverage}{\voxCovFunc} 

\newcommand{\voxInd}[1]{\vox\left[{#1}\right]}
\newcommand{\boundaryTerm}{B}
\newcommand{\emissionTerm}{E}
\newcommand{\scatteringTerm}{S}

\newcommand{\pixSpatAng}{\mathcal{F}}
\newcommand{\transmittancePrefilterFunc}{T}
\newcommand{\transmittancePrefilter}[2]{\transmittancePrefilterFunc({#1},{#2})}

\newcommand{\pseudoScale}{$s$}
\newcommand{\pseudoVox}{$\vox$}
\newcommand{\pseudoInc}{$\volAng_i$}
\newcommand{\pseudoOut}{$\volAng_0$}
\newcommand{\pseudoSVO}{$\setAllVox$}
\newcommand{\pseudoImage}{$L$}
\newcommand{\pseudoWeights}{$\mathcal{W}$}
\newcommand{\pseudoData}{$\mathcal{D}$}
\newcommand{\pseudoLatent}{$\mathcal{L}$}
\newcommand{\pseudoShadow}{$\mathcal{S}$}
\newcommand{\pseudoLatentVec}{$\mathbf{r}$}
\newcommand{\pseudoBeamCoverage}{$\trackedCovFunc$}
\newcommand{\pseudoTransmittance}{$\transmittancePrefilterFunc$}
\newcommand{\pseudoPhaseFull}{$\phasePrefilterFunc$}
\newcommand{\pseudoScattering}{$\scatteringTerm$}
\newcommand{\pseudoEmission}{$\emissionTerm$}
\newcommand{\pseudoEmissionPrefilter}{$\emissionPrefilterFunc$}
\newcommand{\pseudoBoundary}{$\boundaryTerm$}
\newcommand{\pseudoPixel}{$\pix$}
\newcommand{\pseudoColor}{$\pixFlux$}
\newcommand{\pseudoVoxelList}{$\setVox{\pix}$}
\newcommand{\pseudoPhase}{$\voxPhaseFunc$}
\newcommand{\pseudoAlbedo}{$\voxAlbedoFunc$}
\newcommand{\pseudoCoverage}{$\voxCoverage$}

\newcommand{\rangeCompressor}[1]{T({#1})}
\newcommand{\estimate}[1]{\hat{#1}}
\newcommand{\scene}[1]{\textbf{#1}}

\newcommand{\pixFlux}{L_I}
\newcommand{\thetaOut}{\theta_0}
\newcommand{\thetaIn}{\theta_i} 
\newcommand{\phiOut}{\phi_0}
\newcommand{\phiIn}{\phi_i}   

\newcommand{\eggx}{EGGX} 
\newcommand{\sggx}{SGGX} 
\newcommand{\hybrid}{HybridLoD}  
\newcommand{\selfshadow}{MMSS} 
\newcommand{\ours}{DAP} 

\newcommand{\eggxResult}{lodGeoFull}  
\newcommand{\sggxResult}{lodVolFull}  
\newcommand{\hybridResult}{lodHybrid}  
\newcommand{\selfshadowResult}{lodSelfShadow}

\newcommand{\resultsDir}{DAP_Fig8_Complex_Comparisons_Stripped} 
\newcommand{\resultsDirHighRes}{DAP_Fig8_Complex_Comparisons_Stripped} 
\newcommand{\resultsDirSota}{DAP_Fig7_SotA_Comparisons_Stripped} 
\newcommand{\errorViz}{mse}

\newcommand{\suffixLRComplex}{-scale4-res16}
\newcommand{\suffixComplex}{-scale8-res1024}
\newcommand{\oursResult}{net}  

\gdef\useCroppedImages{0}

\gdef\cropInsets{0}

\usepackage[export]{adjustbox}
\usepackage{calc}
\usepackage{tabularx}
\usepackage{tikz}
\usepackage{xstring}

\newlength\beautyHeight
\newlength\beautyPixWidth
\newlength\beautyPixHeight
\newlength\insetvsep
\gdef\useInsetA{0}
\gdef\useInsetB{0}
\gdef\useInsetC{0}

\newcommand{\setInset}[6]{%
    \expandafter\gdef\csname useInset#1\endcsname{1}%
    \expandafter\gdef\csname inset#1Color\endcsname{#2}%
    \expandafter\gdef\csname crop#1X\endcsname{#3}%
    \expandafter\gdef\csname crop#1Y\endcsname{#4}%
    \expandafter\gdef\csname crop#1W\endcsname{#5}%
    \expandafter\gdef\csname crop#1H\endcsname{#6}%
}

\newcommand{\unsetInset}[1]{%
    \expandafter\gdef\csname useInset#1\endcsname{0}%
}

\newcommand{\addBeautyCrop}[8]{%
    \pdfpxdimen=\dimexpr 1 in/72\relax
    \def\beauty{%
        \let\cropR\relax%
        \let\cropB\relax%
        \newlength\cropR%
        \newlength\cropB%
        \setlength\cropR{{#3 px}-{#5 px}-{#7 px}}%
        \setlength\cropB{{#4 px}-{#6 px}-{#8 px}}%
        \sbox0{\includegraphics[width=#2\textwidth,trim={#5px {\cropB} {\cropR} #6px},clip]{#1}}%
        \begin{tikzpicture}
            \node[anchor=north west,inner sep=0] at (0,0) {\usebox0};
            \begin{scope}[x=\wd0/#7, y=\ht0/#8]
            \if\useInsetA1{
                \draw[\insetAColor,thick] (\cropAX-#5,-\cropAY+#6) rectangle + (\cropAW,-\cropAH);
            }\fi
            \if\useInsetB1{
                \draw[\insetBColor,thick] (\cropBX-#5,-\cropBY+#6) rectangle + (\cropBW,-\cropBH);
            }\fi
            \if\useInsetC1{
                \draw[\insetCColor,thick] (\cropCX-#5,-\cropCY+#6) rectangle + (\cropCW,-\cropCH);
            }\fi
            \end{scope}
        \end{tikzpicture}
    }%
    \setlength\beautyHeight{\heightof{\beauty}}%
    \setlength\beautyPixWidth{#3px}%
    \setlength\beautyPixHeight{#4px}%
    \global\beautyHeight=\beautyHeight%
    \global\beautyPixWidth=\beautyPixWidth%
    \global\beautyPixHeight=\beautyPixHeight%
    \begin{adjustbox}{valign=t}
        \beauty
    \end{adjustbox}
}

\newcommand{\addBeauty}[4]{%
    \addBeautyCrop{#1}{#2}{#3}{#4}{0}{0}{#3}{#4}%
}

\newcommand{\trimInset}[6]{%
    \let\cropR\relax%
    \let\cropB\relax%
    \newlength\cropR%
    \newlength\cropB%
    \setlength\cropR{{\beautyPixWidth}-{#3 px}-{#5 px}}%
    \setlength\cropB{{\beautyPixHeight}-{#4 px}-{#6 px}}%
    \color{#2}%
    \fbox{\includegraphics[width=1\linewidth,trim={{#3 px} {\cropB} {\cropR} {#4 px}},clip]{#1}}%
}

\newcommand{\addInset}[2]{%
    \color{#2}%
    \fbox{\includegraphics[width=1\linewidth]{#1}}%
}

\newcommand{\addImage}[3]{%
	\frame{\includegraphics[width=#3\linewidth]{#1}}%
}

\newcommand{\auxtimes}{x}
\newcommand{\auxplus}{+}
\newcommand{\auxspace}{ }

\newcommand{\addInsets}[1]{%
    \begin{adjustbox}{valign=t}
        \StrSubstitute{#1}{.}{-}[\baseFileName]
        \begin{adjustbox}{totalheight=1\beautyHeight,tabular={c}}
            \if\useInsetA1%
                \def\cropfile{\baseFileName-\cropAW\auxtimes\cropAH\auxplus\cropAX\auxplus\cropAY}
                \if\cropInsets1
                    \immediate\write18{convert #1 -crop \cropAW\auxtimes\cropAH\auxplus\cropAX\auxplus\cropAY\auxspace \cropfile.png}
                \fi
                \if\useCroppedImages1
                    \addInset{\cropfile.png}{\insetAColor}
                \else
                    \trimInset{#1}{\insetAColor}{\cropAX}{\cropAY}{\cropAW}{\cropAH}%
                \fi%
            \fi%
            \if\useInsetB1%
                \if\useInsetA1\\[\insetvsep]\fi%
                \def\cropfile{\baseFileName-\cropBW\auxtimes\cropBH\auxplus\cropBX\auxplus\cropBY}
                \if\cropInsets1
                    \immediate\write18{convert #1 -crop \cropBW\auxtimes\cropBH\auxplus\cropBX\auxplus\cropBY\auxspace \cropfile.png}
                \fi
                \if\useCroppedImages1
                    \addInset{\cropfile.png}{\insetBColor}
                \else
                    \trimInset{#1}{\insetBColor}{\cropBX}{\cropBY}{\cropBW}{\cropBH}%
                \fi%
            \fi%
            \if\useInsetC1%
                \if\useInsetB1\\[\insetvsep]\fi%
                \def\cropfile{\baseFileName-\cropCW\auxtimes\cropCH\auxplus\cropCX\auxplus\cropCY}
                \if\cropInsets1
                    \immediate\write18{convert #1 -crop \cropCW\auxtimes\cropCH\auxplus\cropCX\auxplus\cropCY\auxspace \cropfile.png}
                \fi
                \if\useCroppedImages1
                    \addInset{\cropfile.png}{\insetCColor}
                \else
                    \trimInset{#1}{\insetCColor}{\cropCX}{\cropCY}{\cropCW}{\cropCH}%
                \fi%
            \fi%
        \end{adjustbox}
    \end{adjustbox}
}

\newcommand{\addInsetsTwoSource}[2]{%
    \begin{adjustbox}{valign=t}
        \StrSubstitute{#1}{.}{-}[\baseFileName]
        \StrSubstitute{#2}{.}{-}[\baseFileNameTwo]
        \begin{adjustbox}{totalheight=1\beautyHeight,tabular={c}}
            \if\useInsetA1%
                \def\cropfile{\baseFileName-\cropAW\auxtimes\cropAH\auxplus\cropAX\auxplus\cropAY}
                \if\cropInsets1
                    \immediate\write18{convert #1 -crop \cropAW\auxtimes\cropAH\auxplus\cropAX\auxplus\cropAY\auxspace \cropfile.png}
                \fi
                \if\useCroppedImages1
                    \addInset{\cropfile.png}{\insetAColor}
                \else
                    \trimInset{#1}{\insetAColor}{\cropAX}{\cropAY}{\cropAW}{\cropAH}%
                \fi%
            \fi%
            \if\useInsetA1%
                \if\useInsetA1\\[\insetvsep]\fi%
                \def\cropfile{\baseFileNameTwo-\cropBW\auxtimes\cropBH\auxplus\cropBX\auxplus\cropBY}
                \if\cropInsets1
                    \immediate\write18{convert #2 -crop \cropBW\auxtimes\cropBH\auxplus\cropBX\auxplus\cropBY\auxspace \cropfile.png}
                \fi
                \if\useCroppedImages1
                    \addInset{\cropfile.png}{\insetAColor}
                \else
                    \trimInset{#2}{\insetAColor}{\cropAX}{\cropAY}{\cropAW}{\cropAH}%
                \fi%
            \fi%
            \if\useInsetC1%
                \if\useInsetB1\\[\insetvsep]\fi%
                \def\cropfile{\baseFileName-\cropCW\auxtimes\cropCH\auxplus\cropCX\auxplus\cropCY}
                \if\cropInsets1
                    \immediate\write18{convert #1 -crop \cropCW\auxtimes\cropCH\auxplus\cropCX\auxplus\cropCY\auxspace \cropfile.png}
                \fi
                \if\useCroppedImages1
                    \addInset{\cropfile.png}{\insetCColor}
                \else
                    \trimInset{#1}{\insetCColor}{\cropCX}{\cropCY}{\cropCW}{\cropCH}%
                \fi%
            \fi%
        \end{adjustbox}
    \end{adjustbox}
}

\acmJournal{TOG}
\acmVolume{1}
\acmNumber{1}
\acmArticle{1}
\acmYear{2022}
\acmMonth{1}
\acmPrice{}
\acmDOI{10.1145/3570327}

\setcopyright{acmcopyright}

\begin{document}
\title{Deep Appearance Prefiltering}


\author{Steve Bako}
\email{stevebako@ucsb.edu}
\affiliation{%
  \institution{University of California, Santa Barbara}
  \city{Santa Barbara}
  \country{USA}
}
\author{Pradeep Sen}
\email{psen@ucsb.edu}
\affiliation{%
  \institution{University of California, Santa Barbara}
  \city{Santa Barbara}
  \country{USA}
}
\author{Anton Kaplanyan}
\email{kaplanyan@gmail.com}
\affiliation{%
  \institution{Facebook Reality Labs}
  \city{Redmond}
  \country{USA}
}
\renewcommand\shortauthors{Bako, Sen, and Kaplanyan}

\ccsdesc[500]{Computing methodologies~Ray tracing}


\newcolumntype{Y}{>{\centering\arraybackslash}X}
\newcolumntype{P}{>{\raggedleft\arraybackslash}X}
\begin{teaserfigure}
    \center
    \begin{overpic}[width=\textwidth]{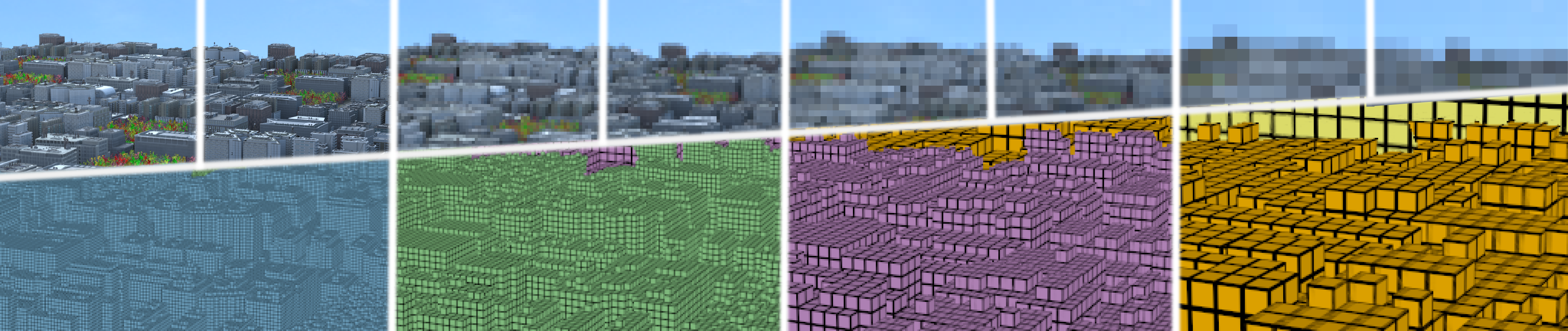}%
	\put(7,197){\color{white}\textbf{Ours}}%
	\put(140,197){\color{white}\textbf{RT}}
	\put(260,197){\color{white}\textbf{Ours}}
	\put(397,197){\color{white}\textbf{RT}}
	\put(510,197){\color{white}\textbf{Ours}}
	\put(643,197){\color{white}\textbf{RT}}
	\put(760,197){\color{white}\textbf{Ours}}
	\put(890,197){\color{white}\textbf{RT}}
    \end{overpic}\\
	\footnotesize%
    \begin{tabularx}{\textwidth}{@{}PPPP@{}}
    		\textbf{City} \hspace{29mm} Scale 8 & Scale 7 & Scale 6 & Scale 5 \\
    	\textbf{Per-pixel memory decrease:} \hspace{4mm} 28.1$\times$ & 44.6$\times$ & 155$\times$ & 528$\times$ \\
	\end{tabularx}
	\vspace{-3mm}
    \caption{We demonstrate the first neural framework for prefiltering the appearance of complex scenes without any access to the original geometry, materials, or textures, whereas a ray tracer (RT) would require the memory cost of the full scene. This allows for a significantly reduced memory footprint across pixels relative to the reference inputs to the ray tracer, as shown here for the \scene{City} scene, which uses the Disney BRDF and consists of over a million voxels in scale 8, the finest resolution. Our method is compared to ray tracing on the top, while the bottom shows the selected voxels \old{shaded}\new{colored} with the scale being accessed. 
    }
    \label{fig:Teaser}
\end{teaserfigure}
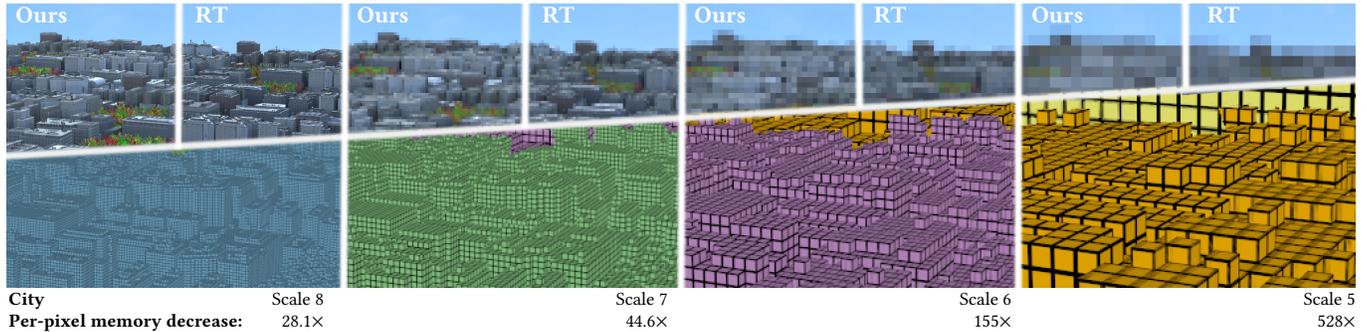

\begin{abstract}


Physically based rendering of complex scenes can be prohibitively costly with a potentially unbounded and uneven distribution of complexity across the rendered image. The goal of an ideal level of detail (LoD) method is to make rendering costs \emph{independent} of the 3D scene complexity, while preserving the appearance of the scene. However, current prefiltering LoD methods are limited in the appearances they can support due to their reliance of approximate models and other heuristics. We propose the first comprehensive multi-scale LoD framework for prefiltering 3D environments with complex geometry and materials (e.g., the Disney BRDF), while maintaining the appearance with respect to the ray-traced reference. Using a multi-scale hierarchy of the scene, we perform a data-driven prefiltering step to obtain an appearance phase function and directional coverage mask at each scale. At the heart of our approach is a novel neural representation that encodes this information into a compact latent form that is easy to decode inside a physically based renderer. Once a scene is baked out, our method requires no original geometry, materials, or textures at render time. We demonstrate that our approach compares favorably to state-of-the-art prefiltering methods and achieves considerable savings in memory for complex scenes.







\end{abstract}

\maketitle

\section{Introduction}

Photorealistic rendering of complex synthetic and captured 3D environments can have an unbounded rendering cost for scenarios such as those in the augmented and virtual reality (AR/VR), gaming, and film industries. The geometry of these scenes can be prohibitively complex, requiring significant amounts of both storage and computation~\cite{Savva19}, especially on immersive mobile devices where compute is limited and missing the performance target is not an option. Furthermore, both captured real-world environments and physically based synthetic scenes typically have sophisticated appearances due to complex materials and light transport. Thus, it is essential for rendering systems to keep costs within budget and make the rendering complexity more predictable.

One strategy is to perform an offline precomputation of the light transport falling within a given pixel to utilize during subsequent renderings and obtain an estimate of the pixel color with fewer samples than before, as is done in the \emph{prefiltering} framework for surfaces by Belcour et al.~\shortcite{Belcour17}. Another option is to apply level of detail (LoD) techniques that replace complex 3D assets with an ideally indistinguishable multi-scale approximation to determine the appearance of surfaces and volumes more efficiently. Seminal work in this area includes geometric simplification~\cite{Hoppe96,Xia97,Cohen97} and texture mip-mapping~\cite{Williams83}, among many others. Recent work takes a more holistic approach by prefiltering multiple scene parameters simultaneously, essentially aiming at preserving the overall appearance of the scene at different scales~\cite{Heitz15,Dong15,Loubet18}. 


However, current approaches have numerous limitations. Firstly, the types of materials that can be accurately prefiltered with existing models are limited. For example, effects like specularities and high-frequency glints are difficult to approximate, as these effects show up as unique shapes or spikes within a narrow solid angle. Moreover, materials with sophisticated BRDFs, such as the Disney BRDF, typically cannot be captured easily since the lobes are often combined in a non-linear way and mapping from arbitrary BRDF parameters to a volumetric representation is non-trivial. Finally, local occlusions are also difficult to fully capture in previous approaches.
%

\begin{figure}[t]
	\centering
	\setlength{\fboxrule}{10pt}%
	\setlength{\insetvsep}{20pt}%
	\setlength{\tabcolsep}{1pt}
	\renewcommand{\arraystretch}{1}%
	\footnotesize%
		

		
	\includegraphics[width=\linewidth]{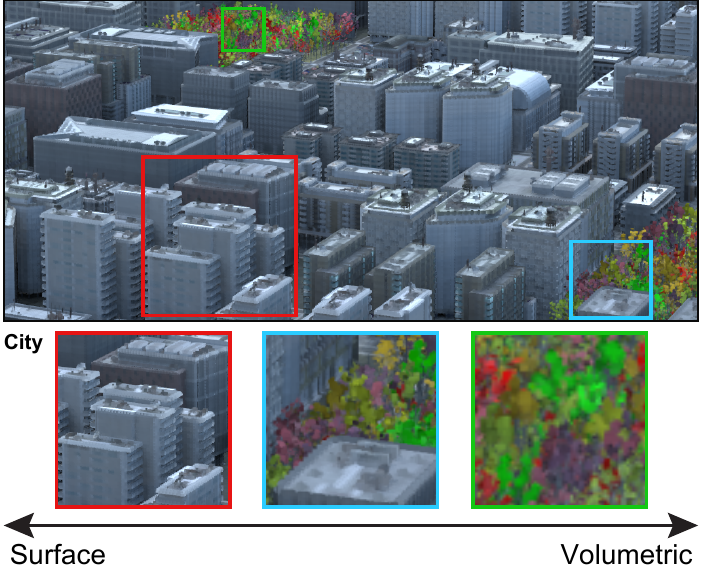}
	\caption{The regions of a scene often exist along a continuum with macrosurfaces corresponding to watertight meshes on one end and an aggregate microgeometry on the other. For prefiltering level of detail, geometric simplification is applied to large surfaces, while volumetric appearance models, such as microflakes, are used for the latter. Although most methods typically have to decide between one or the other, recent hybrid approaches can use a combination of the two for different portions of the same scene. Still, these methods are susceptible to misclassification of regions and are bounded by the limitations of the approximate models. Our approach avoids heuristics and accurately represents the appearance \new{of the scene}, effectively handling all scenarios along the spectrum implicitly.} 
	\label{fig:Motivation}
	\vspace{-6mm}
\end{figure}

Another limitation of existing prefiltering work is that at each particular scale there can be a mixture of both macroscale geometry, more amenable to techniques including edge simplification and microfacet approximations, and aggregate microgeometry, for which a volumetric representation such as microflakes is proven to be more suitable. Ideally, a prefiltering method would not have to choose between different representations. The state-of-the-art hybrid approach by Loubet and Neyret~\shortcite{Loubet17} uses geometric analysis to label regions of a scene as either belonging to macrogeometry or an aggregate microgeometry in order to perform either the geometric or volumetric prefiltering, respectively. Although this provides improvements by enabling a more fine-grained heterogeneous representation, the heuristics used to decide on a label can misclassify regions that do not clearly belong to one or the other representation, which again results in a different appearance. There can be regions within a single asset that straddle both macro and microgeometry and where selecting a single representation could have errors (see Fig.~\ref{fig:Motivation}). Thus, such approaches can neither represent the continuum of in-between cases nor smoothly transition between them, which is important for the practical rendering of complex scenes. 

To overcome the aforementioned issues and to robustly handle a wide range of appearances, we propose a multi-scale hierarchy of local neural representations that can preserve the appearance of scenes containing both complex geometry and materials. We use a standard, multi-scale sparse voxel octree~\cite{Laine10}, or SVO, as the data structure for our representation. This allows us to render without the original geometry or materials once an initial offline data generation step is completed, during which a ray tracer captures the light transport within each voxel. Our network then learns to represent the rendered appearance within a voxel as a lightweight latent representation, which can be efficiently evaluated within a standard physically based renderer. 

To summarize, the precomputation stage of our framework voxelizes a given scene, computes per-voxel reflectance data, and trains a single network to compress this information across all voxels of the specific scene. We note that our network framework is trained per scene on data that takes days to precompute and does not generalize to new scenes (i.e., new scenes require performing data generation and training again). Afterwards, at runtime, a beam tracer is used to find all voxels falling within a pixel's footprint and the corresponding compressed information is efficiently decoded by our network to accurately determine the voxel's contribution and, ultimately, the color of the pixel using only a fraction of the original memory. 


Our current system serves as a proof-of-concept that we hope inspires a new, compelling research direction: machine learning for prefiltering. Although this is a promising area to explore, there are numerous challenges to overcome, and, thus, in this initial prototype, we seek to address some of the most important aspects of the problem and leave other issues as limitations to be addressed in future work. First, we focus on rendering the directly visible appearance, which is often the most sensitive to artifacts and, therefore, quite challenging. However, in Sec.~\ref{sec:TheoreticalOverview}, we present the framework in the context of general light transport to show how the framework can be extended to arbitrary bounces, not just the first bounce shown in the results. In addition, as is common in prefiltering and LoD methods, we assume that the prefiltered portion of the scene is mostly static and can be sufficiently reused across multiple renderings for users to reap the full benefits of our approach.

Thus, our results have limitations include handling only the primary bounce on scenes using simple homogeneous materials (i.e., all objects use the Disney BRDF, albeit with different artist-set parameters). Although our theoretical framework supports global illumination, multiple bounces are omitted from our implementation and results due to the extra computational overhead required during rendering. Our method captures local shadows inherently during voxel data generation, while global occlusions are handled through shadow maps that we generate. \new{Currently,} our unoptimized implementation is bottlenecked by network inference, which can be alleviated by GPU parallelization. In its current form, our approach is only faster than ray tracing at coarser LoD scales. 



In summary, our work makes the following contributions:
\begin{itemize}
	\item We introduce the first neural method for appearance prefiltering of large complex scenes, requiring only data structure traversals and network inference to generate an image without using the original geometry or materials at render time.
	\item Our data-driven representation can preserve hard-to-render appearances resulting from sophisticated materials, complex geometry, local occlusions, and high-frequency, view-dependent effects such as specularities by capturing the phase function more accurately than current approximate models.
	\item We apply a learning-based compression of rendered data to enable its efficient utilization by a physically based renderer.
	\item We preserve the full appearance in a single unified representation rather than using either geometric or volumetric simplification techniques that cannot robustly handle all cases. 
\end{itemize}

\section{Previous Work}

%

Relevant to our approach are the various appearance models and the sophisticated effects they capture, the prefiltered approximations to these models, and recent applications of machine learning.


\subsection{Appearance models}
\label{sec:AppearanceBackground}


\subsubsection*{Background}

For brevity, we only highlight appearance models most relevant to our discussion. For more in-depth information, we refer readers to texts on the subject~\cite{Pharr16,Akenine18} and a survey on volumetric rendering~\cite{Novak18}. 


\subsubsection*{Microfacet materials}

Microfacet BRDF models can be used with explicit geometry to render compelling effects, such as glossy reflections and glints. Yan et al.~\shortcite{Yan14} propose to use high resolution normal maps to render effects such as glints on highly specular materials, and later improve the efficiency by approximating details with Gaussians~\cite{Yan16}. Jakob et al.~\shortcite{Jakob14} demonstrate procedurally generated glints that are efficient, but do not scale to larger structures. Zirr and Kaplanyan~\shortcite{Zirr16} procedurally generate materials to enable real-time rendering of glints and brushed materials at multiple scales, while Raymond et al.~\shortcite{Raymond16} use a spatially varying BRDF model to render scratched metals at multiple scales. 

Recently, there have been improvements in the microfacet-based BRDF for rendering highly specular materials especially at grazing angles~\cite{Chermain19} and also a patch-based extension to better handle glint rendering~\cite{Chermain19b}. Since our framework bakes out the appearance from a standard ray tracer, it can be used in combination with these sophisticated BRDFs and even multi-lobe models such as the Disney BRDF~\cite{Burley12}, which we demonstrate. Furthermore, another important advantage is that our neural framework does not require access to the original scene, allowing us considerable savings in memory.

\subsubsection*{Microflake theory}
 Jakob et al.~\shortcite{Jakob10b} derive the anisotropic radiative transfer equation to allow for physically based rendering of anisotropic participating media. This enables objects to be rendered as scattered particles in a volume rather than with a microfaceted complex geometry, which is more expensive to compute. The model is widely used for items including woven fabric, since it accurately preserves such appearances at fine scales~\cite{Zhao11,Zhao12}. However, some appearances are difficult to model such as glints and self-occlusions that are naturally accounted for in our approach.


\subsubsection*{Correlated transmittance in volumetric light transport}
Properly accounting for transmittance correlation in anisotropic volumes induced by surfaces is a non-trivial and open research topic. Bitterli et al.~\shortcite{Bitterli18} introduce new transmittance models to account for correlated scattering in non-exponential media. Other work by Jarabo et al.~\shortcite{Jarabo18} extend the Generalized Boltzmann Equation to use the Radiative Transfer Equation (RTE) and account for spatially correlated media. Guo et al.~\shortcite{Guo19} propose a new transmittance model to account for correlation in RTE frameworks by using fractional Gaussian fields. Our approach is orthogonal to these works and, in theory, could utilize these more sophisticated models. However, we decided to use the advantage of a data-driven representation and store a simple, spatial coverage mask to account for correlation, similar to Heitz et al.~\shortcite{Heitz12}.


\vspace{-1mm}
\subsection{Prefiltering and level of detail}
\label{sec:PrefilteringBackground}


\subsubsection*{Background}
Until recently, level of detail approaches primarily consisted of either geometric simplification in the case of macrosurfaces or the mapping to a volumetric model for microgeometry. Mixtures of the two have been proposed by recent hybrid approaches. We discuss each in turn, but focus on the most relevant approaches and refer interested readers to other sources for more information~\cite{Luebke01,Luebke03,Bruneton11}.

\vspace{-1mm}
\subsubsection*{Geometric simplification}

Seminal work in this area~\cite{Hoppe96,Xia97,Cohen97,Cohen98,Cohen99,Luebke06,Yoon06} take a mesh as input and reduce their complexity by merging vertices or deleting edges. For assets with large surfaces without a significant amount of sub-resolution detail, these approaches can perform quite well and capture the appearance faithfully. However, it is becoming commonplace that for highly detailed models with complex materials applying a mesh-based simplification will not capture these microscale structures or interesting effects such as glints or sharp specularities. 

Cook et al.~\shortcite{Cook07} use random pruning to simplify scenes with aggregate detail, which are many small, similar objects next to each other that form a complex whole (e.g., sand or foliage), but assume geometric properties are uniformly distributed, an assumption that is frequently violated. Prefiltering occlusions has also been examined~\cite{Lacewell08}, but this approach involves some prior knowledge about the scene and which discrete scales are viable candidates for their occlusion model. Simplification of textured meshes~\cite{Williams03} has also been explored.

There has also been work in hierarchical voxel representations to approximate geometry~\cite{Crassin09,Crassin11,Laine10,Heitz12,Palmer14}. These approaches use a sparse voxel octree (SVO) data structure for prefiltering assets in an appearance-preserving way for LoD by storing surface attributes within each voxel. We use the SVO for storing our neural representations by replacing the simplistic models with our phase functions, base colors, and coverage masks that are precomputed with a ray tracer and compressed/rendered by a neural network. Moreover, we can simplify scenes that have complex materials and microdetail which are difficult to approximate in such frameworks.

\subsubsection*{Material reflectance and aggregate detail prefiltering}

There has been work on prefiltering in surface space to preserve the appearance of complex materials and assets. Dong et al.~\shortcite{Dong15} use an ellipsoid normal distribution function as a fast approximation in a microfacet reflectance model for rendering metals. Kaplanyan et al.~\shortcite{Kaplanyan16} filter the normal distributions on curved surfaces to preserve specular highlights during real-time rendering. Other work explores representing the normal distributions of multi-lobe BRDFs with a mix of Gaussian, cosine, or vMF lobes~\cite{Tan05,Tan08,Xu17}.

Also worth noting is the work by Heitz et al.~\shortcite{Heitz15} that introduces the SGGX microflake distribution to efficiently model anisotropic microflake participating media and prefilter the distribution of visible normals. The recent self-shadow microflake model further builds on SGGX by taking into account occlusions inside a volume for the application of downsampling~\cite{Loubet18,Loubet18b}. Such microflake models are often applied for prefiltering aggregate detail, but with a focus on a specific subset of cases such as foliage~\cite{Max97,Loubet17}, fabric~\cite{Schroder11,Zhao11,Zhao13,Zhao16}, hair~\cite{Moon08}, or granular materials~\cite{Meng15,Muller16}.

A common issue with these models is that, although they are fast, they have a relatively simplistic underlying reflectance or phase function that cannot faithfully capture effects such as specularities or glossy reflections. Meanwhile, our approach better preserves the appearance by more accurately representing the true reflectance obtained through standard ray tracing.

\subsubsection*{Hybrid techniques}
Far voxels~\cite{Gobbetti05} is among the first hybrid approaches and it combines different representations based on scale, often relying on geometry for finer scales and volumetric representations at coarse scales. The recent approach by Loubet and Neyret~\shortcite{Loubet17} is the first fully hybrid system that can do heterogeneous simplification of an asset at a given scale by performing analysis on a surface mesh to label portions as candidates for either geometric simplification or a voxelized volumetric model.

Our framework avoids the difficult problem of trying to unify the two representations~\cite{Dupuy16} or selecting between them (i.e., either a simplified geometry or a volumetric voxel) and, as a result, we can handle the continuum of in-between cases, as shown in Fig.~\ref{fig:Motivation}. Noma~\shortcite{Noma95} also sought to bridge this gap to preserve the appearance of collections of surfaces using volume textures to enable rendering such assets both closeup and far away on multi-resolution displays. Surfaces containing rough detail or holes are difficult to accurately classify with heuristics and cannot be properly represented in an appearance-preserving way by exclusively relying on either prefiltering model. We use only a single representation that bakes out the appearance and bypasses such issues.

\subsubsection*{Visibility}
Accounting for visibility is important for determining occlusions and shadows in level of detail applications. For example, Meyer et al.~\shortcite{Meyer01} utilize a visibility cube map corresponding to each level detail to enable shadowing on trees. Coverage maps~\cite{Lokovic00} have been used to track visibility for anti-aliasing~\cite{Crassin18}. We similarly employ coverage maps to handle the transmittance for the multitude of scenarios that are potentially encountered when rendering a prefiltered asset.

\subsubsection*{Recent approaches}
Some recent approaches target prefiltering for certain materials and effects. For example, Wu et al.~\shortcite{Wu19} jointly prefilters displacement-mapped surfaces and their BRDFs to preserve their appearance along with shadows and interreflections. Meanwhile, Gamboa et al.~\shortcite{Gamboa18} capture high frequency lighting effects for global illumination by using an appearance model that can tractably account for micronormal variation.

\subsection{Machine learning}

\subsubsection*{Background.} Machine learning using neural networks has exploded in popularity since its demonstration as a practical solution for image-based classification~\cite{Krizhevsky12}. Since then, they have achieved state-of-the-art results in countless applications within the fields of computer vision and graphics, among others. See recent texts~\cite{Goodfellow16} for a more detailed background.

\subsubsection*{For graphics applications.}
Learning-based approaches have been successfully applied to Monte Carlo (MC) denoising~\cite{Kalantari15,Bako17,Chaitanya17,Vogels18}. We too apply this concept to denoise our phase functions which are undersampled to save on computation (see Fig.~\ref{fig:Denoising}). Machine learning approaches are used to capture global illumination for relighting~\cite{Ren13,Ren15}, evaluate complex luminaires~\cite{Zhu21}, represent sky models~\cite{Satylmys17}, render clouds~\cite{Kallweit17}, handle multiple scattering in participating media~\cite{Ge18} and subsurface scattering~\cite{Vicini19}, synthesize materials~\cite{Zsolnai-Feher18}\old{or represent mipmapped materials~\cite{Kuznetsov21}}, generate glints~\cite{Kuznetsov19}, and to perform the rendering itself~\cite{Granskog20,Granskog21}. We also utilize networks to render similar anisotropic view-based effects such as glints, but without relying on approximate appearance models. \new{Takikawa et al.~\shortcite{Takikawa21} focus on LoD for the reconstruction of implicit geometry represented through signed distance functions. Kuznetsov et al.~\shortcite{Kuznetsov21} leverage neural networks that are functions of incoming and outgoing directions, but they focus on learning complex, precomputed material responses that can then be mapped onto meshes. On the other hand, our approach learns the entire phase function across voxelized representations of the scene capturing the full appearance including materials, geometry, and occlusions.} 

Finally, neural networks have been applied for lightfield view synthesis~\cite{Kalantari16,Wang17,Mildenhall19} and dense reconstruction of sparse lightfield videos~\cite{Bemana19}. \old{We subsample our 4D phase function across the outgoing direction to save on computation, and thus our network learns to interpolate and synthesize in-between views similarly. However, one fundamental difference of our approach with recent works such as the NeRF architecture of Mildenhall et al.~\shortcite{Mildenhall19} is that we also account for the lighting direction and, therefore, our results can be re-lit. Another difference is that instead of individual rays, we utilize beams to integrate over the entire pixel footprint at once.}\new{In particular, the seminal work by Mildenhall et al.~\shortcite{Mildenhall19} created the NeRF architecture, which poses the view synthesis application as an optimization over a volumetric scene function using sparse RGB inputs. The resulting network can be queried by a 5D coordinate over spatial location and viewing direction. The novel take on view synthesis coupled with its state-of-the-art results inspired a large research field exploring NeRF-like architectures to further improve performance and quality. For example, Liu et al.~\shortcite{Liu20} use a sparse voxel representation to discard irrelevant parts of the scene and learn simpler local properties. The Mip-NeRF system~\cite{Barron21} renders anti-aliased conical frustums across scales rather than individual rays. Our work shares some similarities including the multi-scale, sparse voxel representation that integrates over the pixel footprint, but ours additionally supports relighting. Recent NeRF approaches now support relighting including NeRF-Tex~\cite{Baatz21}, which captures texture information within the networks that can be applied to a mesh. NeRV~\cite{Srinivasan21} also supports relighting but only focuses on large watertight surfaces and cannot represent small microgeometry. Furthermore, the reconstruction does not have high fidelity to the reference nor does it support LoD.}

\subsubsection*{For embedding and compression.}
A subset of approaches, including some in graphics, use networks to encode or compress the input data into a latent representation~\cite{Hinton06}. A decoder network could act on this compressed representation, commonly referred to as a latent feature vector, to reconstruct the data or a subset of the data. For example, Miandji et al.~\shortcite{Miandji13} compress light fields for real-time global illumination, while Chen et al.~\shortcite{Chen18} and Kang et al.~\shortcite{Kang18} encode reflectance capture. Furthermore, encoding networks are used for appearance models for face rendering~\cite{Lombardi18}, image-based relighting from optimal sparse samples~\cite{Xu18}, and appearance maps~\cite{Maximov19}. See a recent survey~\cite{Dong19} for more details.


Recently, a related approach compresses the bidirectional texture function (BTF) with a neural network~\cite{Rainer19} utilizing an encoder/decoder architecture. This method similarly queries a latent vector to directly obtain the appearance for specific view and lighting directions. Inspired by such embedding approaches, we compress/encode our voxel data to keep storage sizes tractable and to practically decode the data during runtime for appearance prefiltering. Rainer et al.~\shortcite{Rainer19} use a compression network per BTF, while in our framework we use a single network per scene, each of which has on the order of one million voxels with different properties and effects captured within each. After training, our decoder networks are subsequently utilized by a beam tracer, rather than a typical ray tracer, to efficiently evaluate the contribution of voxels falling within a pixel's footprint.

\section{Prefiltered Light Transport}
\label{sec:TheoreticalOverview}

\begin{figure}[t]
	\centering
	\includegraphics[width=\linewidth]{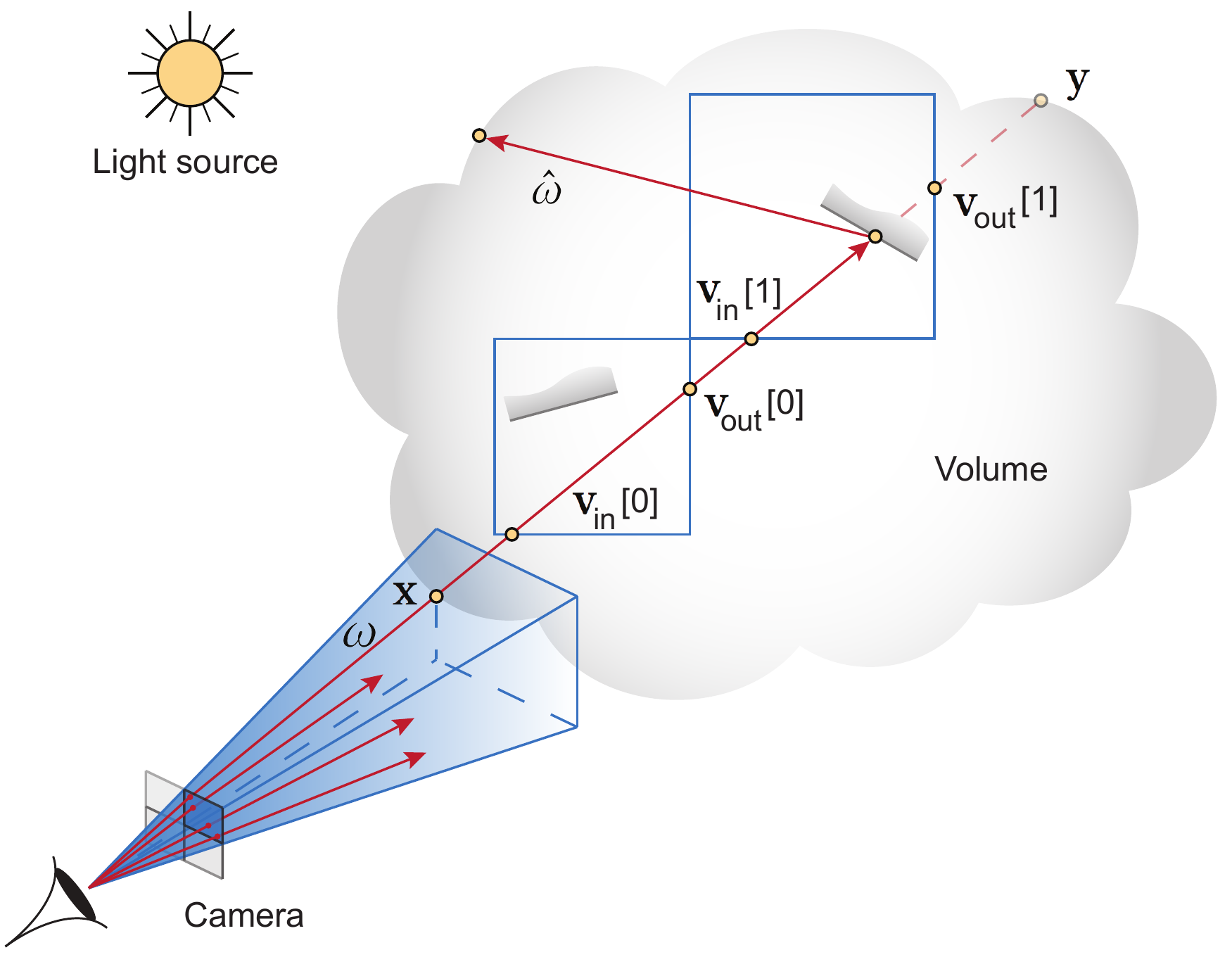}

	\vspace{-3mm}
    \caption{Our approach prefilters the appearance of an asset by treating it as a voxelized volume through which we trace rays to integrate over a pixel's spatio-angular footprint (denoted by the blue beam). We determine the light reaching the camera sensor according to the volume rendering equation, but with precomputed terms used to calculate the amount of light propagated across each voxel's boundary, indicated by the \textit{in} and \textit{out} labels. }
	\label{fig:TheoreticalOverview}
	\vspace{-6mm}
\end{figure}


We now present the theoretical framework for our prefiltering approach. Since we use volumetric structures, we begin with the Radiative Transfer Equation (RTE) ~\cite{Chandrasekhar60} in its integral form for volumetric rendering:

\begin{align}
\label{eq:VolumeRendering}
\radiance{\volPos}{\volAng} &= \int_{\volExit}^{\volPos} \transmittance{\volPosInt}{\volPos} \left[ \emission{\volPosInt} + \scatteringCoef{\volPosInt} \int_{\sphere} \phase{\volPosInt}{\volAng}{\volAngInt} \radiance{\volPosInt}{\volAngInt} \intOver{\volAngInt} \right] \intOver{\volPosInt} \nonumber \\ &+ ~\transmittance{\volExit}{\volPos}\radiance{\volExit}{\volAng},
\end{align}

\noindent where $\radianceFunc$ is the radiance at a point in a particular direction, $\volPos$ is a position in the volume, $\volAng$ is the outgoing direction, $\volExit$ is a point on the volume boundary, $\emissionFunc$ is the emission at every point, $\scatteringCoefFunc$ is the scattering coefficient, $\phaseFunc$ is the phase function, and $\transmittanceFunc$ models the transmittance by accounting for absorption and out-scattering.

After expanding the products out, we can see three distinct terms in the equation:

\vspace{-2mm}
\begin{align}
\label{eq:VolumeRendering3Terms}
\radiance{\volPos}{\volAng} &= \int_{\volExit}^{\volPos} \transmittance{\volPosInt}{\volPos} \emission{\volPosInt} \intOver{\volPosInt} \nonumber \\ &+ \int_{\volExit}^{\volPos} \int_{\sphere} \transmittance{\volPosInt}{\volPos} \scatteringCoef{\volPosInt} \phase{\volPosInt}{\volAng}{\volAngInt} \radiance{\volPosInt}{\volAngInt} \intOver{\volAngInt}  \intOver{\volPosInt} \nonumber \\ &+ ~\transmittance{\volExit}{\volPos}\radiance{\volExit}{\volAng}.
\end{align}

\noindent The first term is the contribution from emission, the second term is the in-scattered light, and the last term corresponds to the volume's boundary condition, which we call $\radianceBoundaryFunc$ for brevity.

If we discretize the volume into a set of voxels, $\setAllVox$, and define the subset $\setVox{\volPos\volExit} = \lbrace \vox \in{\setAllVox} ~\vert ~\volPos \leq \vox_a \leq \volExit \rbrace$ where $a$ is a point inside the voxel (i.e., the set of all voxels that lie in the interval $\left[ \volPos, \volExit \right]$), we can rewrite the previous equation as a sum of integrals: 

\vspace{-2mm}
\begin{align}
\radiance{\volPos}{\volAng} &= \voxSum \int_{\voxExit}^{\voxPos} \transmittance{\volPosInt}{\volPos} \emission{\volPosInt} \intOver{\volPosInt} \nonumber \\ 
&+ \voxSum \int_{\voxExit}^{\voxPos} \int_{\sphere} \transmittance{\volPosInt}{\volPos}  \phaseTerm{\volPosInt}{\volAng}{\volAngInt} \radiance{\volPosInt}{\volAngInt} \intOver{\volAngInt}  \intOver{\volPosInt} \nonumber \\ 
&+ ~\radianceBoundary{\volExit}{\volPos}{\volAng},
\label{eq:VolumeRendering3TermsVoxels}
\end{align}

\noindent where $\phaseTerm{\volPosInt}{\volAng}{\volAngInt} = \scatteringCoef{\volPosInt} \phase{\volPosInt}{\volAng}{\volAngInt}$ is used for conciseness and $\voxExit$ and $\voxPos$ refer to the {\em in} and {\em out} points of voxel $\vox$ along direction $\volAng$.

To introduce the prefiltering of light transport in path space, we follow Belcour et al.~\shortcite{Belcour17} and introduce the notion of a pixel footprint. However, unlike this prior work, we formulate the expression for volumes, instead of surfaces (see Fig.~\ref{fig:TheoreticalOverview}). Thus, we can determine the final flux incident on a sensor's pixel, $\pixFlux$, by integrating the radiance incident on the sensor over the pixel filter, $\pixFilter{0}$:

\vspace{-2mm}
\begin{align}
\label{eq:PixelFlux}
\pixFlux &= \int_{\pixSpatAng} 
        \pixFilter{0}(\volPos,\volAng)\radiance{\volPos}{\volAng}\intOver{\pixAngInt}\intOver{\pixPosInt} = \emissionTerm + \scatteringTerm + \boundaryTerm.
\end{align}

\noindent Here $\pixSpatAng$ is the spatio-angular footprint of pixel $I$, as similarly defined in Belcour et al.~\shortcite{Belcour17}, which is introduced to integrate over image positions and visible directions within the extent of the pixel. Using the three terms of Eq.~\ref{eq:VolumeRendering3TermsVoxels} in place of $\radiance{\volPos}{\volAng}$ yields the filtered emission, scattering, and boundary terms (defined as $\emissionTerm$, $\scatteringTerm$, and $\boundaryTerm$, respectively), which we discuss in turn.

\subsubsection*{Boundary term} 
The propagated radiance received at the volume's boundary is simply multiplied by the pixel filter and is given by: 

\vspace{-2mm}
\begin{align}
\label{eq:BoundaryTerm}
\boundaryTerm = \int_{\pixSpatAng} 
        \pixFilter{0}(\volPos,\volAng) \radianceBoundary{\volExit}{\volPos}{\volAng}\intOver{\pixAngInt}\intOver{\pixPosInt}.
\end{align}

\noindent Therefore, even though our framework prefilters an asset by treating it as a volume, it can be plugged into existing hybrid renderers and will properly transport incoming radiance through the volume. Thus, in practice, the integral in Eq.~\ref{eq:BoundaryTerm} would be evaluated with distributed ray tracing~\cite{Cook84}.

\subsubsection*{In-scattering term}

\begin{figure}[t]
	\centering
	\includegraphics[width=\linewidth]{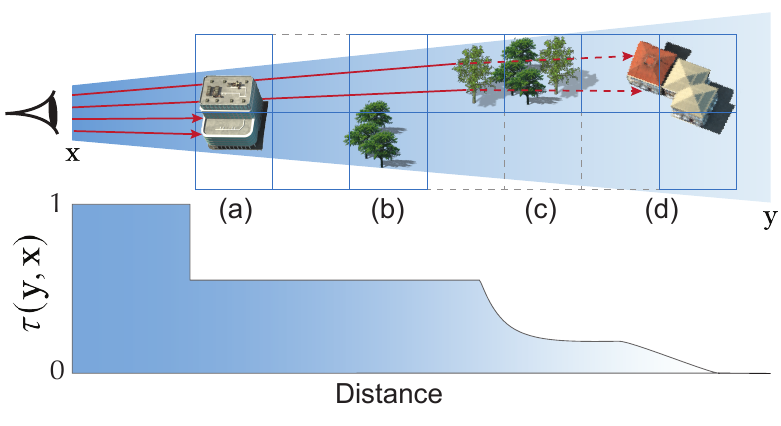}
	\vspace{-8mm}
	\caption{Motivation for our mask-based transmittance model. The top half of the figure shows a top-down view of a beam intersecting with voxels in a scene (non-empty voxels are shown in blue, and ignored (empty) voxels are shown with dashed gray lines), while the bottom half shows the ideal tracked transmittance versus distance along the beam based on the micro and macrogeometry within each voxel. (a) First, we have a large watertight building that occludes half of the beam and the transmittance decreases to 0.5. (b) Next, the tree in this region is completely covered by the previous building, so there is no change in transmittance. (c) The voxels contain trees with many random, small leaves that in aggregate behave similar to a volume, and, thus, an exponential transmittance model is suitable for this region. (d) Finally, there are non-axis-aligned buildings. These macrosurfaces will linearly influence the transmittance. Therefore, in such a typical scenario, no single traditional transmittance model is entirely appropriate, so we instead utilize a spatial coverage mask to track the occlusions along the beam and determine the radiance contribution of each voxel.} 
	\label{fig:BeamTransmittance}
	\vspace{-4mm}
\end{figure}

We next apply our pixel filter to the in-scattering term and rearrange it to get:

\vspace{-2mm}
\begin{align}
\label{eq:ScatteringTerm}
\scatteringTerm = \voxSum \int_{\pixSpatAng} \hspace{-1mm}
        \pixFilter{0}(\volPos,\volAng) \hspace{-1mm} \int_{\voxExit}^{\voxPos} \hspace{-2mm} \int_{\sphere} \hspace{-1mm} \transmittance{\volPosInt}{\volPos}  \phaseTerm{\volPosInt}{\volAng}{\volAngInt} \radiance{\volPosInt}{\volAngInt} \intOver{\volAngInt}  \intOver{\volPosInt}\intOver{\pixAngInt}\intOver{\pixPosInt}. 
\end{align}

\noindent We then apply the \emph{far-field} assumption commonly used in prefiltering (i.e., we assume the voxel is sufficiently far away and the rays in the beam are near parallel), so the incident radiance $\radiance{\volPosInt}{\volAngInt}$ becomes constant across voxel $\vox$ (i.e., in the range $\left[\voxPos, \voxExit \right]$), and we can parameterize it as $\radiance{\vox}{\volAngInt}$, which represents the radiance arriving at the corresponding voxel $\vox$. After rearranging the integral, we have:

\vspace{-2mm}
\begin{align}
\label{eq:ScatteringTermVoxRad}
\scatteringTerm = \hspace{-1mm}\voxSum \int_{\sphere} \hspace{-1mm} \radiance{\vox}{\volAngInt} \hspace{-1mm} \int_{\pixSpatAng} \hspace{-1mm} \pixFilter{0}(\volPos,\volAng) \hspace{-1mm} \int_{\voxExit}^{\voxPos} \hspace{-2mm} \transmittance{\volPosInt}{\volPos}  \phaseTerm{\volPosInt}{\volAng}{\volAngInt}    \intOver{\volPosInt}\intOver{\pixAngInt}\intOver{\pixPosInt}\intOver{\volAngInt}. 
\end{align}

\noindent Note that $\transmittance{\volPosInt}{\volPos}$ is parameterized relative to the point in the volume $\volPos$, \textit{not} the voxel boundary $\voxPos$, so it could be outside the voxel $\vox$. This poses a potential obstacle to our prefiltering application, as $\transmittance{\volPosInt}{\volPos}$ cannot be reasonably precomputed, unless we can factor $\transmittanceFunc$ into the portion inside and outside the voxel:

\vspace{-4mm}
\begin{align}
\label{eq:TransmittanceSplit}
\transmittance{\volExit}{\volPos} = \transmittance{\volExit}{\voxPos} \transmittance{\voxPos}{\volPos}.
\end{align}

\noindent This is a reasonable assumption which holds true for the widely-used exponential transmittance with the extinction coefficient, $\absScatCoefFunc$, given by the sum of the absorption and scattering coefficients (i.e., $\absScatCoef{\volPos} = \absorbCoef{\volPos} + \scatteringCoef{\volPos}$):

\vspace{-2mm}
\begin{align}
\label{eq:ExpTransmittanceSplit}
\transmittance{\volExit}{\volPos} &= \text{exp}(-\int_{\volExit}^{\volPos} \hspace{-2mm} \absScatCoef{\volPosInt} \intOver{\volPosInt}~) \nonumber \\
&= \text{exp}(-\int_{\volExit}^{\voxPos} \hspace{-3mm} \absScatCoef{\volPosInt} \intOver{\volPosInt}~) ~\text{exp}( -\int_{\voxPos}^{\volPos} \hspace{-3mm} \absScatCoef{\volPosInt} \intOver{\volPosInt}~) = \transmittance{\volExit}{\voxPos} \transmittance{\voxPos}{\volPos}.
\end{align}

\noindent Thus, using Eq.~\ref{eq:TransmittanceSplit}, we can rewrite the scattering term as:

\vspace{-2mm}
\begin{align}
\scatteringTerm &= \hspace{-2mm} \voxSum \hspace{-0.5mm} \int_{\sphere} \hspace{-2mm} \radiance{\vox}{\volAngInt} \hspace{-1mm} \int_{\pixSpatAng} \hspace{-1.5mm} \pixFilter{0}(\volPos,\volAng) \transmittance{\voxPos}{\volPos} \hspace{-1mm} \int_{\voxExit}^{\voxPos} \hspace{-3.5mm} \transmittance{\volPosInt}{\voxPos} \phaseTerm{\volPosInt}{\volAng}{\volAngInt} \intOver{\volPosInt}\intOver{\pixAngInt}\intOver{\pixPosInt}\intOver{\volAngInt} \nonumber \\
&\approx \hspace{-2mm} \voxSum \hspace{-0.5mm} \int_{\sphere} \hspace{-2mm} \radiance{\vox}{\volAngInt} ~\transmittancePrefilter{\voxPos}{\volPos} ~\phasePrefilter{\vox}{\pixAngInt}{\volAngInt} \intOver{\volAngInt}.
\label{eq:ScatteringTermPrefilter}
\end{align}

\noindent The approximation stems from splitting the integral of the product as the product of the integrals to obtain our \textit{prefiltered transmittance function}, $\transmittancePrefilterFunc$, and our \textit{prefiltered phase function},\footnote{This is a slight abuse of notation as it is really the phase function, $\phaseFunc$, multiplied by the transmittance, $\transmittanceFunc$, and scattering coefficient, $\scatteringCoefFunc$, across a voxel, but it can still be thought of as the 4D throughput over incoming outgoing directions.} $\phasePrefilterFunc$, given by:

\vspace{-3mm}
\begin{align}
\label{eq:TransmittancePrefilter}
\transmittancePrefilter{\voxPos}{\volPos} &= \int_{\pixSpatAng} \hspace{-1.5mm} \pixFilter{0}(\volPos,\volAng) \transmittance{\voxPos}{\volPos} \intOver{\pixAngInt}\intOver{\pixPosInt},
\\
\label{eq:PhasePrefilter}
~\phasePrefilter{\vox}{\pixAngInt}{\volAngInt} &= \int_{\pixSpatAng} \hspace{-1.5mm} \pixFilter{0}(\volPos,\volAng) \hspace{-1mm} \int_{\voxExit}^{\voxPos} \hspace{-3.5mm} \transmittance{\volPosInt}{\voxPos} \phaseTerm{\volPosInt}{\volAng}{\volAngInt} \intOver{\volPosInt}\intOver{\pixAngInt}\intOver{\pixPosInt}.
\end{align}

\noindent Since the prefiltered phase function is integrated over the in and out points of the voxel, $\voxExit$ and $\voxPos$, and the spatial footprint, $\volPos$, we parameterize it over voxel $\vox$ and the incoming and outgoing directions, $\volAngInt$ and $\volAng$. Note, by splitting the integral product and applying our transmittance separation (Eq.~\ref{eq:TransmittanceSplit}), the prefiltered transmittance, Eq.~\ref{eq:TransmittancePrefilter}, includes the transmittance from $\voxPos$ to $\volPos$, while the prefiltered phase, Eq.~\ref{eq:PhasePrefilter}, accounts for the transmittance through a point in the voxel, $\volPosInt$, to the corresponding voxel boundary, $\voxPos$. Although it is possible to avoid this assumption and preserve the integral of the product of transmittance and the prefiltered phase function, this adds substantial costs to the precomputation, as we would have to store the prefiltered phase function across two extra dimensions corresponding to the pixel's spatial footprint in order to evaluate the exact integral during runtime. Furthermore, we found this approximation worked well in practice (see Fig.~\ref{fig:ToyExample}) and use Eq.~\ref{eq:ScatteringTermPrefilter} in our system. At runtime, we simply sum over all the voxels in a pixel's footprint and attenuate the incoming radiance over the sphere by the scalar voxel-to-voxel tracked transmittance, $\transmittancePrefilter{\voxPos}{\volPos}$, and the stored prefiltered phase function, $\phasePrefilter{\vox}{\pixAngInt}{\volAngInt}$. 

Note that this scattering term is a general equation that works recursively to propagate indirect illumination back to the camera sensor, similar to Belcour et al.'s~\shortcite{Belcour17} surface formulation. Although multiple bounces are theoretically supported with our prefiltering framework, we focus on demonstrating results for the first bounce, as that is typically the most susceptible to objectionable artifacts. 

\subsubsection*{Emission term}
The pixel filter applied to the emission term yields:

\vspace{-2mm}
\begin{align}
\label{eq:EmissionTermPrefilter}
\emissionTerm &= \voxSum \int_{\pixSpatAng} 
        \pixFilter{0}(\volPos,\volAng) \int_{\voxExit}^{\voxPos} \transmittance{\volPosInt}{\volPos} \emission{\volPosInt} \intOver{\volPosInt} \intOver{\pixAngInt}\intOver{\pixPosInt} \nonumber \\
        &\approx \voxSum \transmittancePrefilter{\voxPos}{\volPos}
        ~\emissionPrefilter{\vox}{\pixAngInt},
\end{align}

\noindent where we apply the split in transmittance, $\transmittanceFunc$, as before using Eq.~\ref{eq:TransmittanceSplit}. Similarly to our scattering term, we approximate the integral product over the pixel footprint to obtain $\emissionPrefilterFunc$, our \textit{prefiltered emission}: 

\vspace{-2mm}
\begin{align}
 \label{eq:EmissionTermPrefilterExplicit}
\emissionPrefilter{\vox}{\pixAngInt} = \int_{\pixSpatAng} \pixFilter{0}(\volPos,\volAng) \int_{\voxExit}^{\voxPos} \transmittance{\volPosInt}{\voxPos} \emission{\volPosInt} \intOver{\volPosInt}\intOver{\pixAngInt}\intOver{\pixPosInt}.
\end{align}

\noindent This means that we can precompute the contribution from internal emitters and then track the voxel-to-voxel transmittance during runtime to determine the final contribution at the pixel sensor. Although our framework supports prefiltered emissions, this is less common and our scenes did not contain any internal emitters, so we omit this term from subsequent discussion.

\subsubsection*{Transmittance}

Since an exponential transmittance is not suitable for watertight surfaces and neither is a simple linear one for aggregates (see Fig.~\ref{fig:BeamTransmittance}), we utilize a more appropriate numerical transmittance model. Specifically, in the spirit of deep shadow maps~\cite{Lokovic00} and previous SVO-based prefiltering approaches~\cite{Heitz12}, we use a 2D spatial coverage mask to numerically model different transmittance modes along the beam:

\vspace{-2mm}
\begin{align}
\label{eq:Transmittance}
\transmittancePrefilter{\volExit}{\volPos} = \int_{\pixSpatAng} \hspace{-1.5mm} \pixFilter{0}(\volPos,\volAng) \transmittance{\voxPos}{\volPos} \intOver{\pixAngInt}\intOver{\pixPosInt} \approx \frac{1}{N} \sum_{j=0}^{N-1} \trackedCov{j}{\volExit}{\volPos},
\end{align}


\noindent where $N$ is the number of pixels in our coverage mask and $\trackedCov{j}{\volExit}{\volPos}$ is the tracked coverage from $\volPos$ to $\volExit$ at the $j^{\text{th}}$ pixel. The general tracked coverage is then given by:

\vspace{-2mm}
\begin{align}
\label{eq:TrackedCoverage}
\trackedCov{j}{\volExit}{\volPos} = \left[ 1 - \sum_{\vox\in{\setVox{\volPos\volExit}}} \trackedCov{j}{\voxPos}{\volPos} \right] \voxCov{j}{\voxInd{\volExit}}{\volPos},
\end{align}

\noindent where $\setVox{\volPos\volExit}$ is the ordered set (closest to $\volPos$ first) of all voxels that overlap the interval $\volPos$ to $\volExit$ and $\trackedCov{j}{\volPos}{\volPos} = 0$. Meanwhile, $\voxCov{j}{\voxInd{\volExit}}{\volPos}$ is the $j$th pixel of the coverage mask corresponding to the voxel at $\volExit$ in the direction towards $\volPos$, and which lies in the range from 0 (transparent) to 1 (blocked). Fortunately, we are able to precompute the coverage mask at each voxel, $\voxCovFunc$, during the preprocessing step, so only the tracked coverage, $\trackedCovFunc$, needs to be computed at runtime. Essentially, we sum over all of the contributions of the voxels in order up until the edge of the current voxel containing $\volExit$. Since $\sum \trackedCov{j}{\voxPos}{\volPos} \in{\left[0,1\right]}$, we subtract this from 1 to determine the maximum amount of contribution this voxel can have. Finally, we multiply this weight by the coverage of the current voxel, $\voxCov{j}{\voxInd{\volExit}}{\volPos}$, to determine its final contribution.



\section{Neural Prefiltering}
\label{sec:NeuralPrefiltering}

\begin{figure*}[t]
\centering
\includegraphics[width=\linewidth]{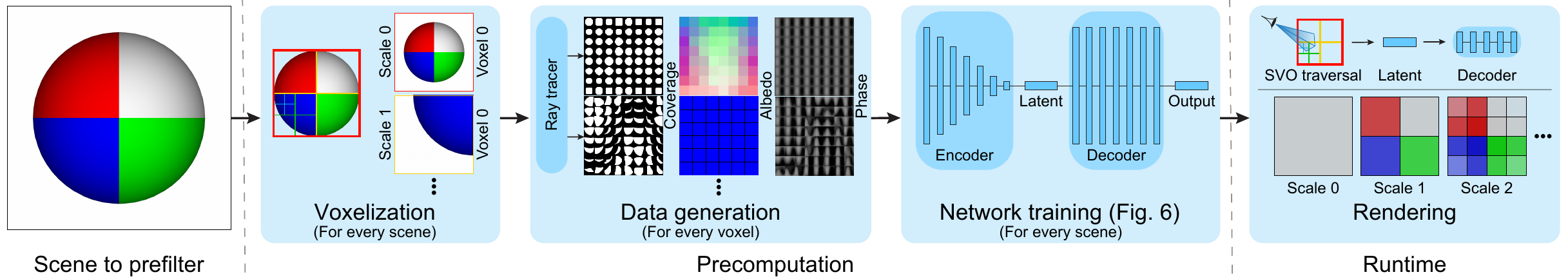}
\vspace{-7mm}
\caption{Overview of our algorithm. We first voxelize the scene at multiple scales and store non-empty voxels in a sparse voxel octree (SVO). Next, in the data-generation step, we process the portion of the scene within each voxel in isolation and sample incident and outgoing directions using a ray tracer to record the phase (i.e., the throughput), coverage, and albedo information across the volume. We then train a single network to efficiently compress this per-voxel data into small latent feature vectors that can be unpacked by a lightweight decoder. During rendering, for each voxel intersected by our beam along a pixel's footprint, we evaluate the phase, albedo, and coverage by decoding the precomputed latent representations for specific incident and outgoing query directions from a light transport algorithm to generate an image. Note that during rendering, we use only  SVO traversals and inference of our pretrained networks and have no reliance on the original geometry or materials of the scene.} 
\label{fig:Overview}
\vspace{-4mm}
\end{figure*}

\subsection{Framework overview}
\label{sec:FrameworkOverview}

This section discusses our implementation based on the theory from Sec.~\ref{sec:TheoreticalOverview}. We describe our general pipeline (see Fig.~\ref{fig:Overview}) starting with the original scene as an explicit mesh and materials to the intermediate prefiltered representation and, finally, the rendered image.

\subsubsection*{Spatial representation}

We chose a sparse hierarchical voxel representation that directly corresponds to the level of detail (LoD) scales. Specifically, for a given scale, we perform uniform discretization in world space (i.e., $x$, $y$, and $z$ coordinates) and then we double the resolution of our voxelization in each dimension for each subsequent, higher scale. In this way, LoD scale $i$, contains $8^i$ total voxels (before considering sparsity). During rendering, the filtering kernel (such as the pixel footprint and pixel filter) determines which LoD scale is used so that the relative size of a voxel does not exceed the bandlimiting frequency of the filtering kernel.\footnote{Typically, LoD approaches even use the next finer scale to so that the voxels are sub-Nyquist resolution.} In other words, similarly to texture filtering, the further away the camera is from the asset, the coarser the LoD scale that is used since a pixel will map to an increased portion of the scene and larger voxels. 

As in Heitz et al.~\shortcite{Heitz12}, we use a sparse voxel octree (SVO) representation at each scale, which discards empty voxels and allows for efficient voxel lookups. Each voxel contains both the prefiltered phase function, $\phasePrefilterFunc$, and 2D coverage mask, $\voxCovFunc$, introduced in Sec.~\ref{sec:TheoreticalOverview}, and an albedo term, which are each described in turn.

\subsubsection*{Phase function}
\label{subsec:PhaseFunc}

In general, a phase function is a dimensionless measure of how light is scattered at a point in a volume, typically parameterized by incoming and outgoing directions. 
Thus, each one of our voxels has a different phase function based on how light propagates through it. We represent the prefiltered phase function, $\phasePrefilterFunc$ from Eq.~\ref{eq:ScatteringTermPrefilter}, as a high-resolution uniform grid that maps to a 4D parameterization of input and output directions as spherical coordinates: $\volAng_i=(\thetaIn,\phiIn)$ and $\volAng_0=(\thetaOut,\phiOut)$. Each index of this table is a single monochromatic value of the voxel's phase function corresponding to direction pairs that map to that index. This information can be queried during rendering and multiplied with a similarly queried base color, described next, to find the radiance contribution of each voxel based on view and lighting directions.

\subsubsection*{Albedo}
\label{subsec:Albedo}

In addition to the monochromatic phase function, we record the albedo, the average RGB base color (i.e., the average diffuse color across the geometry's projected cross-section) of the phase function in a specific outgoing direction. The 2D directional albedo is decoupled from the phase to reduce storage costs of a voxel during precomputation and also as a means of helping the network optimization. By learning a single value per direction for the albedo, the phase sub-network does not have to learn to reproduce the same base color for a given outgoing direction at all the incident directions and instead it just learns to reproduce the more coherent monochromatic data. The albedo sub-network predicts a single RGB color per direction, which then multiplies the phase. Thus, our full phase function with RGB values is given by:

\vspace{-4mm}
\begin{align}
\label{eq:FullPhase}
\phasePrefilter{\vox}{\pixAngInt}{\volAngInt} = \voxPhase{\vox}{\pixAngInt}{\volAngInt}\voxAlbedo{\vox}{\pixAngInt},
\end{align}
\vspace{-4mm}

\noindent where $\voxPhaseFunc$ and $\voxAlbedoFunc$ are the scalar monochromatic phase and RGB base albedo, respectively. Note that $\voxAlbedoFunc$ is parameterized over the outgoing direction only (2D spherical coordinates), unlike $\voxPhaseFunc$ which uses both outgoing and incoming directions (4D spherical coordinates). Moreover, we chose to represent only the diffuse albedo and leave explorations in additionally utilizing a specular color for future work. By using only a single scalar per direction for albedo, we do not account for different contributions of spatially varying textures, as these contributions are averaged out in each outgoing direction (see limitations in Sec.~\ref{sec:Discussion}). Furthermore, we do not track color correlations from voxel to voxel, but we do spatially track transparency.
\vspace{-5mm}
\subsubsection*{Coverage mask to approximate non-exponential transmittance}
\label{subsec:CoverageMask}
Our transmittance model is used to determine the contribution of a voxel based on its correlation with preceding voxels along the path to the camera. If two voxels are $100\%$ negatively correlated, it means that neither voxel is blocked by the other and the contribution of each is simply added. If the voxels are $100\%$ positively correlated, it means that one voxel completely blocks another one behind it, and, thus, the second voxel has no contribution. Note, there can be varying degrees of correlation that create a spectrum of cases (see Fig.~\ref{fig:BeamTransmittance}). 


We determine each voxel's coverage with a fixed-resolution 2D spatial mask, $\voxCovFunc$, per particular viewing direction and we compute this for the same outgoing directions as our phase function, resulting in a 4D table. In other words, the elements of our table are indexed by a particular view, and the element itself is the voxel's 2D coverage mask. The 2D tracked coverage, $\trackedCovFunc$, is determined during runtime based on the 2D precomputed, per-voxel coverage mask, $\voxCovFunc$, as shown in Eq.~\ref{eq:TrackedCoverage}. To determine the contribution of a given voxel, we compute the proportion of its coverage mask that remains unoccluded based on the beam's tracked coverage mask so far. \new{We found our approach was sufficient for the scenes we evaluated, but it would be interesting to utilize recent models in our framework~\cite{Vicini21}.}
\vspace{-2mm}
\subsubsection*{Rendering with a prefiltered representation}

We implement a single-bounce renderer (direct lighting only) to demonstrate the directly visible quality of our prefiltered scene representation. We use a beam tracer that traverses our SVO and determines the ordered set of voxels using the intersection distance of a beam originating at each pixel. From the phase and albedo, it determines the RGB throughput in a given incoming/outgoing direction based on the scene's camera and lighting setup (Eq.~\ref{eq:FullPhase}) and evaluates the prefiltered scattering term in Eq.~\ref{eq:ScatteringTermPrefilter}. The 2D coverage mask, $\voxCovFunc$, is used to track correlations across voxels to compute the transmittance along the beam with Eq.~\ref{eq:Transmittance}, while Eq.~\ref{eq:PixelFlux} determines the final radiance accumulated along the beam to a given pixel. Note, our current scenes have no emissive surfaces, so the filtered emission term, $\emissionTerm$, is ignored.

\vspace{-1mm}
\subsection{Prefiltering and data generation}
\vspace{-0.5mm}
\label{subsec:DataGeneration}
During data generation, we compute the phase function, albedo, and coverage mask for each voxel at every LoD scale. We normalize the scene by the largest dimension and then consider a unit volume bounding box around the scene. We then perform voxelization and construct our SVO before proceeding to compute each voxel's data.

We represent our phase function as a 4D table indexed by spherical coordinates corresponding to a pair of incoming/outgoing directions and where each element in a bin corresponds to a scalar monochromatic throughput. In other words, if viewing the scene from a given outgoing direction, each bin corresponds to how much light is reflected towards the camera for a given incoming direction. 

To compute the value at each bin (a 4D index of incoming/outgoing directions) of a given voxel, we measure the throughput by sending many samples with a ray tracer in a specific view direction and recording each sample's eventual exit direction from the voxel's volume, as well as its accumulated radiance. Specifically, we first set up a constant environment map with a value of one everywhere (i.e., a white furnace) and sample the projected cross-section of the scene from various outgoing directions using an orthographic camera, so that all samples are intersecting the voxel with the same direction. The samples bounce around in the voxel until they ultimately exit by hitting the environment map. We use MIS sampling within a voxel and record the corresponding incident radiance and direction for the light sample at each bounce.

Once a sample terminates, we accumulate the radiance in the corresponding bins based on the view (outgoing direction) and the exit or light sample directions (incoming direction). After all the samples for this outgoing direction are recorded, we normalize each bin by the total sample count. We then calculate this for all outgoing directions based on our table's resolution. Note we sample outgoing directions uniformly along the sphere rather than uniformly in the grid to avoid bias from bins mapping to different solid angles. To ensure that the entire voxel gets properly sampled at each outgoing direction, we sample the entire projected cross section and discard any samples that do not go through the bounding box of the voxel. 

At the same time we generate samples of our phase function, we also construct our 2D coverage mask, $\voxCovFunc$, and directional albedo, $\voxAlbedoFunc$. When processing each sample and binning them, we can use the cross-section sample location to find the corresponding texel in our fixed-resolution coverage mask (e.g., $16 \times 16$ in our implementation) and deposit either a 1 or a 0, based on whether the sample intersected any geometry within the voxel. We average this visibility information at each texel to generate the final supersampled 2D coverage mask corresponding to the processed outgoing direction. To compute the albedo, we average the base color from each phase function sample that intersected voxel geometry (i.e., those samples with a value of 1 for visibility). In other words, we integrate the albedo over the area of the voxel's projected cross-section along a given view direction and record a single RGB value at the 2D index.


\vspace{-1mm}
\subsection{Neural compression}
\label{sec:Compression}

Although baking out the phase function, albedo, and coverage mask into our tables should be sufficient to render an image with high fidelity to the original, there are various considerations that make this straightforward approach impractical. First, since these tables need to have a reasonably high resolution to be able to reproduce the original appearance, each voxel has a significant amount of data associated with it (i.e., on the order of MBs). Moreover, typical pipelines utilize several scales of LoDs for their assets which can amount to millions of voxels at the finer scales, even when considering sparsity (i.e., we can ignore empty voxels), totaling TBs of data for a single scene, which would undermine the savings relative to a brute-force ray tracer. Another issue is that the tables must be computed for directions at a high resolution to minimize approximation errors stemming from interpolating the non-computed views.

To be able to get a high compression rate and to more accurately interpolate non-calculated views, we propose to use a deep neural network solution. Specifically, the network takes in the full generated dataset and encodes it into a latent representation, a common strategy in the machine learning community, that is orders of magnitude smaller than the dataset size. The network includes a decoder that extracts the desired elements from the latent vector at runtime. A network-based solution allows substantial compression as is shown in previous work~\cite{Rainer19}. Furthermore, we have the added benefit that network can perform more accurate, non-linear interpolation for views that were not precomputed.

After the network is trained, we incorporate it into our pipeline with a preprocessing step that encodes all the voxels in a scene at all the scales and generates their latent feature vectors, which are stored on disk. Prior to rendering, we load all of these vectors into memory, along with the decoder network and its trained weights. During rendering, for every intersected voxel, the beam tracer evaluates the decoder using the voxel's latent vector and a specific query direction based on the outgoing and incident directions to generate the corresponding element. Finally, we combine all the voxel data to synthesize the final image, as described in the previous sections.

We now describe the specifics of our system's neural networks.






\section{Network design and utilization}

\begin{figure*}[t]
	\setlength{\fboxrule}{10pt}%
	\setlength{\insetvsep}{20pt}%
	\setlength{\tabcolsep}{0pt}
	\renewcommand{\arraystretch}{1}%
	\footnotesize%
	\includegraphics[width=\linewidth]{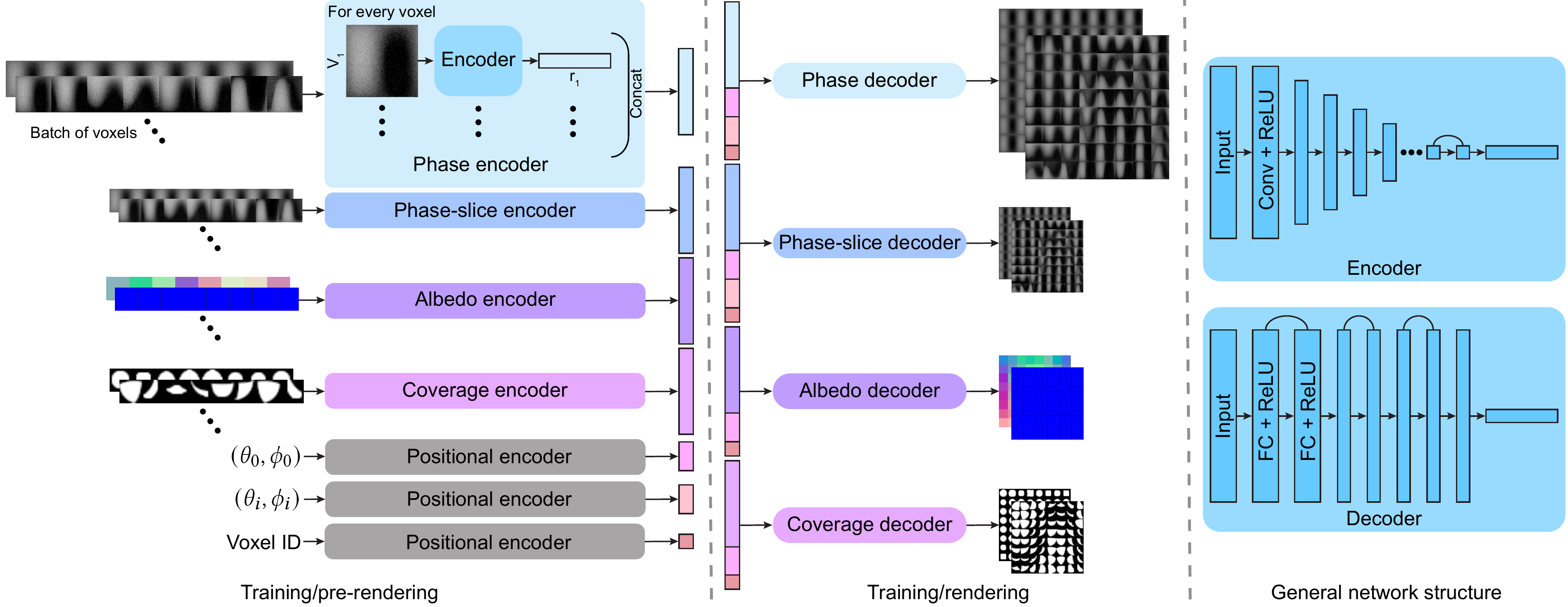}
	\caption{An overview of our four networks: \textit{phase}, \textit{phase-slice}, \textit{albedo}, and \textit{coverage}. The positional encoder is an explicit function~\cite{Mildenhall20} and not a learned one. All our learned encoder/decoder  nets compress the input data for each voxel into latent representations during a preprocess step. During rendering, a decoder takes the latent feature vectors for active voxels along with directional queries from light transport to evaluate these functions. The phase net returns a scalar throughput at a specific 4D direction, while the phase-slice net returns a 2D image corresponding to the throughput for incident directions conditioned by a particular view direction. The coverage net and albedo net return a 2D coverage mask and RGB base color, respectively, conditioned by outgoing direction. On the right is the general structure of the encoders/decoders used within the different network modules.} 
	\label{fig:NetworkOverview}
	\vspace{-2mm}
\end{figure*}

Our networks first must be trained offline on the precomputed data described in Sec.~\ref{sec:NeuralPrefiltering}. The networks are trained across all the voxels for a specific scene. During prefiltering, compressed representations in the form of latent vectors are precomputed for all the voxels using the converged, frozen weights of the networks. Finally, during rendering, we evaluate a subset of the networks and apply the equations from Sec.~\ref{sec:TheoreticalOverview} to generate the final image. In this section, we describe our network architecture and optimization, followed by its use during runtime, and conclude with implementation details.

\subsection{Compression networks}
\vspace{-0.5mm}
We use four networks in our system, as shown in Fig.~\ref{fig:NetworkOverview}. In general, all networks act as compressors/decompressors with a ``butterfly,'' or U-Net style, encoder and a lightweight decoder architecture. The networks are trained per scene, {\em not} per voxel. In other words, these four networks are trained to handle the encoding/decoding of the millions of voxels across all of the scales of a given scene. We now describe their specific uses and subtle differences. 

\vspace{-1mm}
\subsubsection*{Encoders}
The goal of the encoders is to generate a compact latent representation of the various prefiltered input data to be able to efficiently store and access this information in each voxel during rendering. Using the raw prefiltered input can quickly become intractable. Fortunately, the input plenoptic data has plenty of coherence that can be effectively compressed with a network. 

The encoders are made up of convolutional neural networks (CNNs) that operate on 2D slices of the original data. The use of convolutional encoders is also present in the architecture of Rainer et al.~\shortcite{Rainer19}. For example, for the phase encoder and a given outgoing direction, we take the throughput for all incoming directions and arrange them into a uniform 2D grid corresponding to their spherical coordinates, $\volAng_i=(\thetaIn,\phiIn)$. These 2D grids can be viewed as separate images, one for each outgoing direction, that are all sent to the encoder to be processed one at a time. Thus, our input is $N \times N \times k$, where $N$ is the grid resolution of our table in a single dimension and $k$ is the number of outgoing views that are fed in. The encoders separately process the $k$ slices from each viewing direction and then concatenate each of the $k$ resulting latent vectors together at the bottleneck. We found concatenating latent vectors performed significantly better than averaging when using equal vector sizes.

In order to support both single directional queries (e.g., point sampling light sources) and range queries to efficiently handle area light sources (e.g., environment maps), we use two different kinds of phase networks. During rendering, with a single inference, we can determine the throughput for multiple directions simultaneously, which can be used to perform the dot product with a large light source, such as an environment map. Without such a network, we would have to point sample a range query, which could result in noise and would require many calls of the point query network.

\vspace{-1mm}
\subsubsection*{Decoders}

The goal of the decoders is to accurately access elements from the original data during rendering using the latent vectors produced by the encoders, directional queries (both outgoing and incoming), and voxel ID. Of course, these queries can exist anywhere within the plenoptic data, including regions outside the available data, requiring the decoder to be able to interpolate. At render time, none of the original scene data (e.g., geometry and materials) is directly available to the decoder, which requires the encoder to compress the essential information into the latent representation.

The decoders consist of fully connected layers, as in Rainer et al.~\shortcite{Rainer19} and Mildenhall et al.~\shortcite{Mildenhall19}, that operate on flat vectors, rather than 2D images typical of convolutional networks, which can be efficiently evaluated during runtime. It is worth emphasizing that the decoders only return a small subset (e.g., a single value) of the original data corresponding to the query direction, in contrast with autoencoders that reconstruct the full input, allowing the memory footprint to remain minimal when rendering. Thus, the phase and albedo decoders output a scalar and RGB value, respectively. Meanwhile, the phase-slice and coverage decoders both generate 2D images. Still, these images are conditioned by a single outgoing direction query.

\vspace{-2mm}
\subsubsection*{Phase networks}

The \textit{phase network}, takes both outgoing and incident directions as input to the decoder and provides a single scalar throughput value, which is useful for sampling light sources. The range query network, called the \textit{phase-slice network}, will take in only the outgoing query direction for the decoder to produce a uniform grid containing the throughput for all incident directions at once. 

The phase and phase-slice encoders take in the same data, except we use a downsampled version of the original phase data as input to the phase-slice encoder, since reconstructing the original resolution at the decoder backend requires a larger, more expensive network and would be intractable during rendering. Furthermore, we found that using separate encoders was more effective than having the decoders share the latent vector of a single encoder.

\subsubsection*{Input parameterization}

First, we reparametrize all our input data so that the coordinate frame is in a local space of the view direction instead of a global world-space parameterization. This bypasses the additional complexity of having the network learn how to shift phase functions based on the outgoing direction. With this scheme, all the views of a diffuse sphere would have the same orientation for the 2D throughput across incident directions, whereas globally they would be translated versions of each other. To facilitate training, we employ standard practice and normalize the incident and outgoing directions and voxel ID to be between [-1,1]. Next, we apply positional encoding on all three vectors, as in the recent NeRF architecture~\cite{Mildenhall20}, since it helped the network recognize patterns in the data and improved convergence time. 

The latent vectors do not undergo any additional processing with the exception that the albedo compressed representation has the original raw albedo that was input to the encoder concatenated to it, as we found this helped interpolation and had a low memory overhead (only $k$ additional RGB values need to be stored or, in other words, one RGB base color for each encoded view). The raw coverage and albedo values are used as provided by the rendering system since they are already low dynamic range (LDR).

However, our phase data is high dynamic range (HDR), which presents two issues:~1)~the influence of dark regions is lower than that of brighter ones when using a typical $\ell_1$ or $\ell_2$ norm, and~2)~high values can produce large gradients that make network updates unstable. To combat this, learning-based methods have applied logarithmic transformations~\cite{Bako17,Vogels18} and gamma compression~\cite{Lehtinen18} to reduce the gap in range. For our system, using another alternative, the range compressor, performed best. This function was applied in recent works for high dynamic range imaging~\cite{Kalantari17,Kalantari19} and importance sampling~\cite{Bako19}, and is written as: 


\vspace{-3mm}
\begin{align}
\label{eq:RangeCompressor}
\rangeCompressor{x} = \frac{log(1 + \mu x)}{log(1 + \mu)},
\end{align}

\noindent where $\mu = 2$ controls the degree of compression. We use our range compressor only to preprocess the inputs to the network and leave the reference data for training in the linear domain. In general, having network inputs with large values negatively impacts training and can cause exploding gradients. Indeed, we observed significantly more stable behavior and a lower converged error after applying the range compressor to the inputs. A similar observation was described in Lehtinen et al.~\shortcite{Lehtinen18}, where the authors used tonemapping, albeit using a different function, only on the inputs of their denoising network for the optimization benefits and similarly kept the reference in the linear domain.

\subsection{Training}
\label{subsec:Training}
Training is performed end-to-end, with both the encoders and decoders trained simultaneously. 

\subsubsection*{Data}

To sample the outgoing directions, we use stratified sampling to better ensure coverage over the entire sphere. This means that, within each stratum, we generate a uniform random sample. In the encoding step, we send $k$ representative views for determining the latent encoding. For generating query samples, we sample from all available views, including those not provided to the encoder, which enables the decoder to generate accurate results for any direction. 

We employ a continuous sampling of the space and have the data generator running during training, which will periodically update the data stored at each voxel using different outgoing directions. Without this, we found the network would easily overfit to the static views in the table and would not interpolate well to other non-computed views. Note, generating data on the fly can mitigate storage costs (i.e., the networks can consume new training data as it is produced instead of placing it on disk), but this does not forego our discretized tables for the coverage and phase-slice networks. 

\subsubsection*{Loss function}
\label{sec:Loss}

For our loss, we optimize:
\begin{align}
\label{eq:NetLoss}
\mathcal{L} = \lVert \estimate{\voxCoverage} - \voxCoverage \rVert_2 + \mathbf{1}(\voxCoverage) \left[ \frac{\left( \estimate{\voxPhaseFunc} - \voxPhaseFunc \right)^2}{ \left( \estimate{\voxPhaseFunc} + \epsilon \right) ^2}  + \lVert \estimate{\voxAlbedoFunc} - \voxAlbedoFunc \rVert_1 + \lVert \estimate{\voxPhaseSlice} - \voxPhaseSlice \rVert_2 \right].
\end{align}

Here $\voxCoverage$ and $\estimate{\voxCoverage}$ denote the reference and network estimate of the voxel coverage. Similar definitions follow for phase, $\voxPhaseFunc$, albedo, $\voxAlbedoFunc$, and phase slice, $\voxPhaseSlice$. We use $\epsilon=0.01$ to avoid division by zero. The $\mathbf{1}(\voxCoverage)$ is an indicator function that is on if the reference coverage, $\voxCoverage$, has any nonzero pixels for the current slice. There are certain views of non-empty voxels where no coverage is detected, such as the case of a triangle viewed along its edge. In these cases, the network estimate of the coverage should still match the all-zero coverage mask of the reference, but we no longer need to force the phase and albedo networks to predict zero. Since a zero-coverage view should not have any contribution, the indicator function allows the network flexibility in not needing to accurately predict the remaining terms. 

For the phase term, we use the relative MSE metric as defined in Noise2Noise~\cite{Lehtinen18}, which is more robust in handling HDR data by being less susceptible to outliers compared to an $\ell_2$ norm. Note, as described in that work, we use the network prediction, $\estimate{\voxPhaseFunc}$, instead of the typical reference, $\voxPhaseFunc$, in the denominator to avoid biasing the expected value through a nonlinearity. For this same reason, we avoid using the range compressor (Eq.~\ref{eq:RangeCompressor}) on the reference data, which would introduce a nonlinearity that can map small errors in the compressed domain to large errors in the linear domain, creating noticeable problems in the final image. Thus, as in Lehtinen et al.~\shortcite{Lehtinen18}, we train in the linear domain. Although both the phase ($\voxPhaseFunc$) and phase slice ($\voxPhaseSlice$) are HDR, we found that the typical $\ell_2$ norm performed better when generating a 2D image and averaging across pixels, as in the phase slice $\voxPhaseSlice$, whereas the relative metric was more robust for the single point queries of the phase, $\voxPhaseFunc$. 



\subsection{Prerendering and runtime}
\label{subsec:Runtime}
After training, we must encode voxel data into latent variables to be utilized during the subsequent runtime rendering. 

\subsubsection*{Data}

First, we found it necessary to render out inference data for each voxel, which differs slightly from our training data. We use random views generated with our stratified sampling strategy, but only create a single set of random outgoing directions and use this same set for computing the data at every voxel. Using random views at each voxel in the same manner as our training data caused objectionable noise in the final image due to slight variations in the latent vectors, and hence the output values, at adjacent pixels. We also attempted using data generated with a fixed uniform stride but found this resulted in unacceptable structural artifacts/patterns.

\subsubsection*{Encoding and prerendering}

Afterwards, the encoding happens as a preprocess step before rendering to generate and store the latent features for each voxel, which can be interpreted as compressed representations of the input data. After saving this vector for all voxels, we can discard the encoder. Next, the trained weights of the decoder are frozen and exported to inference within the renderer.

\subsubsection*{Rendering}

Finally, during rendering, we look up the latent vector and append directional information from the current render to query with the decoder-only portion of our network to generate a specific entry from the original tabular data. Note, the original geometry and materials are no longer used at this stage, and all values are computed through SVO lookups and network inference.

\subsection{Implementation details}
\label{sec:Implementation}

\subsubsection*{Data generation}
Our framework was implemented using our own GPU ray tracer developed with CUDA and OptiX~\cite{Parker10} and we used a cluster of NVIDIA Volta GPUs (256 in total for computing data and training). For each scene, a full cycle of data generation took \textasciitilde{0.5 to 2} days (depending on the number of voxels).

At any given time, each voxel had 16 slices of data (where a slice is a 2D image of the 4D light field obtained by fixing the outgoing direction and having each pixel correspond to a different incoming direction), each computed with 256K samples and corresponding to different outgoing directions. Note, we found this sampling rate sufficient for all voxels in the multi-scale hierarchy, even those from the coarse scales despite them accounting for a larger portion of the entire scene and potentially more complex light transport. For each voxel, we also save out the bounding sphere around the actual geometry (4 floats total, 1 float for radius and 3 floats for the offset). This is another form of regularization for the data so that the coverage masks are more uniform across slices and so the network does not have to learn translations. Moreover, it facilitates training by making the coverage masks larger for voxels with tiny geometry allowing for more gradients in the coverage regions and, thus, better optimization. Note that during runtime, we re-scale and translate the network's predicted coverage mask using this information to accurately compute the transmittance.

Our uniform grid sizes are $N = 128$ for the phase encoder, $N = 16$ for the coverage and phase-slice encoders, and $N = 1$ for the albedo encoder (i.e., $128\times128$ and $16\times16$ images and a $1\times1$ scalar, respectively) and they correspond to spherical coordinates for the incoming direction. These dimensions are used for both training and runtime. In all encoders, we use $k=8$ random views, one from each octant of the sphere, but chosen so that samples uniformly cover the sphere. Each of these corresponds to a different viewing direction that is constantly updated to a new viewing direction by the data generator to allow for a continuous sampling of the space. Note, these views are randomly selected for each iteration and are only used to encode the latent variable, while the point query training samples can come from all the data available on disk for that voxel at any given time (e.g., 16 views per voxel), not only from the $k=8$ random views fed to the encoder. For example, for the phase tabular data, each voxel would have $16\times128\times128$ floats saved on disk, which are constantly being updated with random views by the data generator during training. We downsample the phase-slice encoder's input images using OpenCV's resize function with the area filter option. 

\subsubsection*{Beam tracer}
Another challenge when rendering with a beam tracer is the common case where beams are not axis-aligned with the SVO, so multiple voxels can fall within a beam at varying depths and locations and with only partial contributions. To solve this issue, we perform beam marching across sets of voxels ordered by distance, which we call \textit{wavefronts}. A wavefront consists of all the voxels in an interval (e.g., the size of a voxel length) that overlap with the beam. Within a wavefront group, we essentially stitch together the coverage masks from each voxel at a given depth and splat them onto the beam's coverage mask, which we continue to track as we march further along our beam. We weight each voxel's contribution by the proportion of area it adds to the beam. Note, the coverage network robustly interpolates the 2D mask across continuous viewing directions due to our constantly updated training set (described later in this section). This implementation, while not trivial, will be part of our full code release upon publication. Pseudocode of our overall algorithm can be found in Appendix~\ref{appendix:AppendixB}.

Furthermore, we use discrete LoDs in our system, but this can lead to popping artifacts from abruptly switching between one scale and another. To avoid this issue, we use standard trilinear interpolation between the voxels in the wavefront of the current scale and those in the next coarser scale, as is commonly done for LoD prefiltering~\cite{Loubet17}. Note, we train a single set of networks across all of the voxels in all of the discrete scales of a given scene, so this explicit interpolation helps avoid popping between scales. A more systematic approach that perhaps blends latent vectors of voxels in different scales is left for future research. 


\subsubsection*{Shadow maps}
To enable shadows in our current system, we capture intra-voxel and inter-voxel shadowing. For the former, we can save each voxel's precomputed data with self-shadowing accounted for. For example, if two triangles are inside a given voxel, certain lighting and view directions can result in one triangle occluding the other. We incorporate this local shadowing into our data generation step. 

On the other hand, to capture global shadowing, we make use of the common strategy of shadow maps~\cite{Williams78}. Specifically, we first render the depth of the directly visible voxels from the direction of each light source as a pre-process. In the case of environment maps, we compute a shadow map for each direction corresponding to the bins of the low resolution grid used in our phase-slice network (center of the bin was sufficient in practice). All surfaces that are directly seen from the light source can potentially have a radiance contribution, while those surfaces that are not visible will be in shadow. Then during final rendering, while accumulating the contribution of each voxel during beam marching, we use the voxel's world coordinates to project to each light's coordinate system and look up the voxel's depth at the corresponding pixel of the light's shadow map. If the current voxel's depth is greater than the depth in the shadow map then the voxel is occluded and does not have a radiance contribution at the current pixel being rendered. 

\subsubsection*{Networks}
We implemented our networks in TensorFlow~\cite{Abadi15} and used Xavier normal initialization~\cite{Glorot10} and the Adam~\cite{Kingma14} optimizer with a learning rate of $3.0 \times 10^{-4}$ and mini-batches of 1 voxel with 4096 queries. Training took \textasciitilde{2} days, on average. For more efficient inference, we used the TensorRT library within our renderer to evaluate the decoders with the trained weights at the query directions. 

For the CNN encoders, we start with a relatively small number of feature channels in the input layer (e.g., 8 for the phase encoder), but then double the layer size after each subsequent reduction of the spatial resolution (clipped to be a maximum of 256 channels). The spatial downsampling was performed with strided convolutions, which performed better than average or max pooling, and ReLU was used for the activation function on all layers except the output layer, which was linear. After concatenating the encodings of all the views, each encoder outputs its final latent vector of 256 floats. 

The decoder networks all consisted of three residual blocks (each block has two layers of non-linearity) with ReLU activation functions and differed only in the output layer. The coverage decoder used a sigmoid output activation, whereas the other networks used a linear activation, as is common practice. 

All networks employed residual connections where possible (when layer count and spatial dimension matched) to mitigate general optimization pitfalls, such as vanishing gradients, which thereby helped reduce convergence time. Finally, we train the networks from scratch, as we found no noticeable improvements when pretraining an autoencoder architecture to stabilize the encoder weights before using transfer learning with the final decoder architecture. 

Regarding our memory requirements, we store voxel ID (1 float), bounding sphere information (4 floats), and a latent vector (792 floats total consisting of 256, 256, and 280 floats from the phase or phase-slice, coverage, and albedo encoders, respectively) at each voxel, making the memory size 797 floats, just over 3 KB per voxel.\footnote{There is also a fixed cost that we include for the decoder network weights and data structures (e.g., the SVO), but these are relatively small and negligible.} If we used all four sub-networks (i.e., both the phase and phase-slice networks, not just one), for example when using point light sources together with an environment map, then the vector would be 1048 floats or 4 KB. All of the scenes in the results use only environment maps or point light sources, so our memory estimates are based on using 3 KB per voxel.  Note, it could be possible to utilize half-floats (i.e., 16-bit floats) instead of the 32-bit floats at various stages throughout our pipeline for a potential $2\times$ savings, but we leave this optimization for future work. 


\section{Results}
\label{sec:Results}

\begin{table}[t]
	\centering
	\setlength{\fboxrule}{30pt}%
	\setlength{\insetvsep}{60pt}%
	\setlength{\tabcolsep}{8pt}
	\begin{tabular}{l|c|l|c}
		\hline
		\textbf{Cutlery} & 598{,}032 & \textbf{Oak} & 12{,}589{,}056  \\
		\hline
		\scene{Mossy Rock} & 5{,}996{,}334 & \textbf{Forest} & 53{,}948{,}550  \\
		\hline
		\scene{Parking Lot} & 24{,}165{,}771 & \textbf{Stormtrooper} & 175{,}986  \\
        \cline{1-2}
		\scene{City} & 87{,}433{,}857 & \textbf{Army} &  \\
		\hline
		
	\end{tabular}
 	\vspace{0.25mm}
	\caption{Number of triangles for each scene in the paper.}
	\label{table:TriangleCount}
	\vspace{-8mm}
\end{table}

\begin{figure*}[tp]
	\setlength{\fboxrule}{10pt}%
	\setlength{\insetvsep}{20pt}%
	\setlength{\tabcolsep}{1pt}
	\renewcommand{\arraystretch}{1}%
	\footnotesize%
	\begin{tabular}{ccccccc}
		
		& \eggx & \sggx & \selfshadow & \hybrid & Ours & Reference (16K spp) \\
        \begin{sideways} \hspace{10mm}\textbf{Cutlery} \end{sideways} &
		\frame{\begin{overpic}[width=\imgW\textwidth]{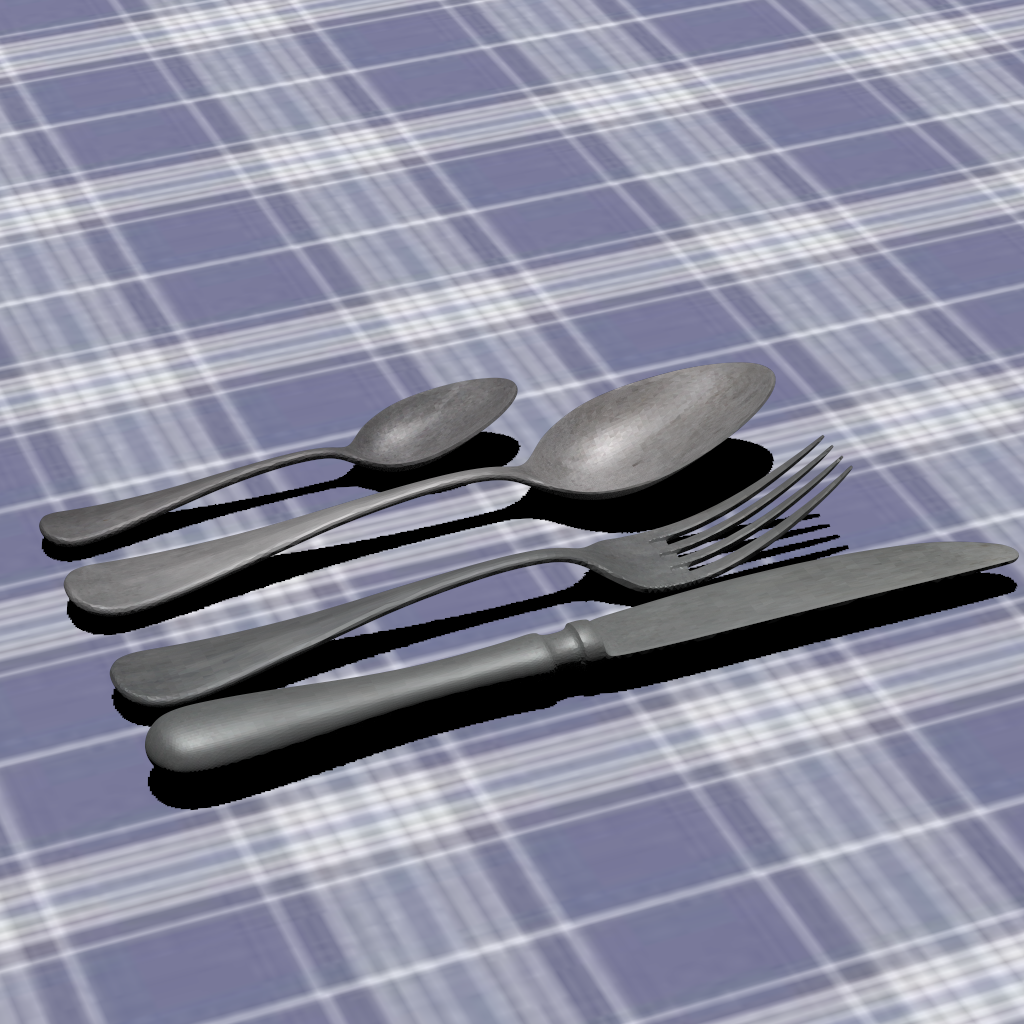}%
        \put(\putX,\putY){\frame{\includegraphics[width=\putW\linewidth]{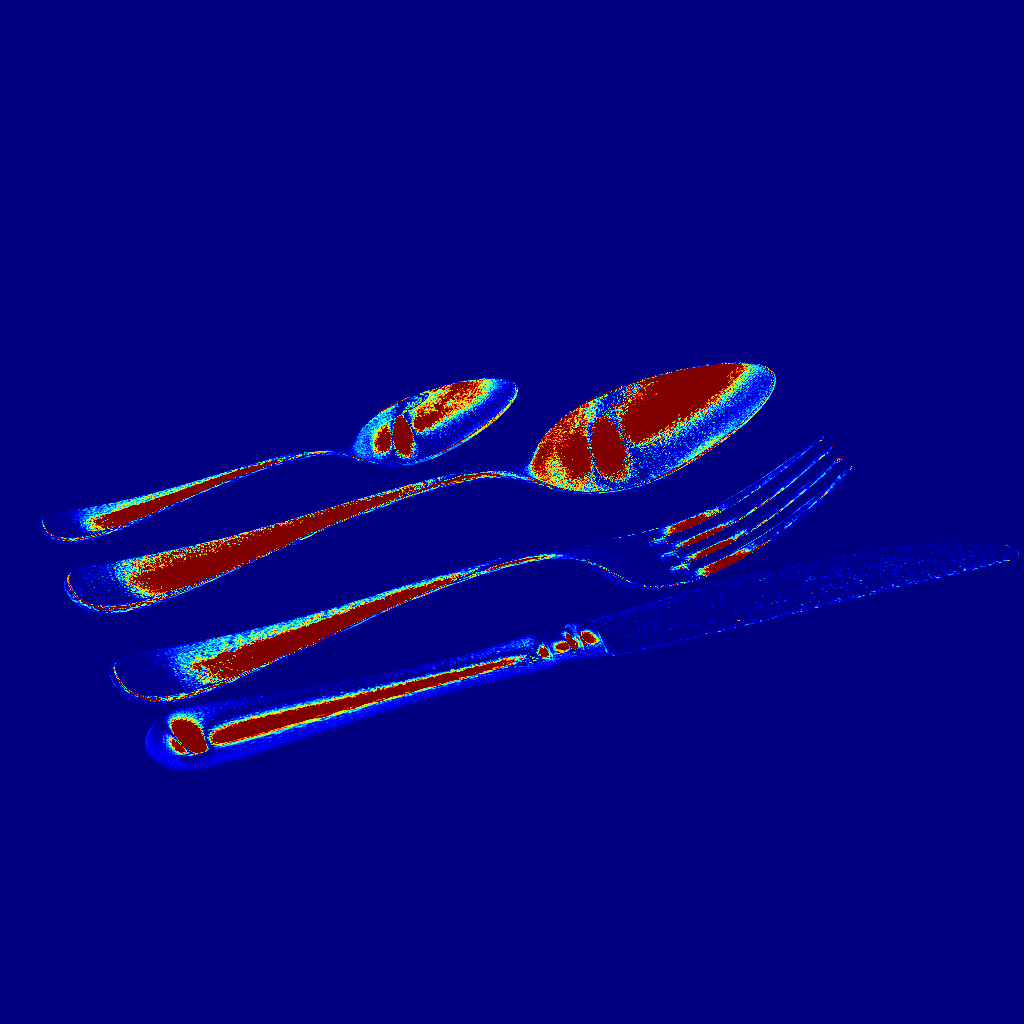}}}
        \end{overpic}} &
		\frame{\begin{overpic}[width=\imgW\textwidth]{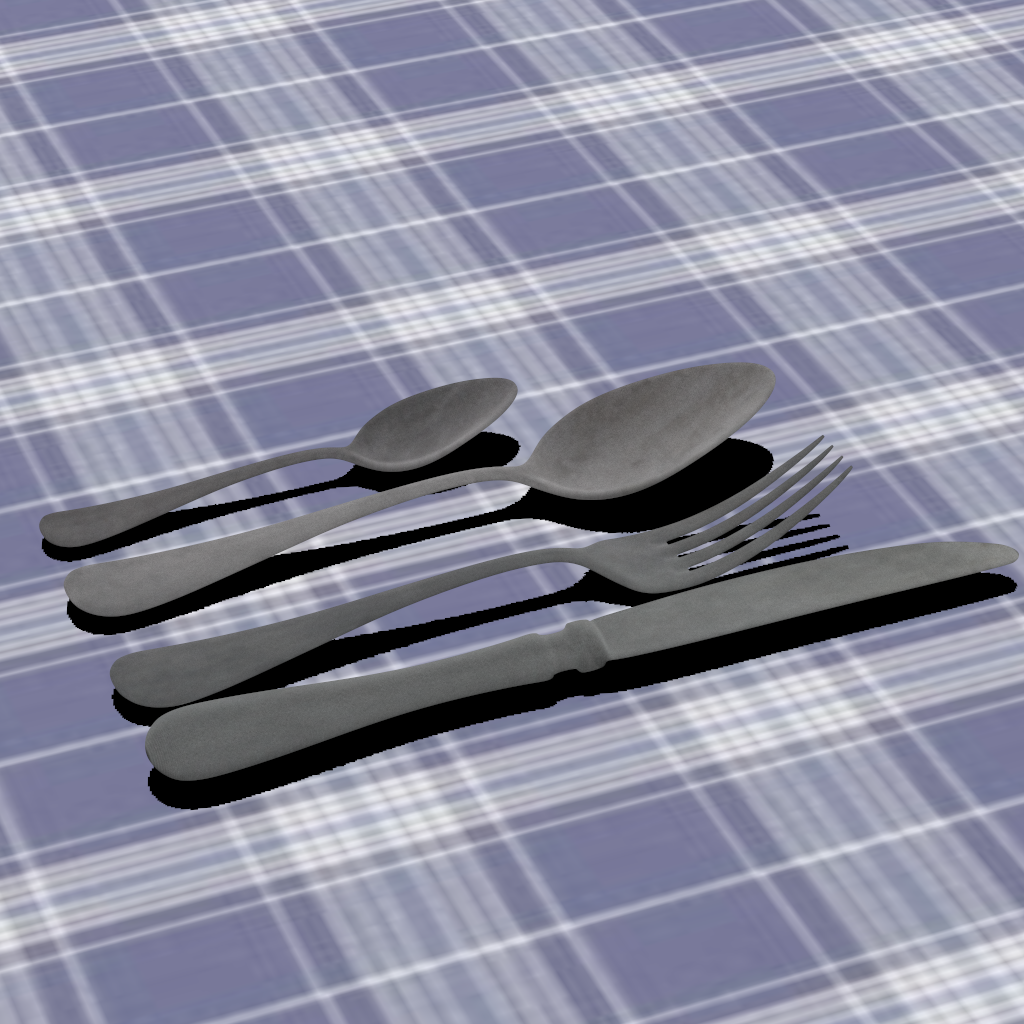}%
        \put(\putX,\putY){\frame{\includegraphics[width=\putW\linewidth]{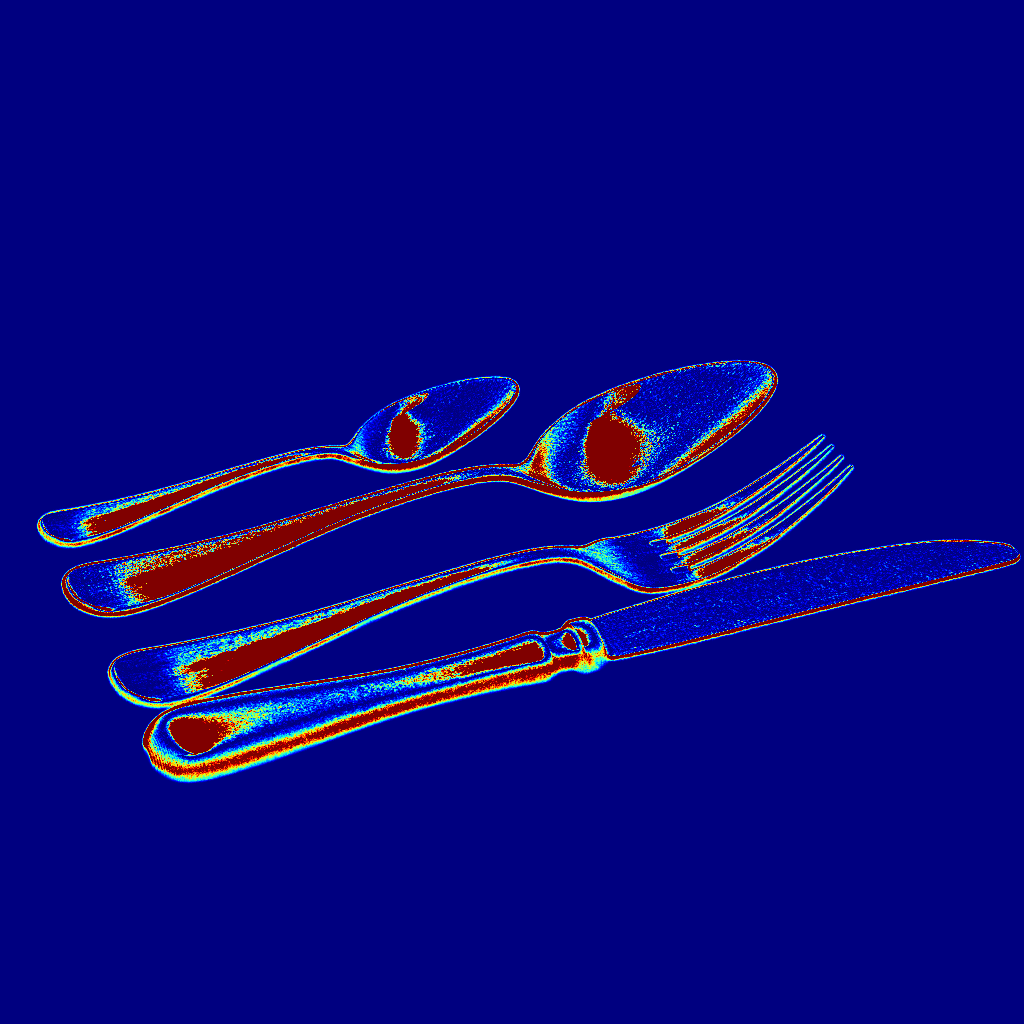}}}
        \end{overpic}} &
        \frame{\begin{overpic}[width=\imgW\textwidth]{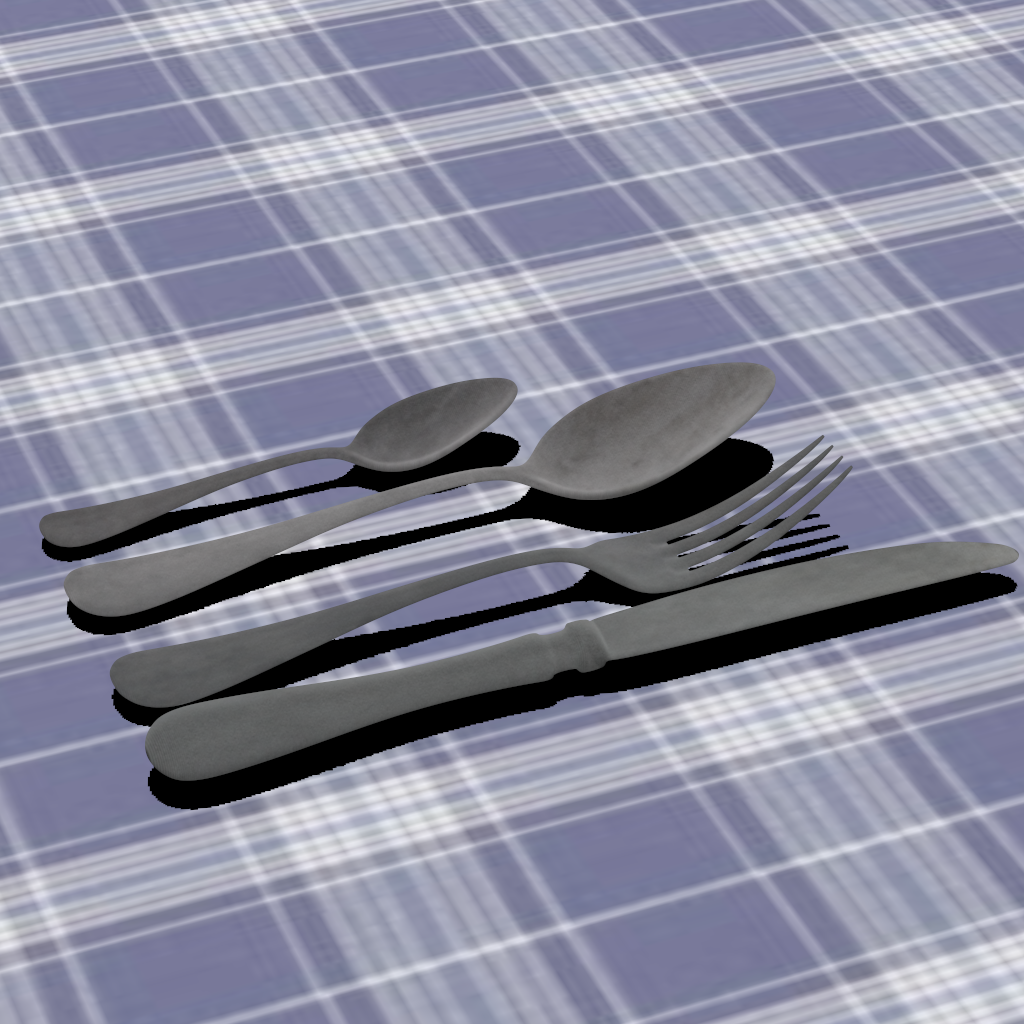}%
        \put(\putX,\putY){\frame{\includegraphics[width=\putW\linewidth]{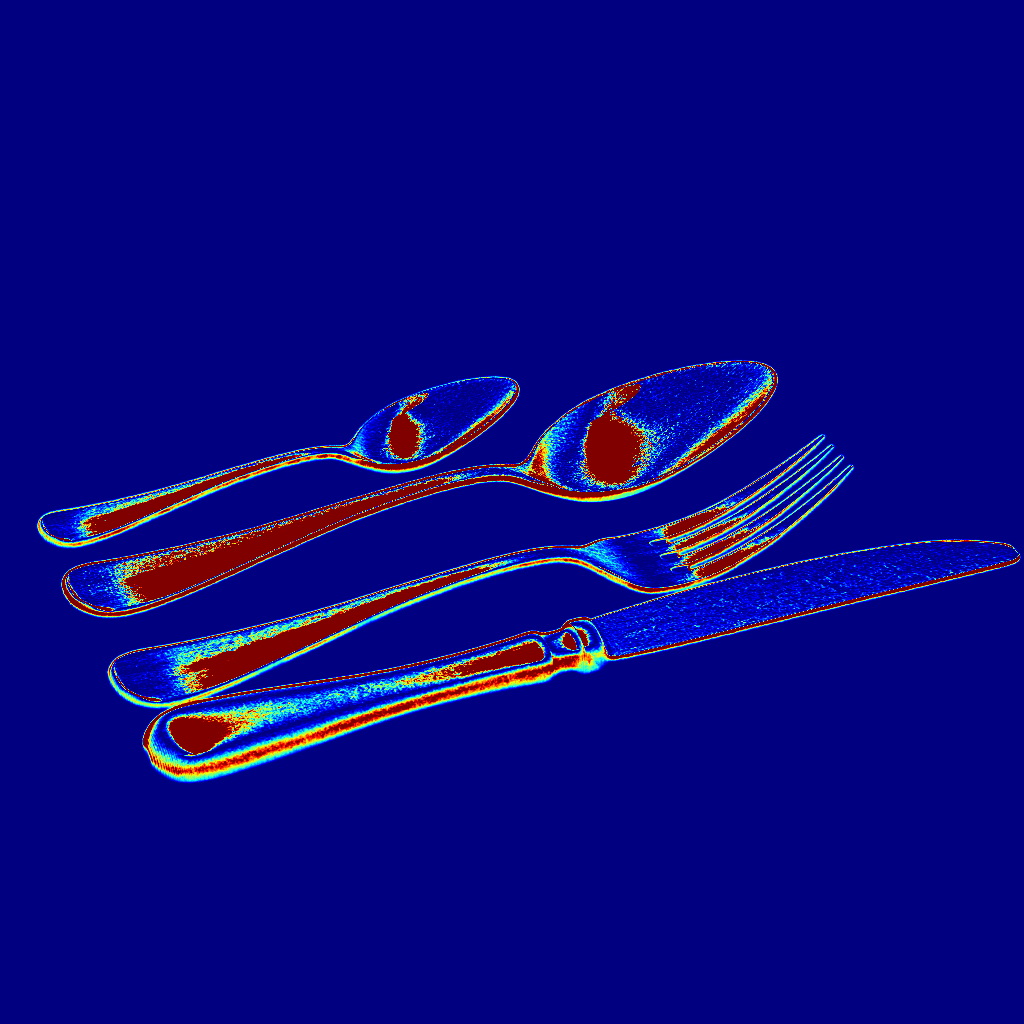}}}
        \end{overpic}} &
        \frame{\begin{overpic}[width=\imgW\textwidth]{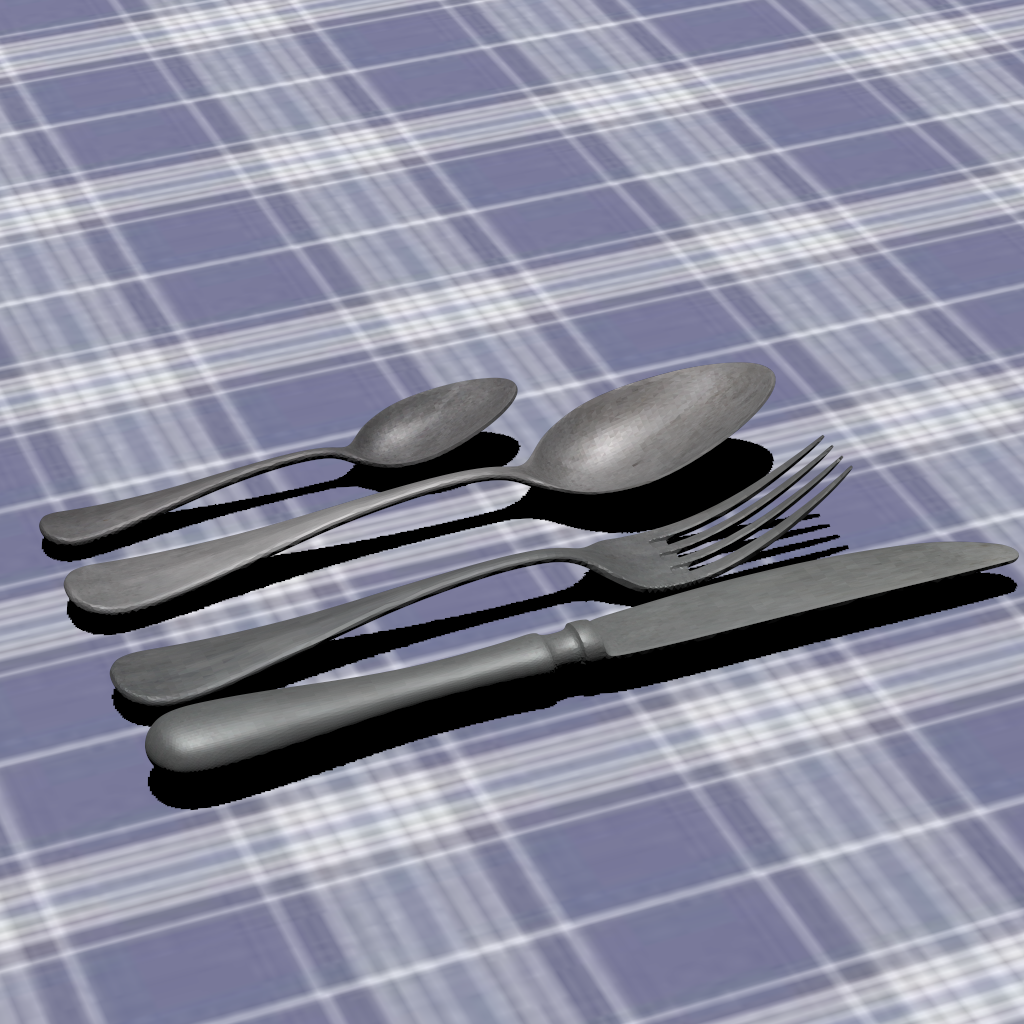}%
        \put(\putX,\putY){\frame{\includegraphics[width=\putW\linewidth]{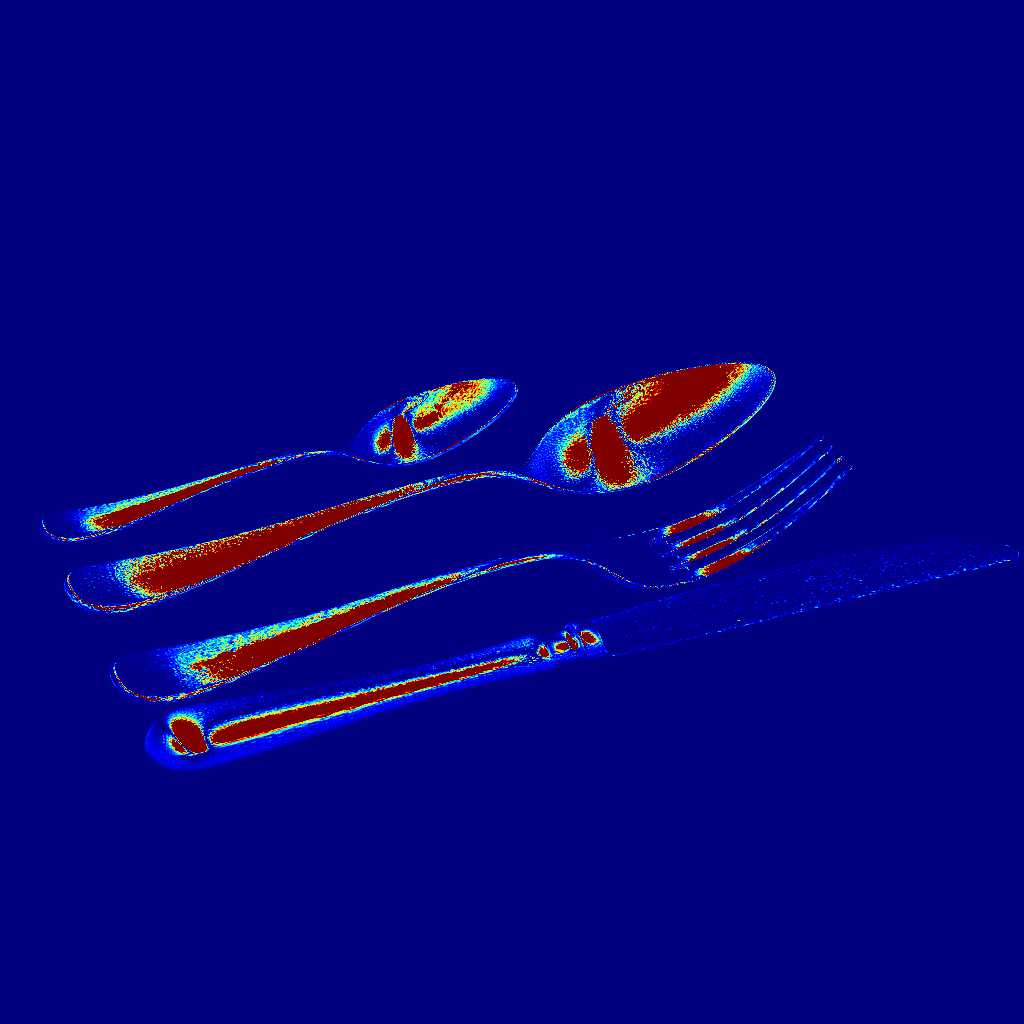}}}
        \end{overpic}} &
        \frame{\begin{overpic}[width=\imgW\textwidth]{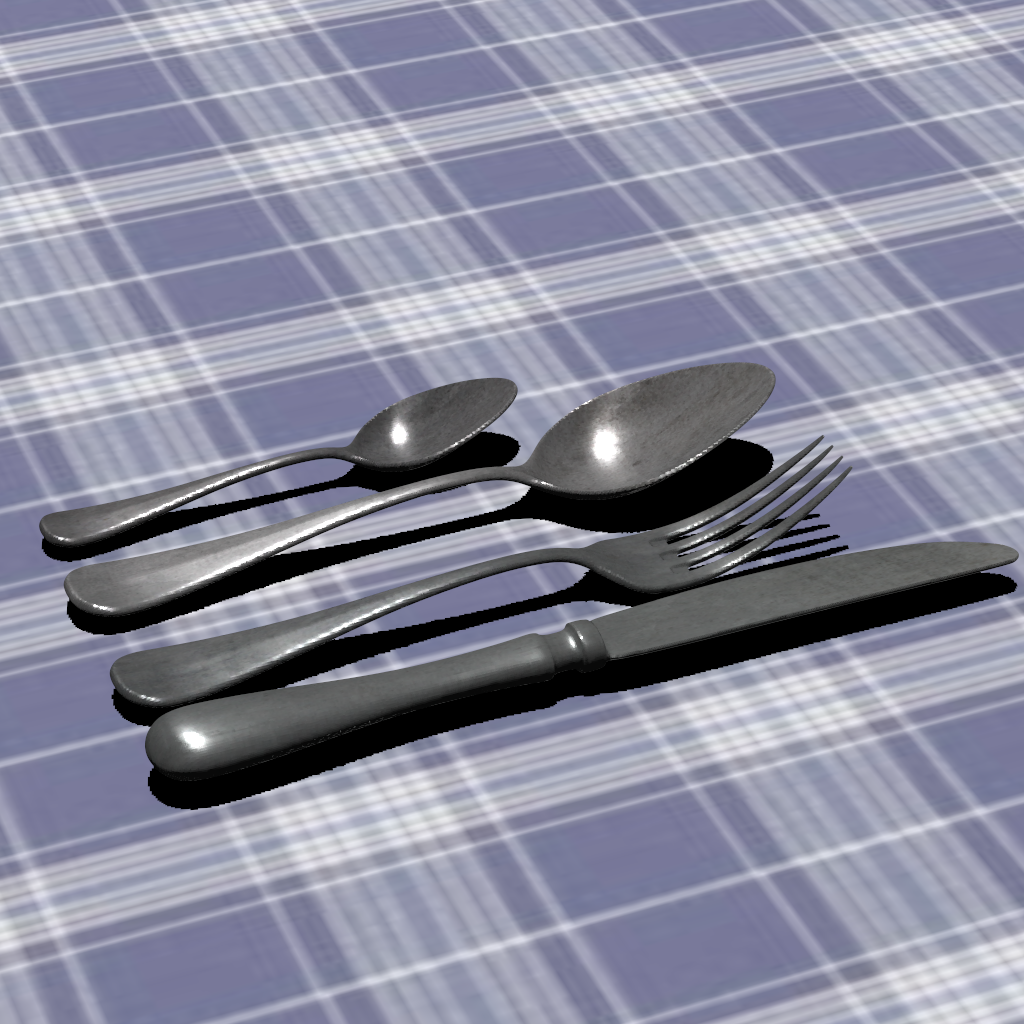}%
        \put(\putX,\putY){\frame{\includegraphics[width=\putW\linewidth]{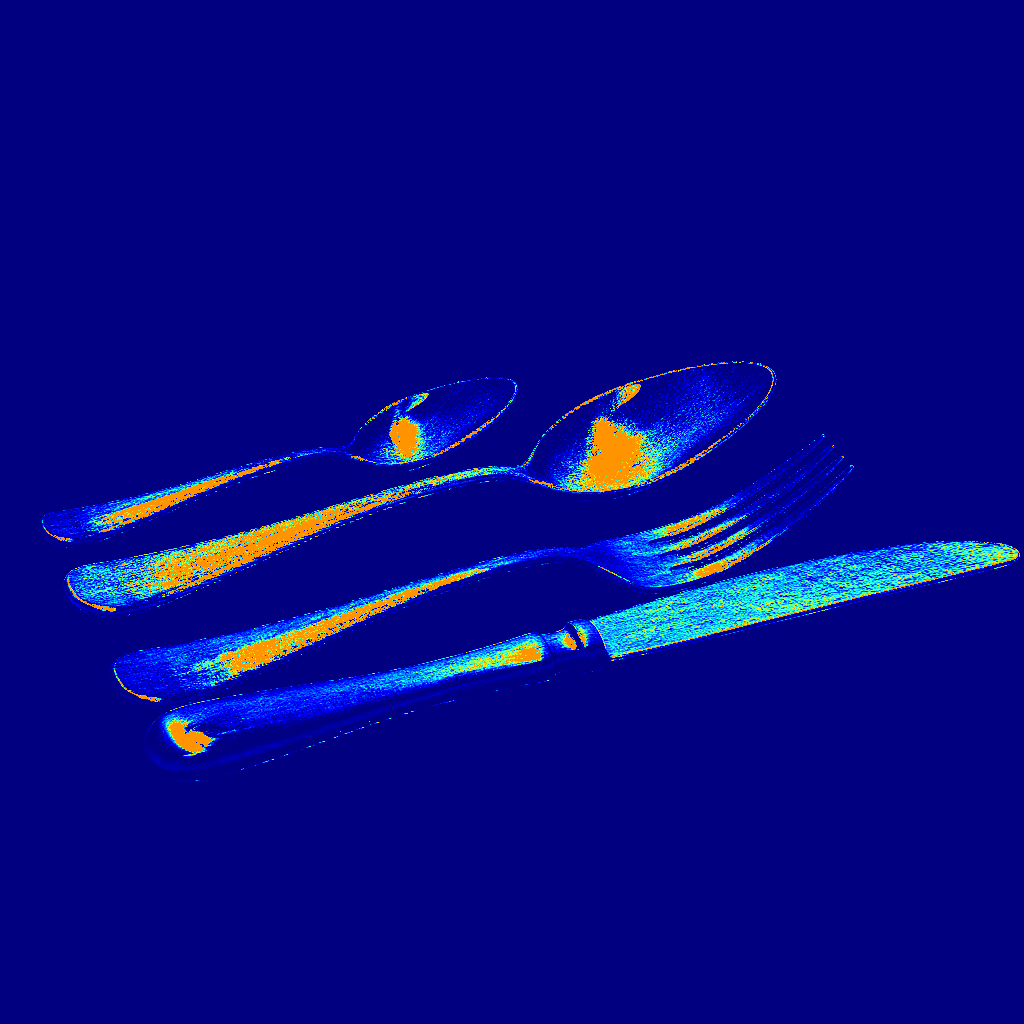}}}
        \end{overpic}} &
		\frame{\includegraphics[width=\imgW\textwidth]{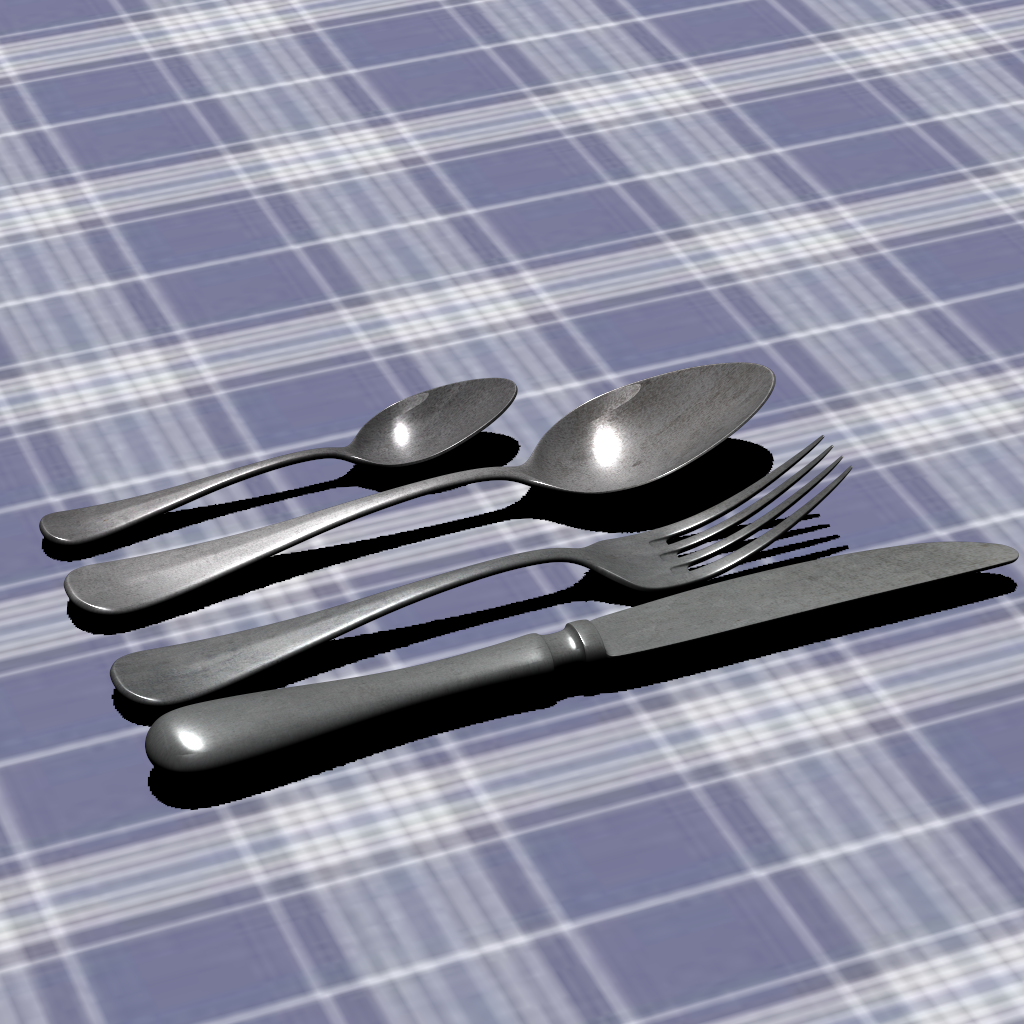}}\\
		&\multicolumn{1}{l}{\textbf{MSE:} \hspace{11mm} 14.01$\times10^{-3}$}& \multicolumn{1}{r}{18.07$\times10^{-3}$}& \multicolumn{1}{r}{18.14$\times10^{-3}$}& \multicolumn{1}{r}{14.15$\times10^{-3}$}& \multicolumn{1}{r}{4.62$\times10^{-3}$} & \\
								
		\begin{sideways} \hspace{10mm}\textbf{Oak} \end{sideways} &
		\frame{\begin{overpic}[width=\imgW\textwidth]{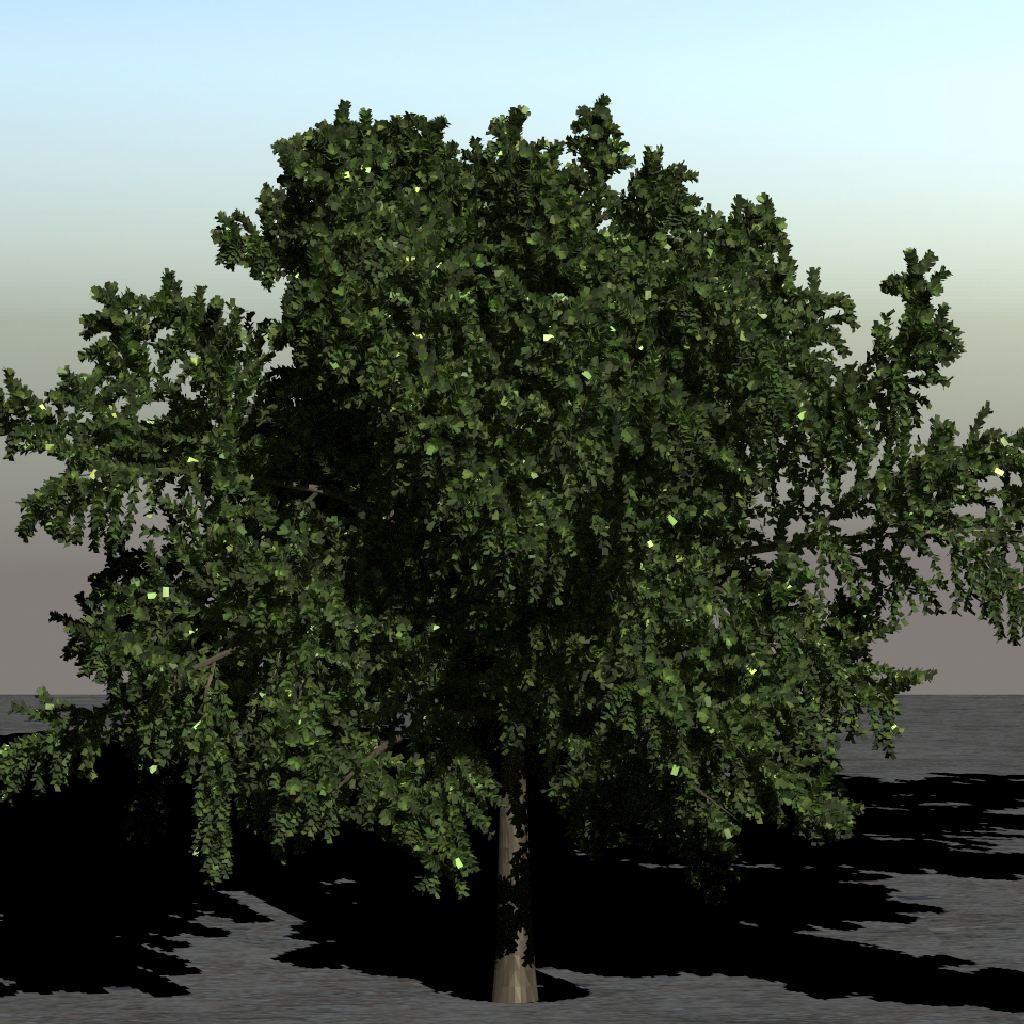}%
        \put(\putX,\putY){\frame{\includegraphics[width=\putW\linewidth]{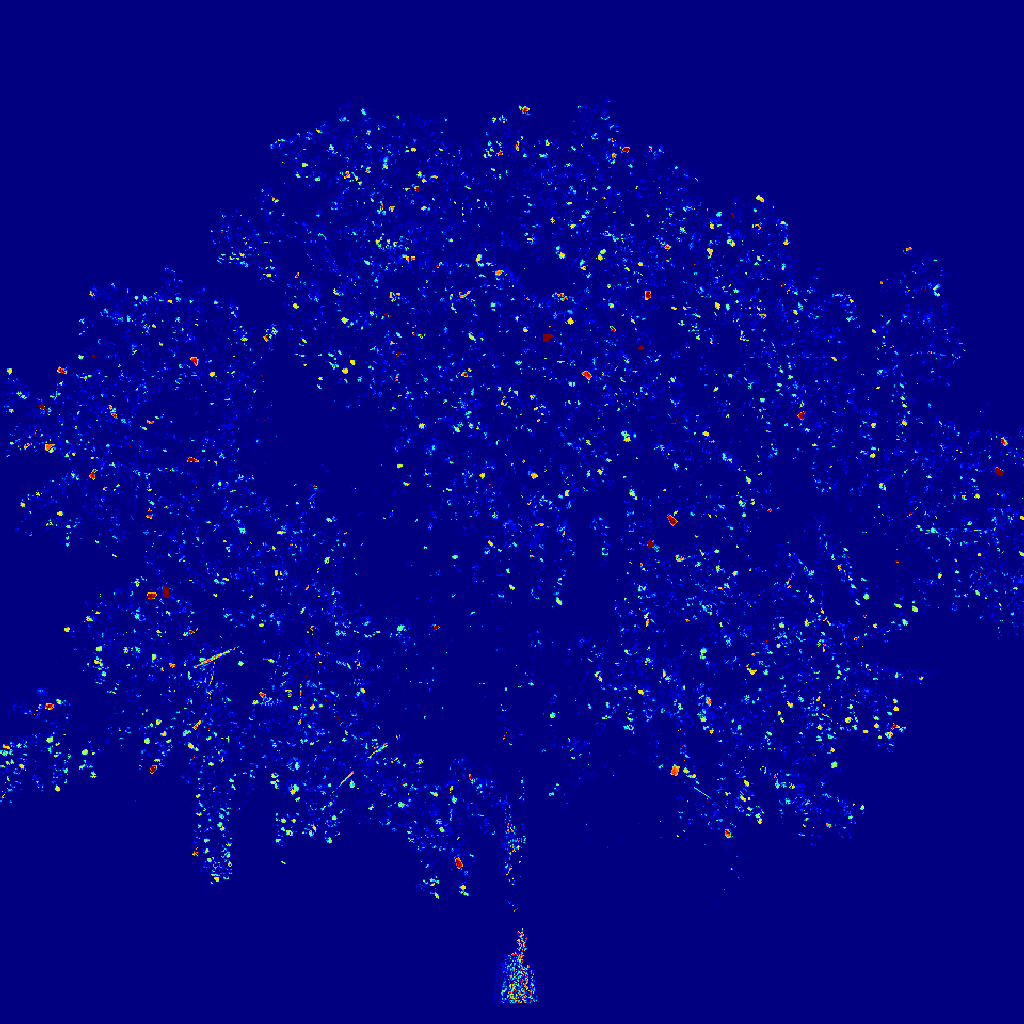}}}
        \end{overpic}} &
		\frame{\begin{overpic}[width=\imgW\textwidth]{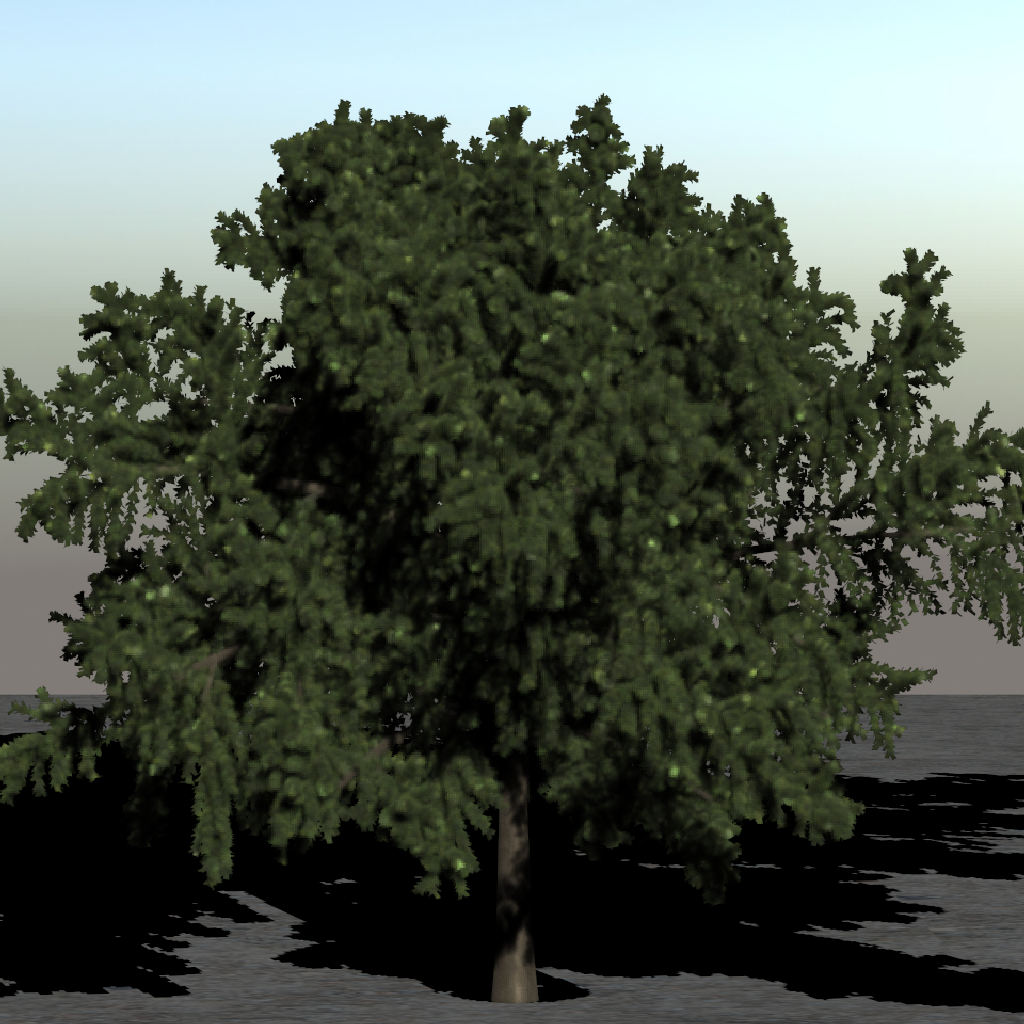}%
        \put(\putX,\putY){\frame{\includegraphics[width=\putW\linewidth]{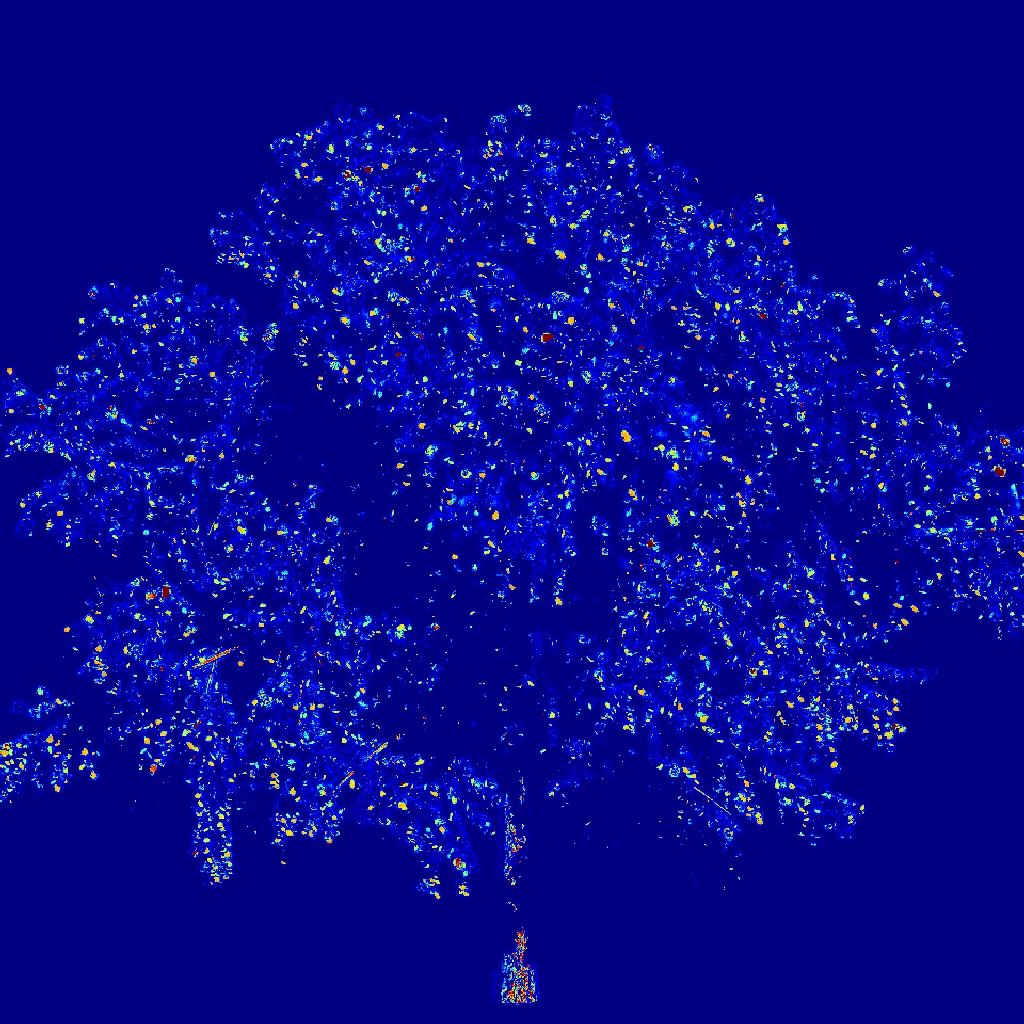}}}
        \end{overpic}} &
        \frame{\begin{overpic}[width=\imgW\textwidth]{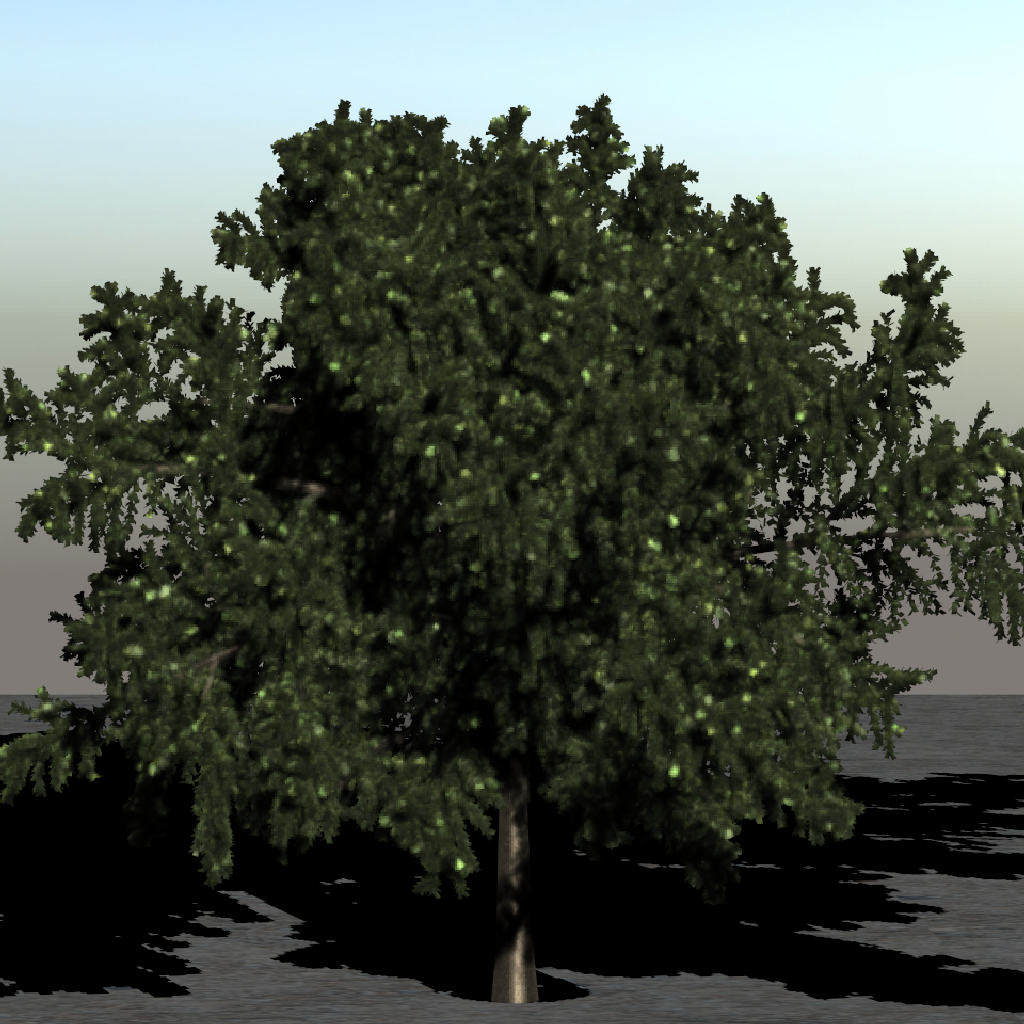}%
        \put(\putX,\putY){\frame{\includegraphics[width=\putW\linewidth]{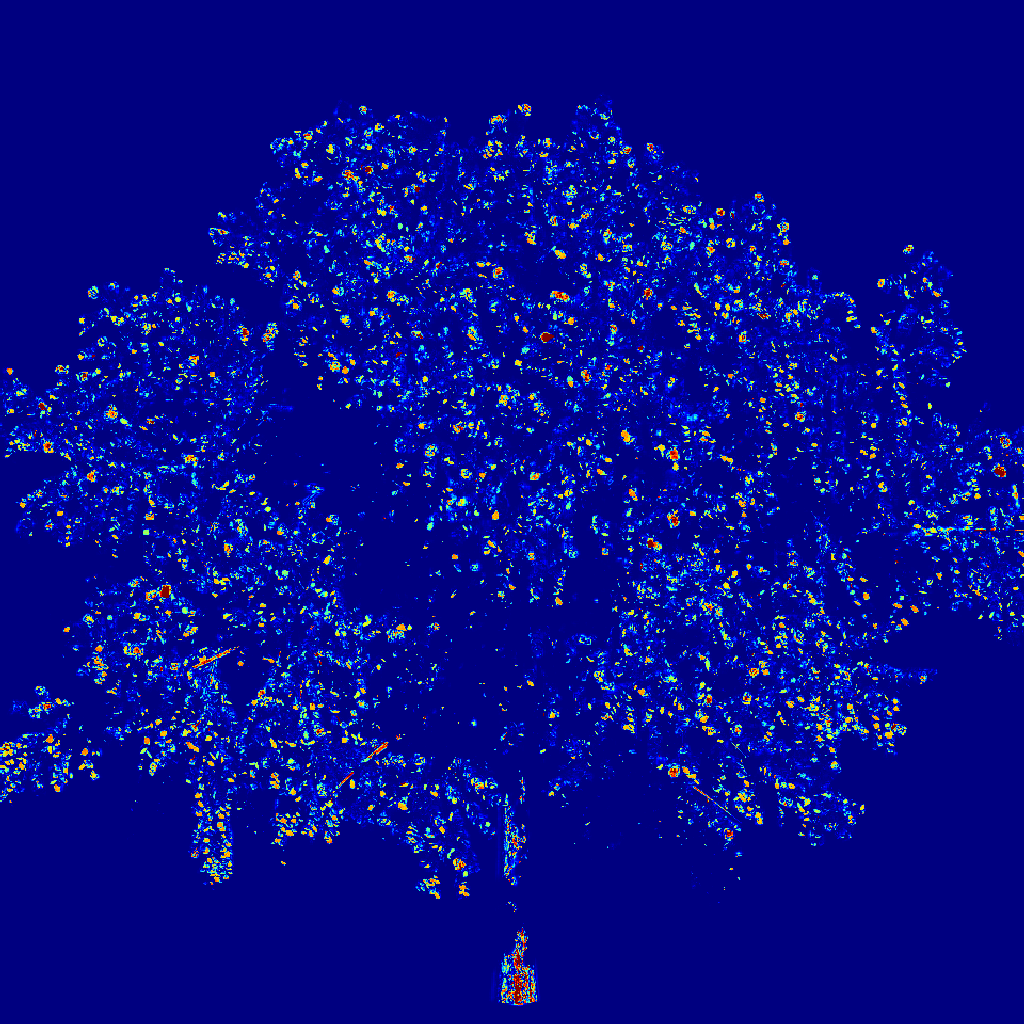}}}
        \end{overpic}} &
        \frame{\begin{overpic}[width=\imgW\textwidth]{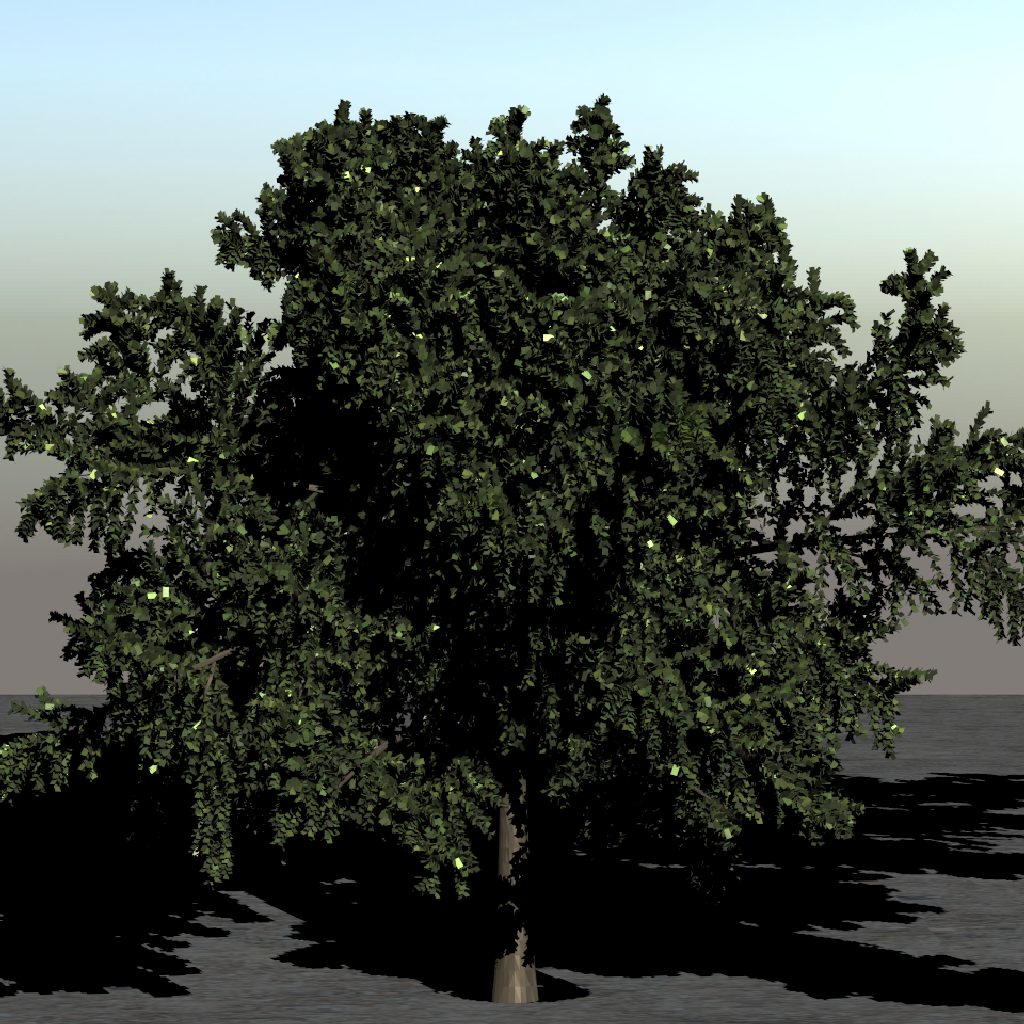}%
        \put(\putX,\putY){\frame{\includegraphics[width=\putW\linewidth]{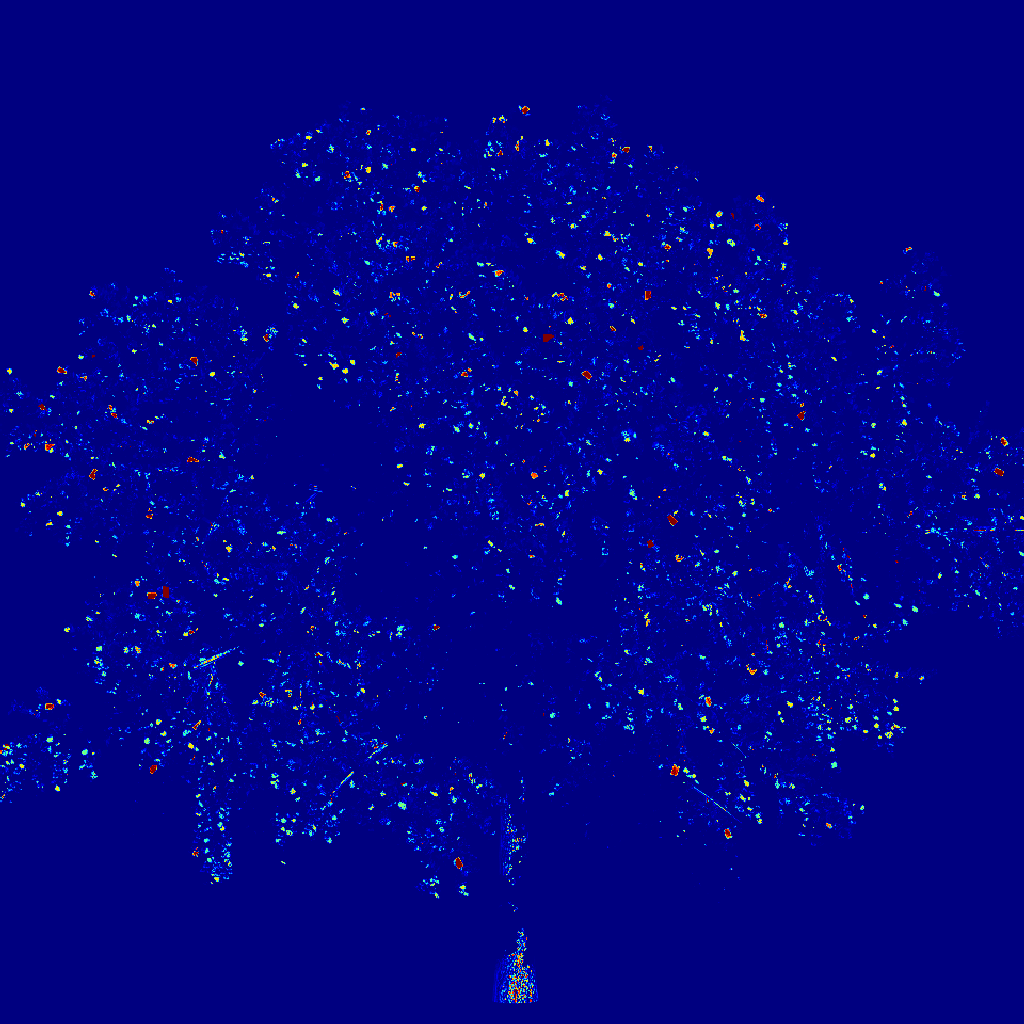}}}
        \end{overpic}} &
        \frame{\begin{overpic}[width=\imgW\textwidth]{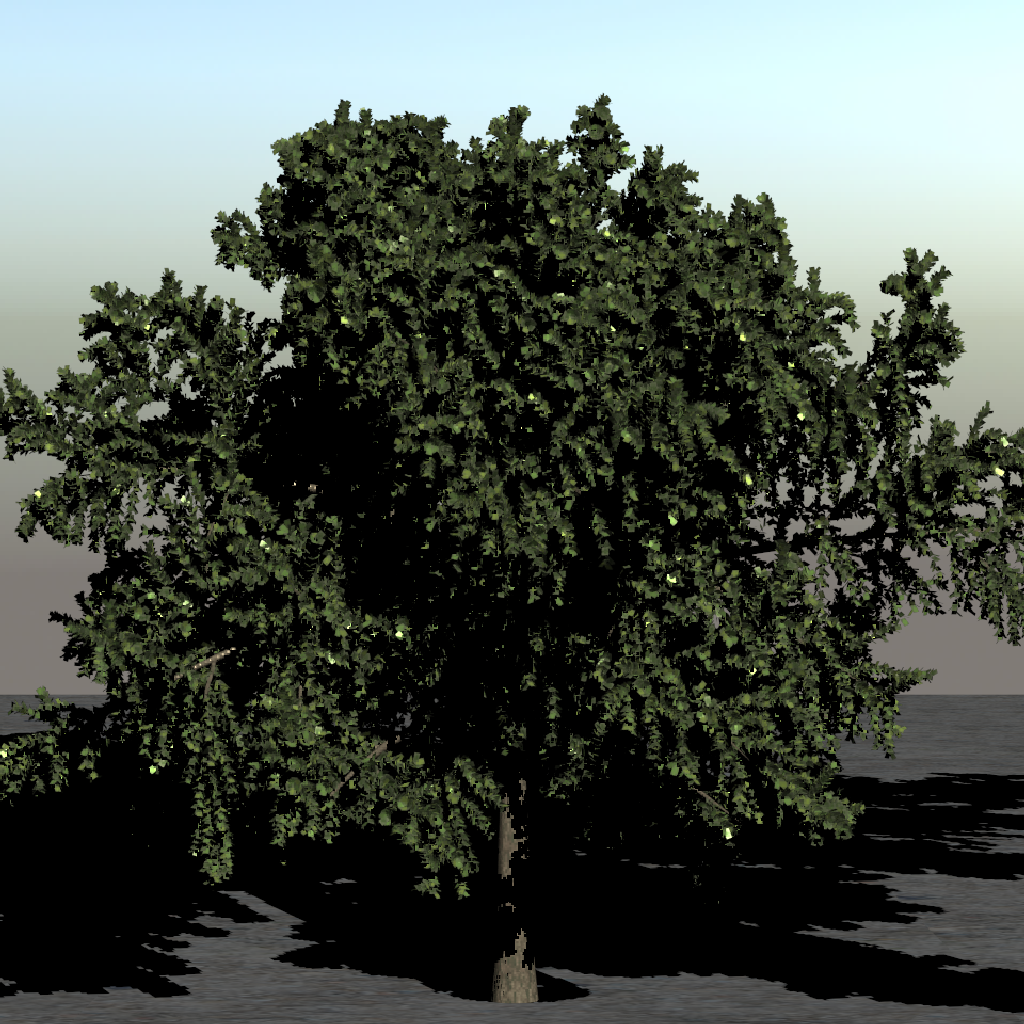}%
        \put(\putX,\putY){\frame{\includegraphics[width=\putW\linewidth]{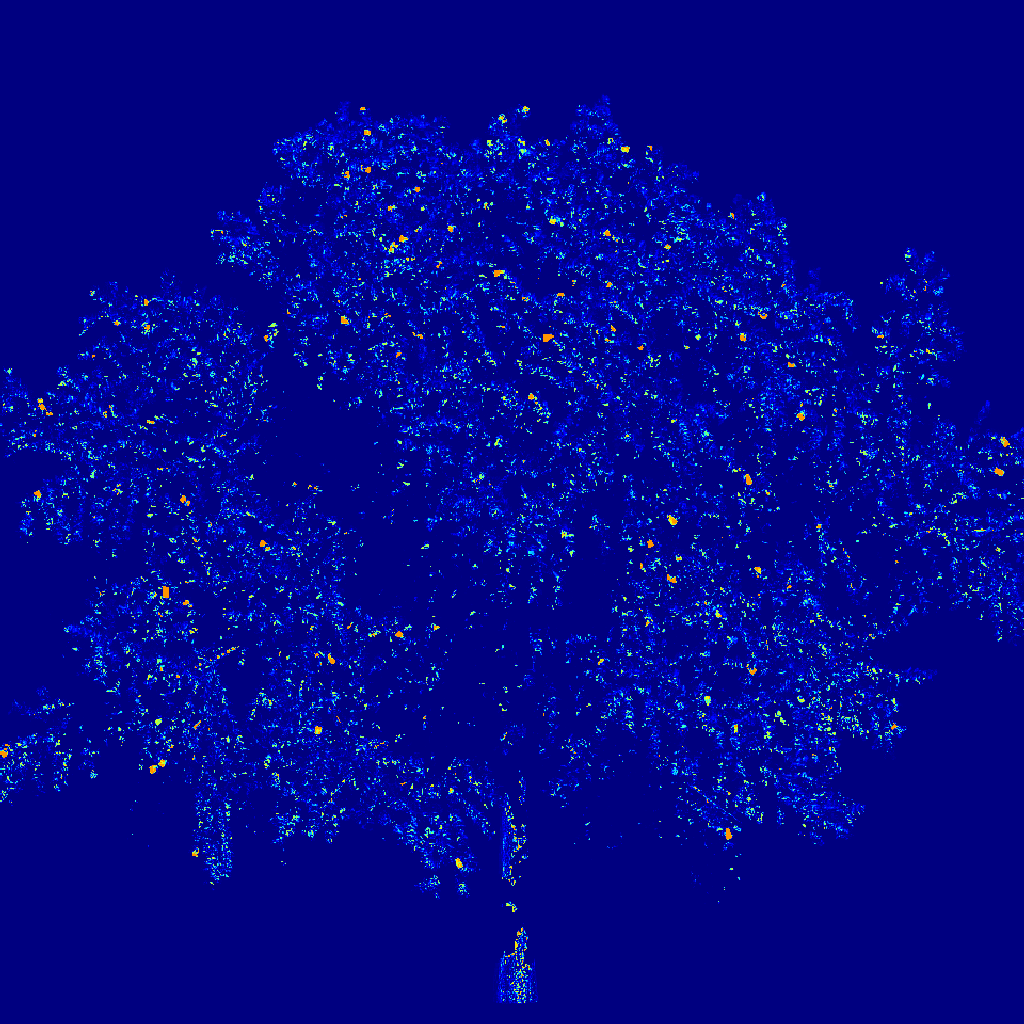}}}
        \end{overpic}} &
		\frame{\includegraphics[width=\imgW\textwidth]{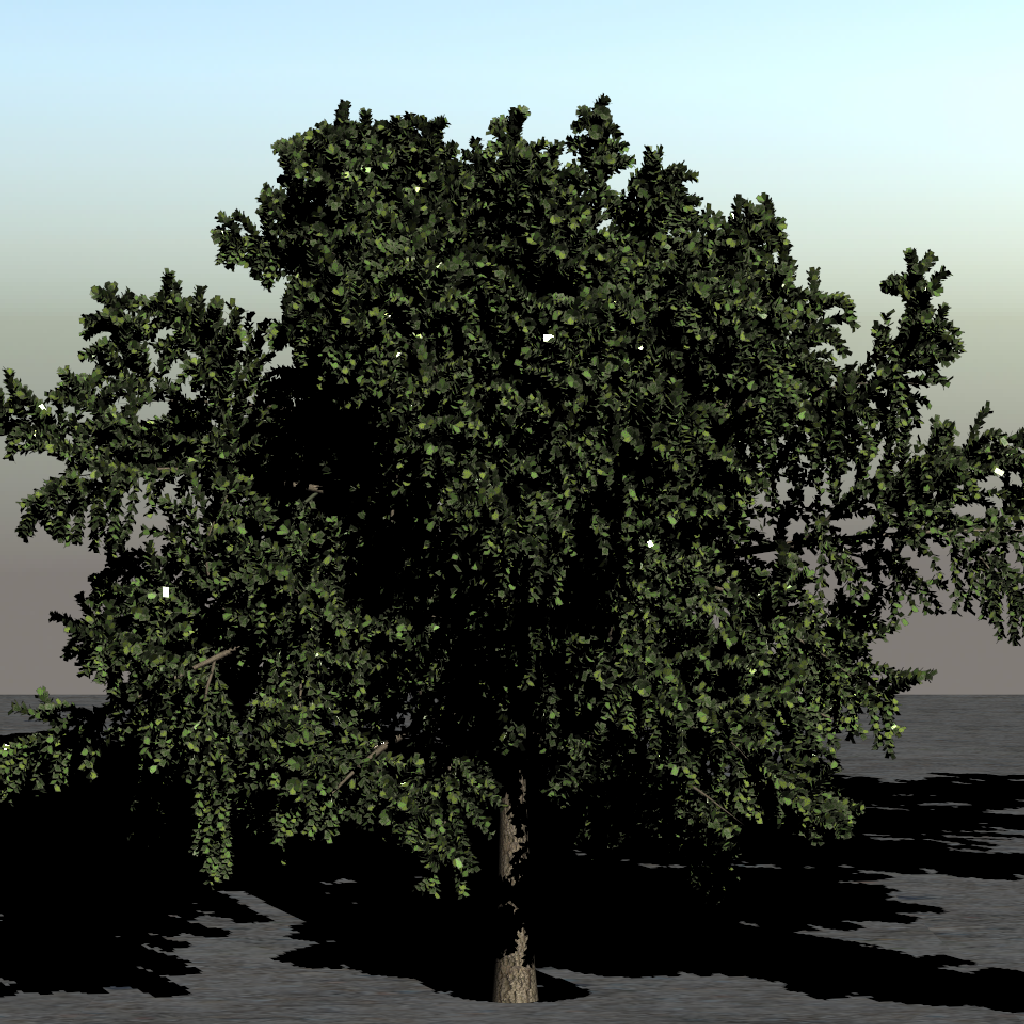}}\\
		&\multicolumn{1}{l}{\textbf{MSE:} \hspace{12mm} 6.28$\times10^{-3}$}& \multicolumn{1}{r}{7.67$\times10^{-3}$}& \multicolumn{1}{r}{7.55$\times10^{-3}$}& \multicolumn{1}{r}{6.05$\times10^{-3}$}& \multicolumn{1}{r}{5.81$\times10^{-3}$}& \\	
        		
		\begin{sideways} \hspace{7.5mm}\textbf{Mossy Rock} \end{sideways} &
		\frame{\begin{overpic}[width=\imgW\textwidth]{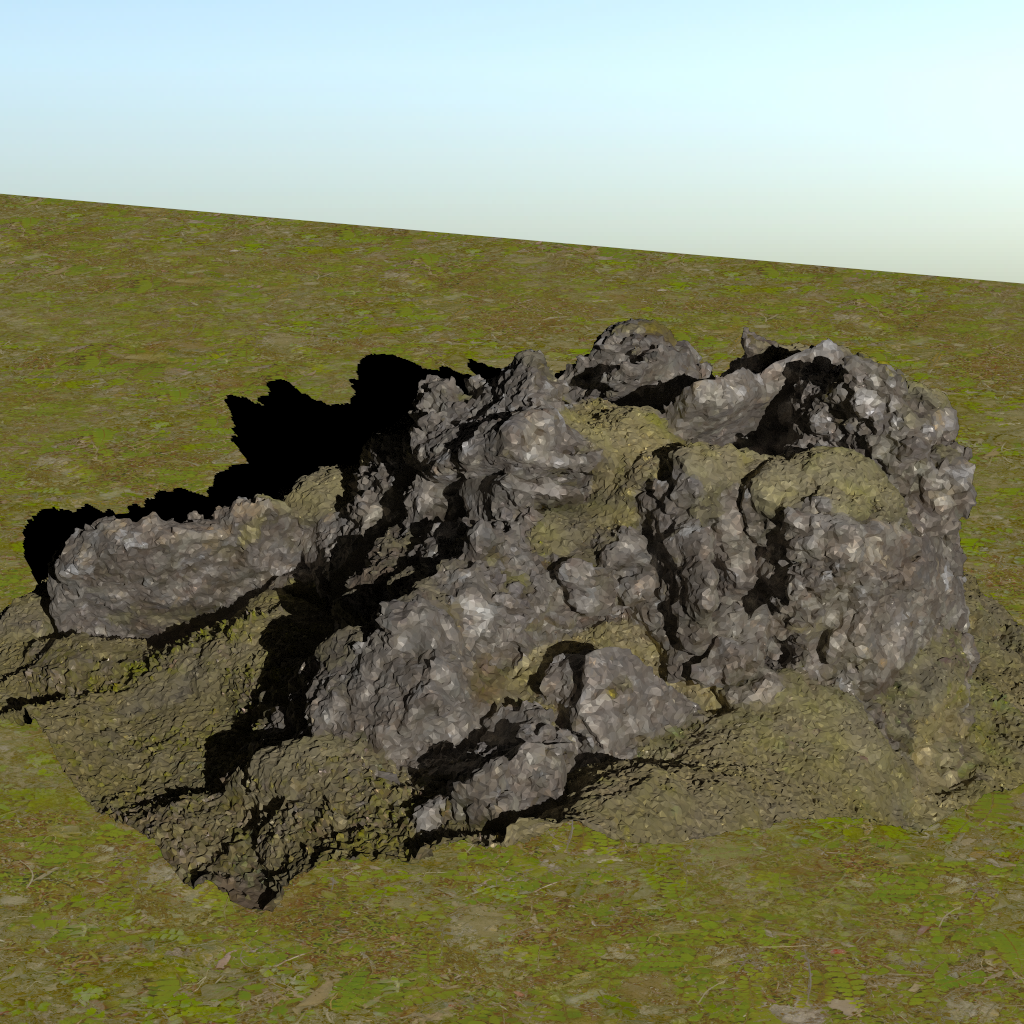}%
        \put(\putX,\putY){\frame{\includegraphics[width=\putW\linewidth]{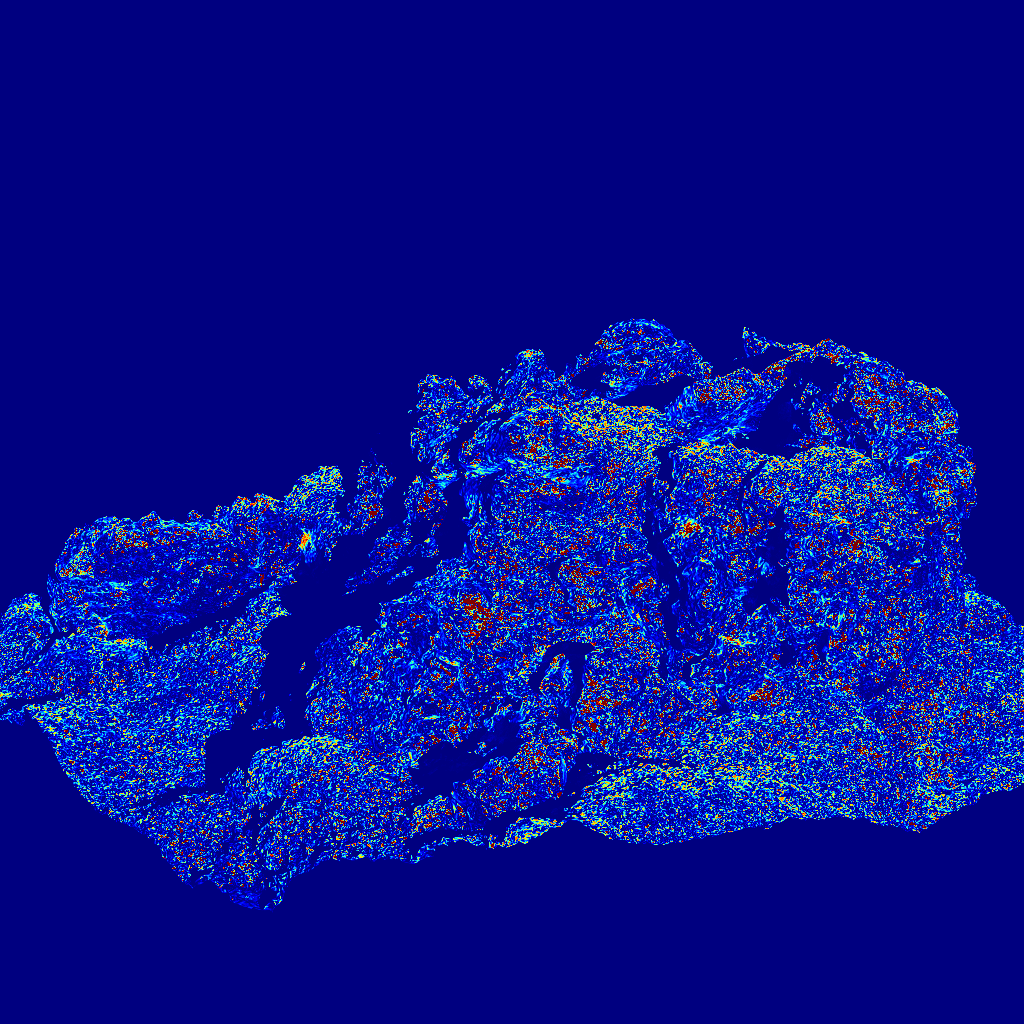}}}
        \end{overpic}} &
		\frame{\begin{overpic}[width=\imgW\textwidth]{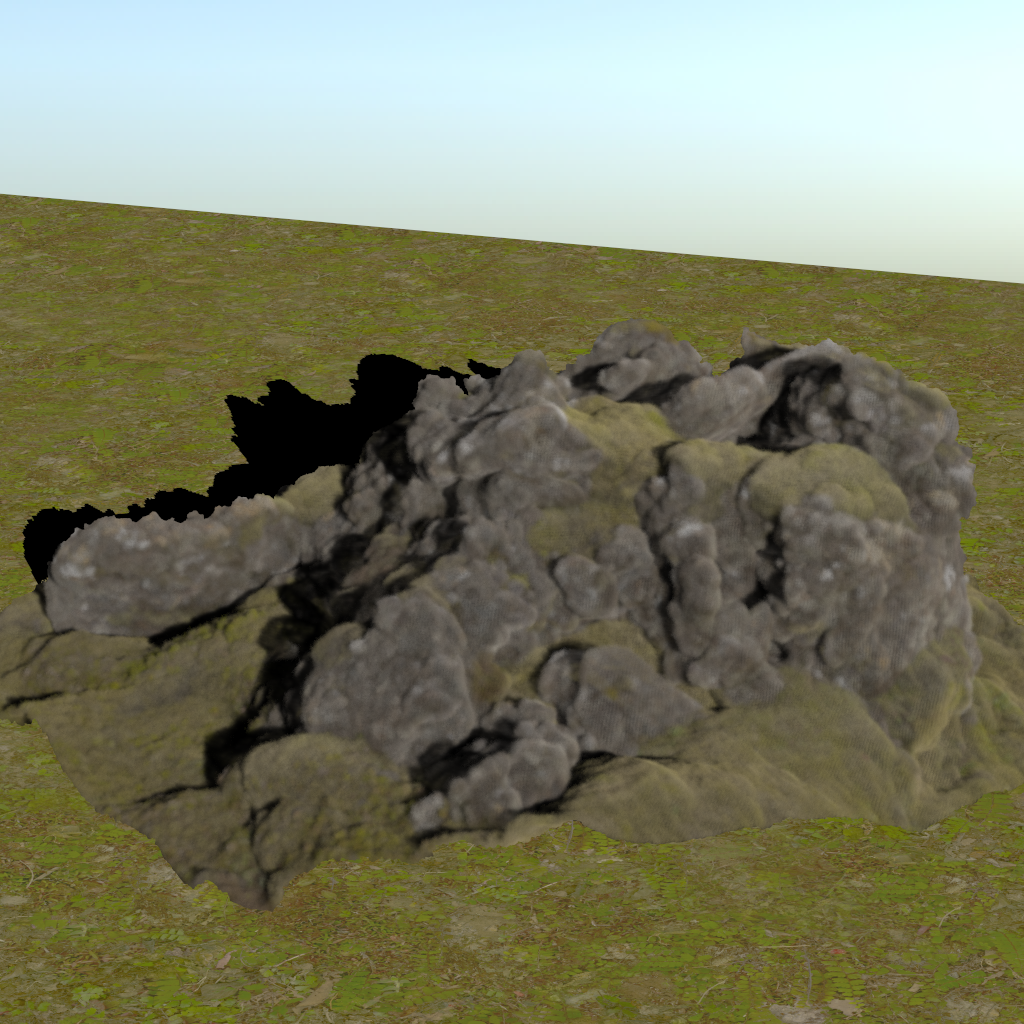}%
        \put(\putX,\putY){\frame{\includegraphics[width=\putW\linewidth]{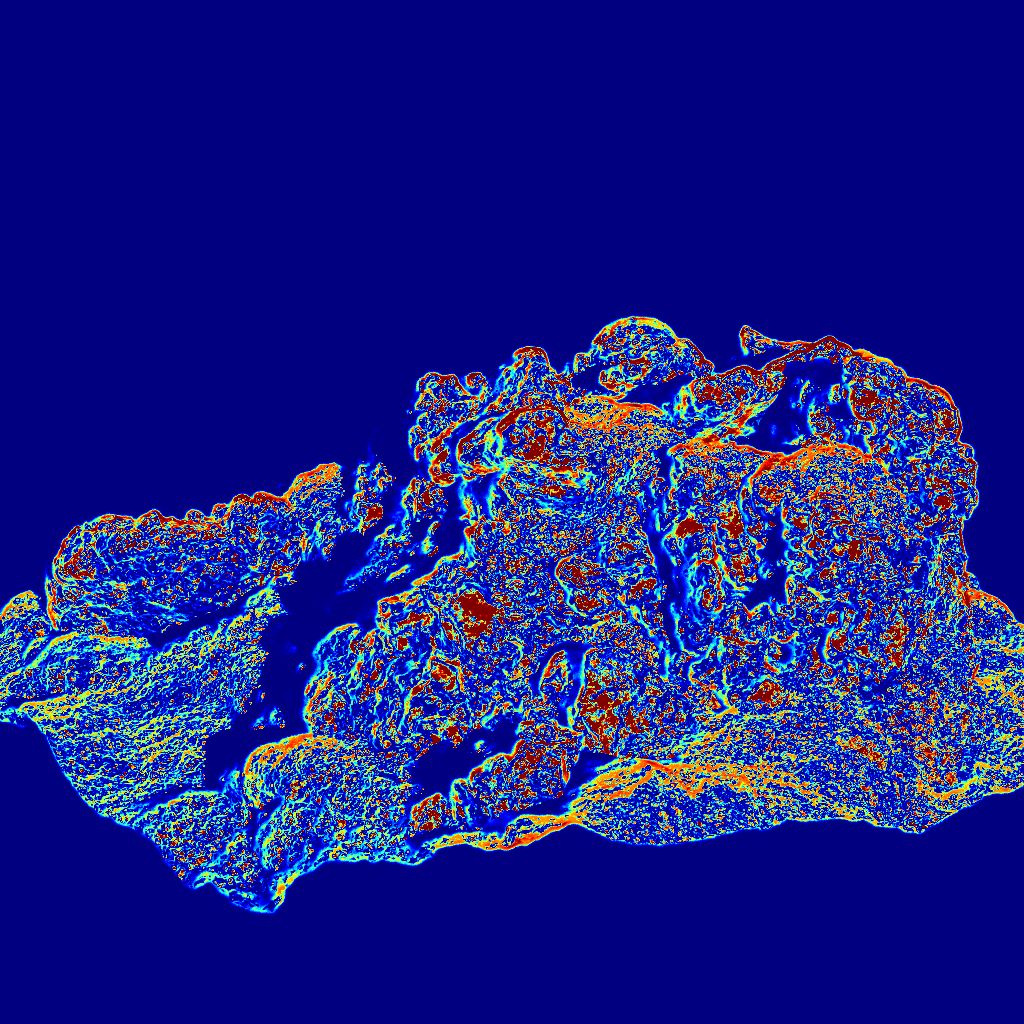}}}
        \end{overpic}} &
        \frame{\begin{overpic}[width=\imgW\textwidth]{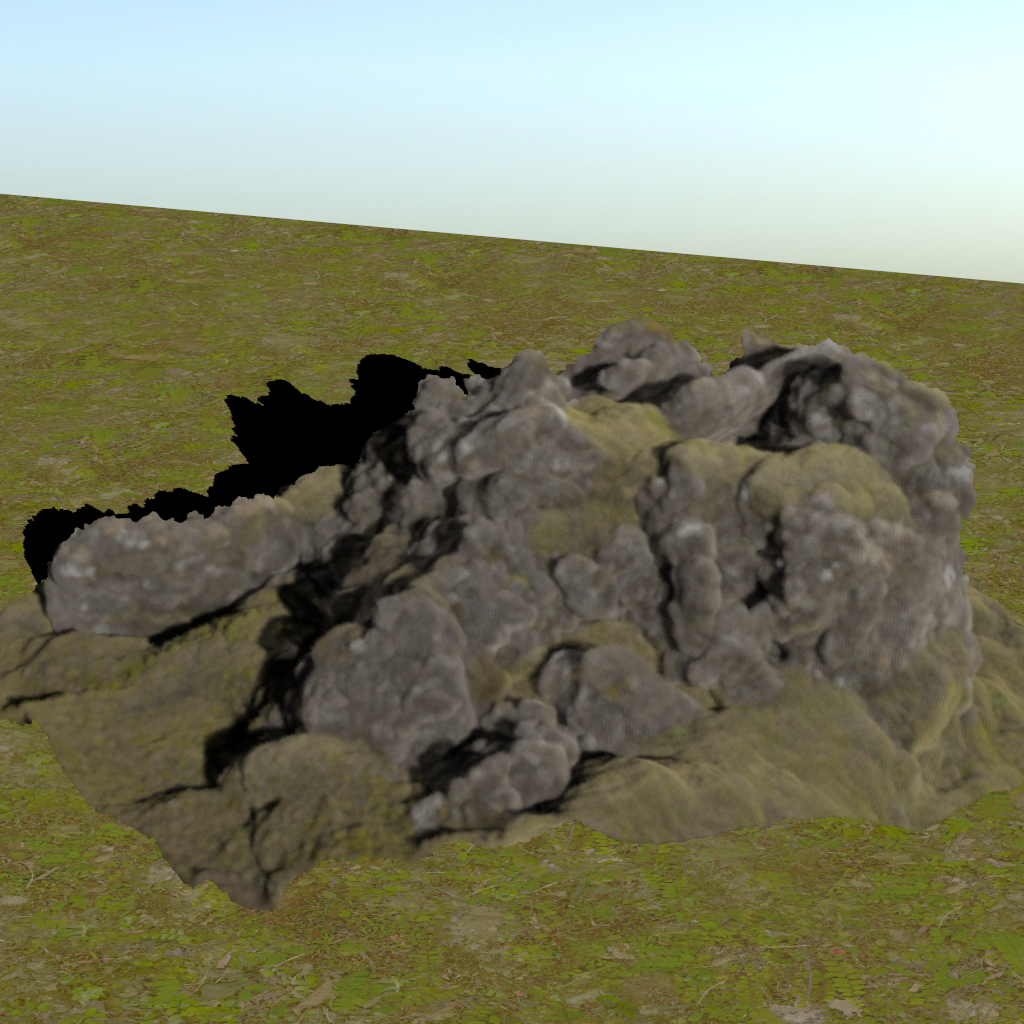}%
        \put(\putX,\putY){\frame{\includegraphics[width=\putW\linewidth]{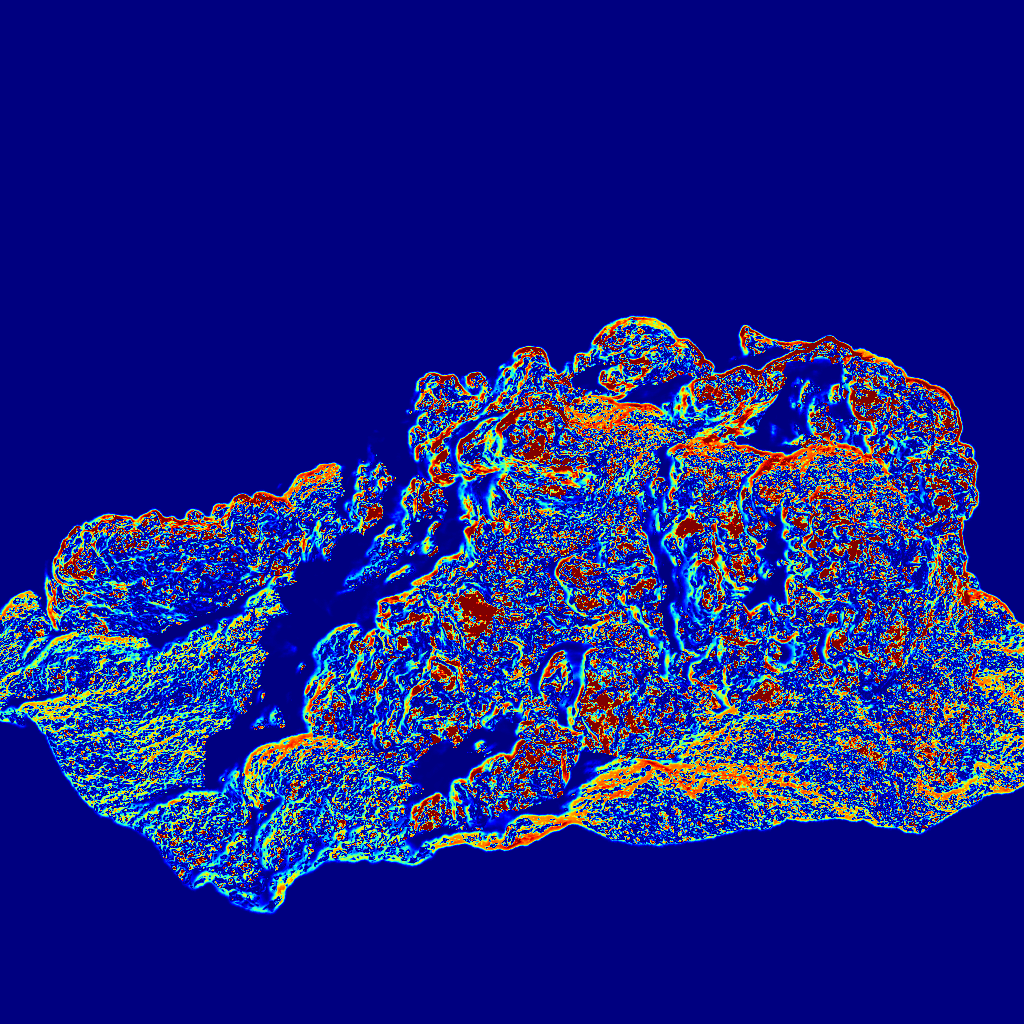}}}
        \end{overpic}} &
        \frame{\begin{overpic}[width=\imgW\textwidth]{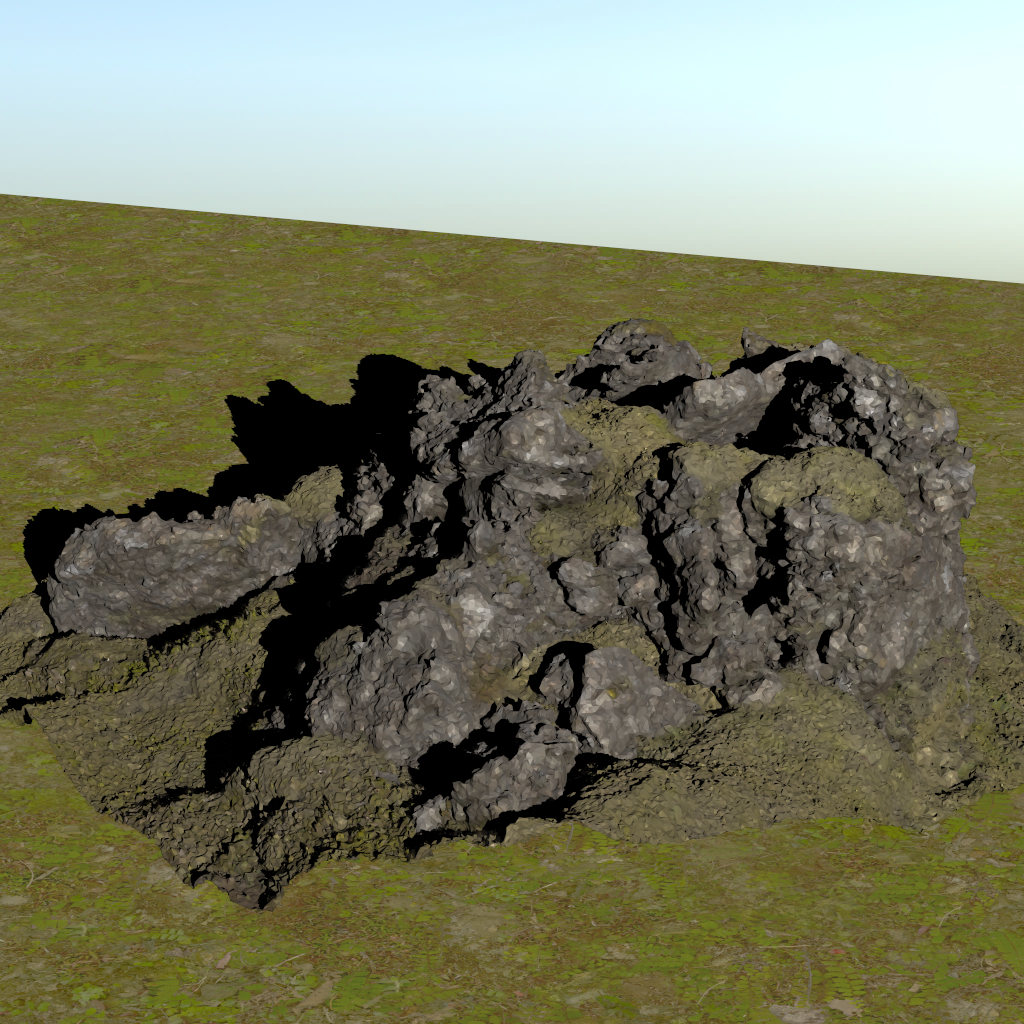}%
        \put(\putX,\putY){\frame{\includegraphics[width=\putW\linewidth]{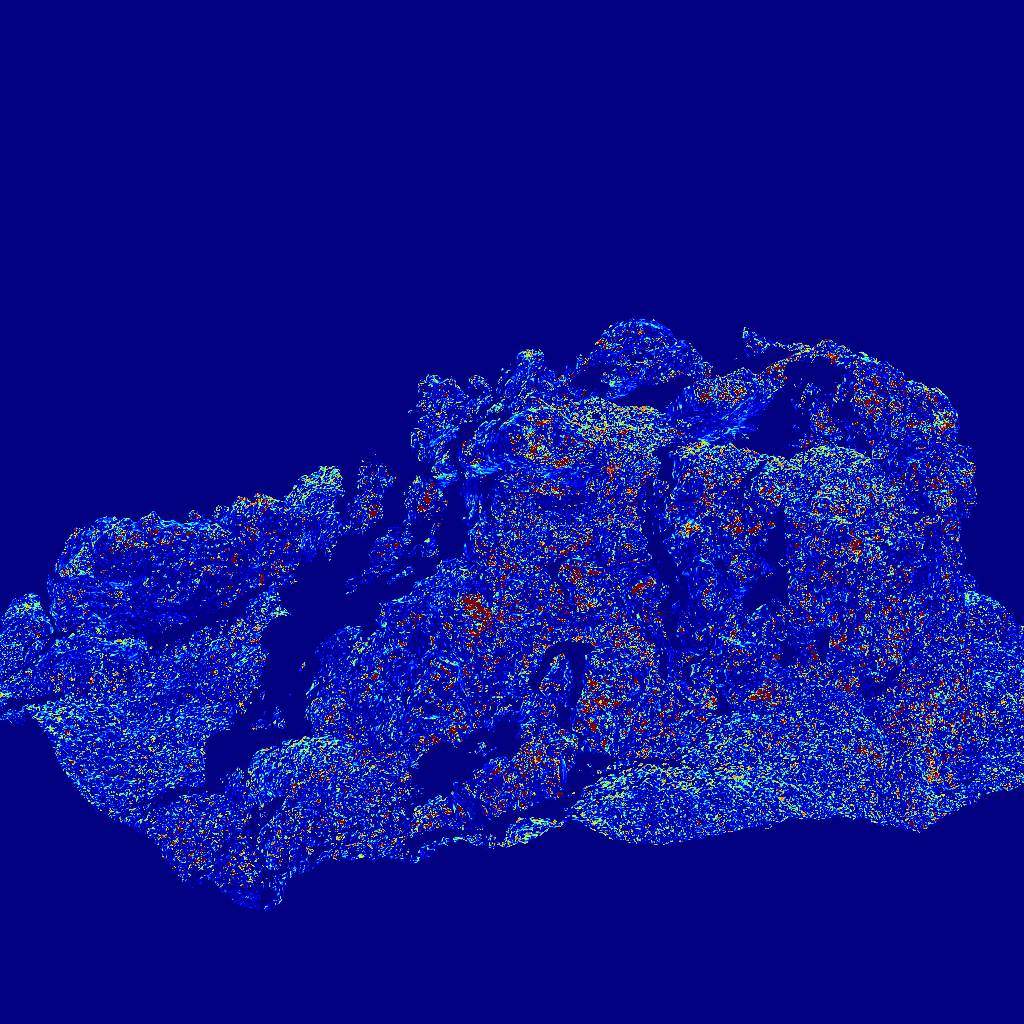}}}
        \end{overpic}} &
        \frame{\begin{overpic}[width=\imgW\textwidth]{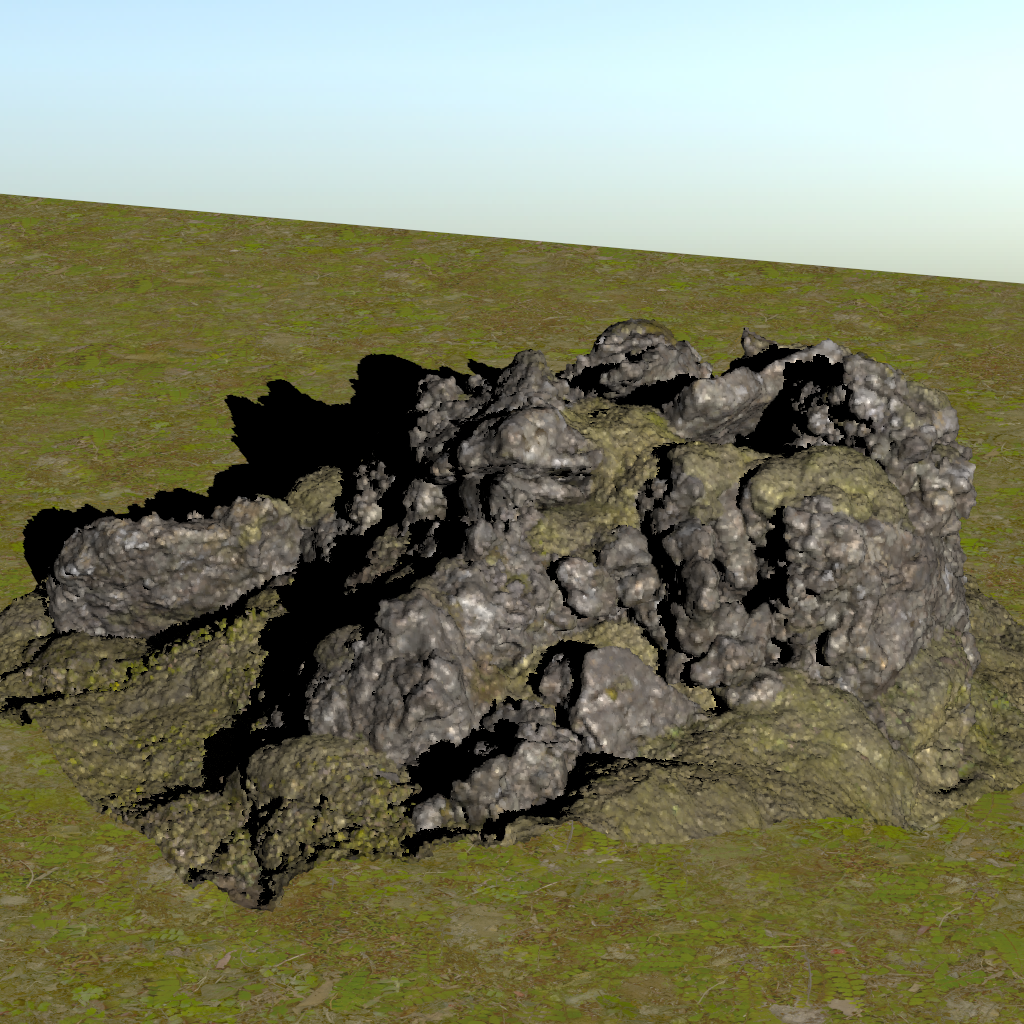}%
        \put(\putX,\putY){\frame{\includegraphics[width=\putW\linewidth]{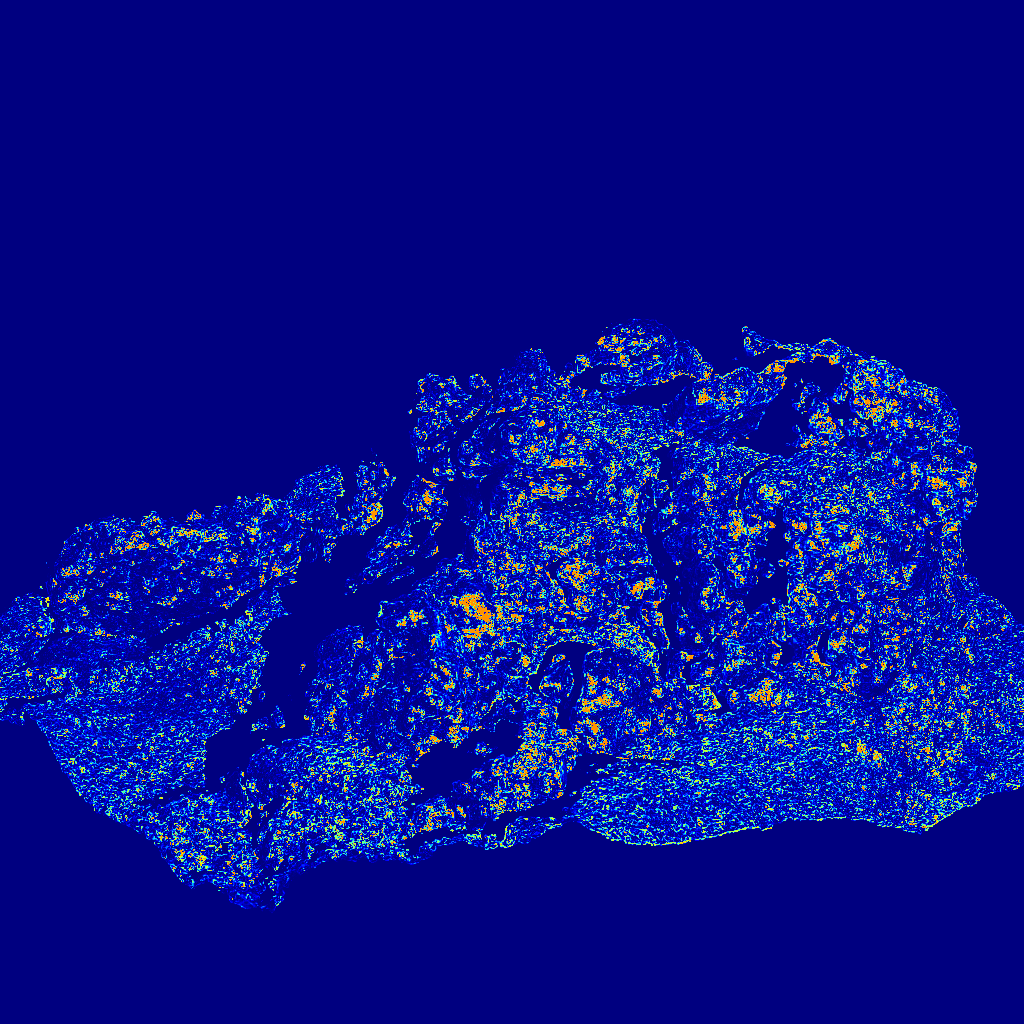}}}
        \end{overpic}} &
		\frame{\includegraphics[width=\imgW\textwidth]{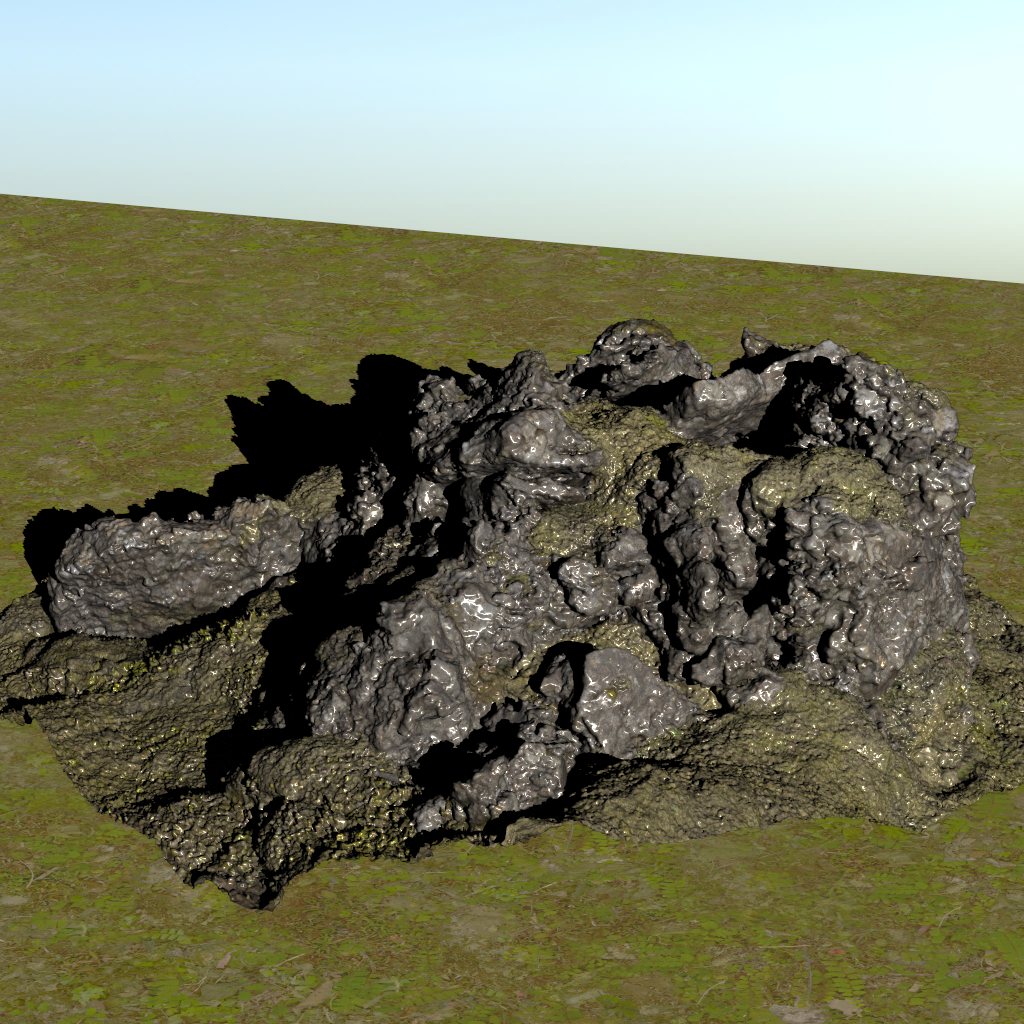}}\\
		&\multicolumn{1}{l}{\textbf{MSE:} \hspace{12mm} 3.60$\times10^{-3}$}& \multicolumn{1}{r}{5.56$\times10^{-3}$}& \multicolumn{1}{r}{5.56$\times10^{-3}$}& \multicolumn{1}{r}{3.65$\times10^{-3}$}& \multicolumn{1}{r}{2.25$\times10^{-3}$} & \\
	\end{tabular}
	\vspace{-3mm}
	\caption{Comparisons with state-of-the-art approaches \eggx~\cite{Dong15}, \sggx~\cite{Heitz15}, \selfshadow~\cite{Loubet18}, and \hybrid~\cite{Loubet17}. For all scenes, we demonstrate higher fidelity to the ray traced reference, as can be seen in the difference images (mapped from blue to red with increasing per-pixel squared error) and the lower MSE error. Note, only \sggx ~and our method do not rely on any explicit geometry or materials for rendering. Full results can be found in the supplemental along with videos.}

	\label{fig:SotaComparisons}
	\vspace{-2mm}
\end{figure*}

\begin{figure*}[t]
	\setlength{\fboxrule}{10pt}%
	\setlength{\insetvsep}{20pt}%
	\setlength{\tabcolsep}{1pt}
	\renewcommand{\arraystretch}{1}%
	\footnotesize%
	\begin{tabular}{lcccccc}
		
		& \multicolumn{3}{c}{Scale 4 ($16^3$ grid)}  & \multicolumn{3}{c}{Scale 8 ($256^3$ grid)} \\
        & Ours & Reference (16K spp) & Difference & Ours & Reference (16K spp) & Difference \\
        \begin{sideways}  \hspace{10mm}\textbf{Forest} \end{sideways} &
		\addImage{Figures/\resultsDir/Forest/data/\oursResult\suffixLRComplex-img.png}{black}{\imgWr} &
		\addImage{Figures/\resultsDir/Forest/groundtruth\suffixLRComplex.png}{black}{\imgWr} &
		\addImage{Figures/\resultsDir/Forest/data/\oursResult\suffixLRComplex-\errorViz.png}{black}{\imgWr} &
		\addImage{Figures/\resultsDirHighRes/Forest/data/\oursResult\suffixComplex-img.png}{black}{\imgWr} &
		\addImage{Figures/\resultsDirHighRes/Forest/groundtruth\suffixComplex.png}{black}{\imgWr} &
		\addImage{Figures/\resultsDirHighRes/Forest/data/\oursResult\suffixComplex-\errorViz.png}{black}{\imgWr} \\
		& \multicolumn{1}{l}{\textbf{Per-pixel mem. (KB):} 249.2} & \multicolumn{1}{r}{84{,}368.2} & & \multicolumn{1}{r}{5.4} & \multicolumn{1}{r}{361.2} & \\
		
        \begin{sideways}  \hspace{7mm}\textbf{Parking Lot} \end{sideways} &
		\addImage{Figures/\resultsDir/ParkingLot/data/\oursResult\suffixLRComplex-img.png}{black}{\imgWr} &
		\addImage{Figures/\resultsDir/ParkingLot/groundtruth\suffixLRComplex.png}{black}{\imgWr} &
		\addImage{Figures/\resultsDir/ParkingLot/data/\oursResult\suffixLRComplex-\errorViz.png}{black}{\imgWr} &
		\addImage{Figures/\resultsDirHighRes/ParkingLot/data/\oursResult\suffixComplex-img.png}{black}{\imgWr} &
		\addImage{Figures/\resultsDirHighRes/ParkingLot/groundtruth\suffixComplex.png}{black}{\imgWr} &
		\addImage{Figures/\resultsDirHighRes/ParkingLot/data/\oursResult\suffixComplex-\errorViz.png}{black}{\imgWr} \\
		& \multicolumn{1}{l}{\textbf{Per-pixel mem. (KB):} 138.2} & \multicolumn{1}{r}{4{,}825{,}832.7} & & \multicolumn{1}{r}{47.4} & \multicolumn{1}{r}{20{,}493.3} & \\
		
        \begin{sideways}  \hspace{2mm}\textbf{Stormtrooper Army} \end{sideways} &
		\addImage{Figures/\resultsDir/StormtrooperSquad/data/\oursResult\suffixLRComplex-img.png}{black}{\imgWr} &
		\addImage{Figures/\resultsDir/StormtrooperSquad/groundtruth\suffixLRComplex.png}{black}{\imgWr} &
		\addImage{Figures/\resultsDir/StormtrooperSquad/data/\oursResult\suffixLRComplex-\errorViz.png}{black}{\imgWr} &
		\addImage{Figures/\resultsDirHighRes/StormtrooperSquad/data/\oursResult\suffixComplex-img.png}{black}{\imgWr} &
		\addImage{Figures/\resultsDirHighRes/StormtrooperSquad/groundtruth\suffixComplex.png}{black}{\imgWr} &
		\addImage{Figures/\resultsDirHighRes/StormtrooperSquad/data/\oursResult\suffixComplex-\errorViz.png}{black}{\imgWr} \\
		& \multicolumn{1}{l}{\textbf{Per-pixel mem. (KB):} \hspace{1mm} 67.2} & \multicolumn{1}{r}{17{,}213.8} & & \multicolumn{1}{r}{16.0} & \multicolumn{1}{r}{81.4} & \\
		
        \begin{sideways}  \hspace{10mm}\textbf{City} \end{sideways} &
		\addImage{Figures/\resultsDir/City/data/\oursResult\suffixLRComplex-img.png}{black}{\imgWr} &
		\addImage{Figures/\resultsDir/City/groundtruth\suffixLRComplex.png}{black}{\imgWr} &
		\addImage{Figures/\resultsDir/City/data/\oursResult\suffixLRComplex-\errorViz.png}{black}{\imgWr} &
		\addImage{Figures/\resultsDirHighRes/City/data/\oursResult\suffixComplex-img.png}{black}{\imgWr} &
		\addImage{Figures/\resultsDirHighRes/City/groundtruth\suffixComplex.png}{black}{\imgWr} &
		\addImage{Figures/\resultsDirHighRes/City/data/\oursResult\suffixComplex-\errorViz.png}{black}{\imgWr} \\
		& \multicolumn{1}{l}{\textbf{Per-pixel mem. (KB):} 107.9} & \multicolumn{1}{r}{189{,}711.8} & & \multicolumn{1}{r}{27.5} & \multicolumn{1}{r}{772.8} & \\
		
	\end{tabular}
	\vspace{-3mm}
	\caption{We compare to standard ray tracing on large scenes with significant geometric and material complexity for a coarse scale (scale 4 on the left) and a finer scale (scale 8 on the right). For example, at scale 4 the voxels divide up the scene as $16\times16\times16$ or coarser and scale 8 divides up the scene as $256\times256\times256$ or coarser. Our method's output closely matches the reference, demonstrated by blue (lower error) to red (higher error) MSE difference images, at only a fraction of the memory cost, despite not using any of the scene's original geometry and materials. The memory consumption is reported per-pixel and is computed for ray tracing by dividing the size of the scene (e.g., the size of the mesh and materials) by the number of active pixels for a given scale. Similarly, for our method, we divide the total size of all of the active voxels' latent variables by the number of active voxels (the latter is typically larger than the number of active pixels since multiple voxels often map to a given pixel). The scenes here highlight hard-to-capture scenarios including anisotropic specularities using the multi-lobe Disney BRDF as well as mixtures of macrosurfaces and microgeometry illuminated with an environment map. See the supplemental for the full images, as well as videos of these scenes demonstrating our network's ability to render temporally-smooth, anti-aliased sequences.}
	\label{fig:ComplexScenes}
	\vspace{-2mm}
\end{figure*}

\begin{figure*}[!t]
  \centering
  \setlength{\tabcolsep}{1pt}
  \footnotesize%
  \begin{tabular}{ccc}
    
    \includegraphics[width=\plotW\linewidth]{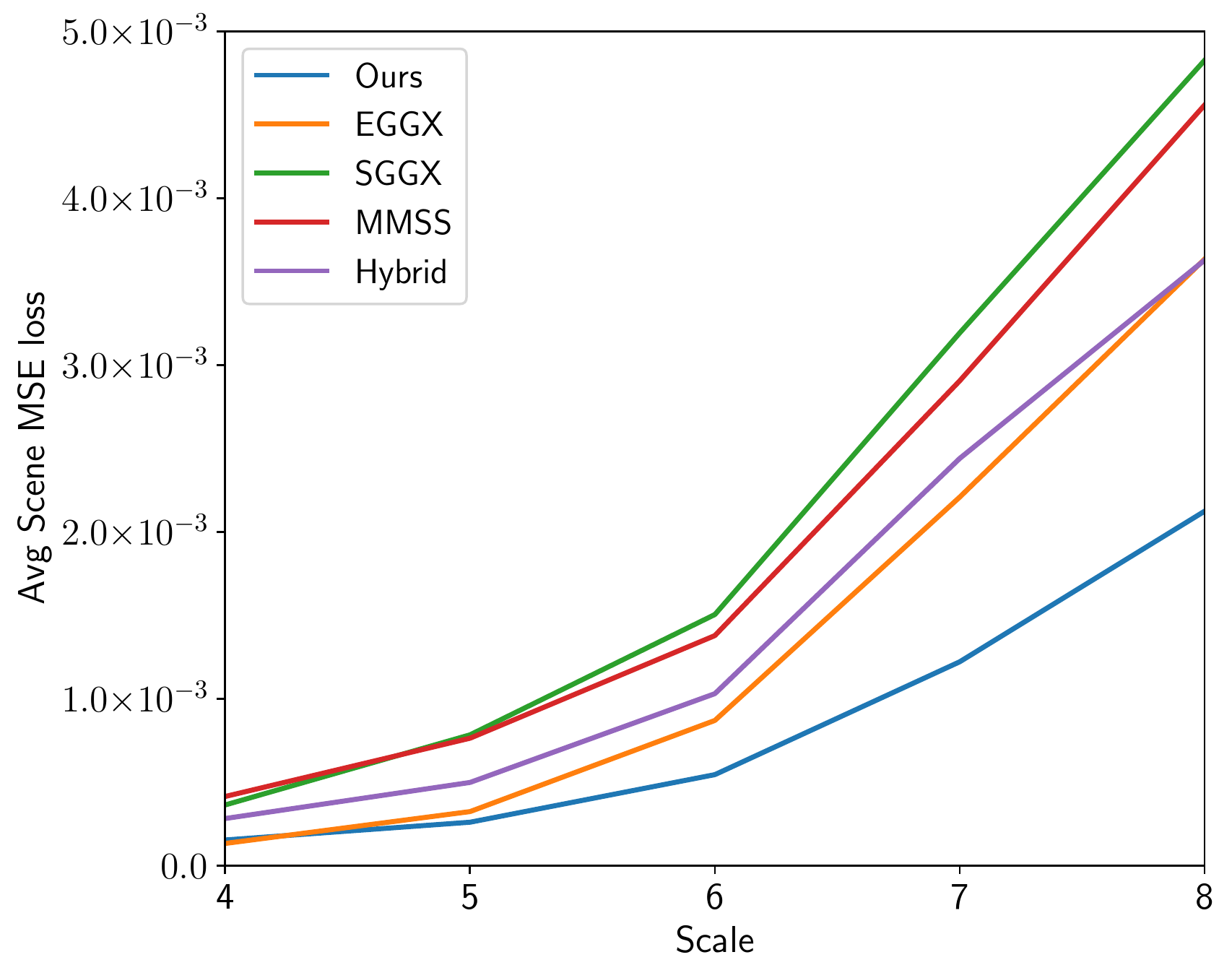} &
    \includegraphics[width=\plotW\linewidth]{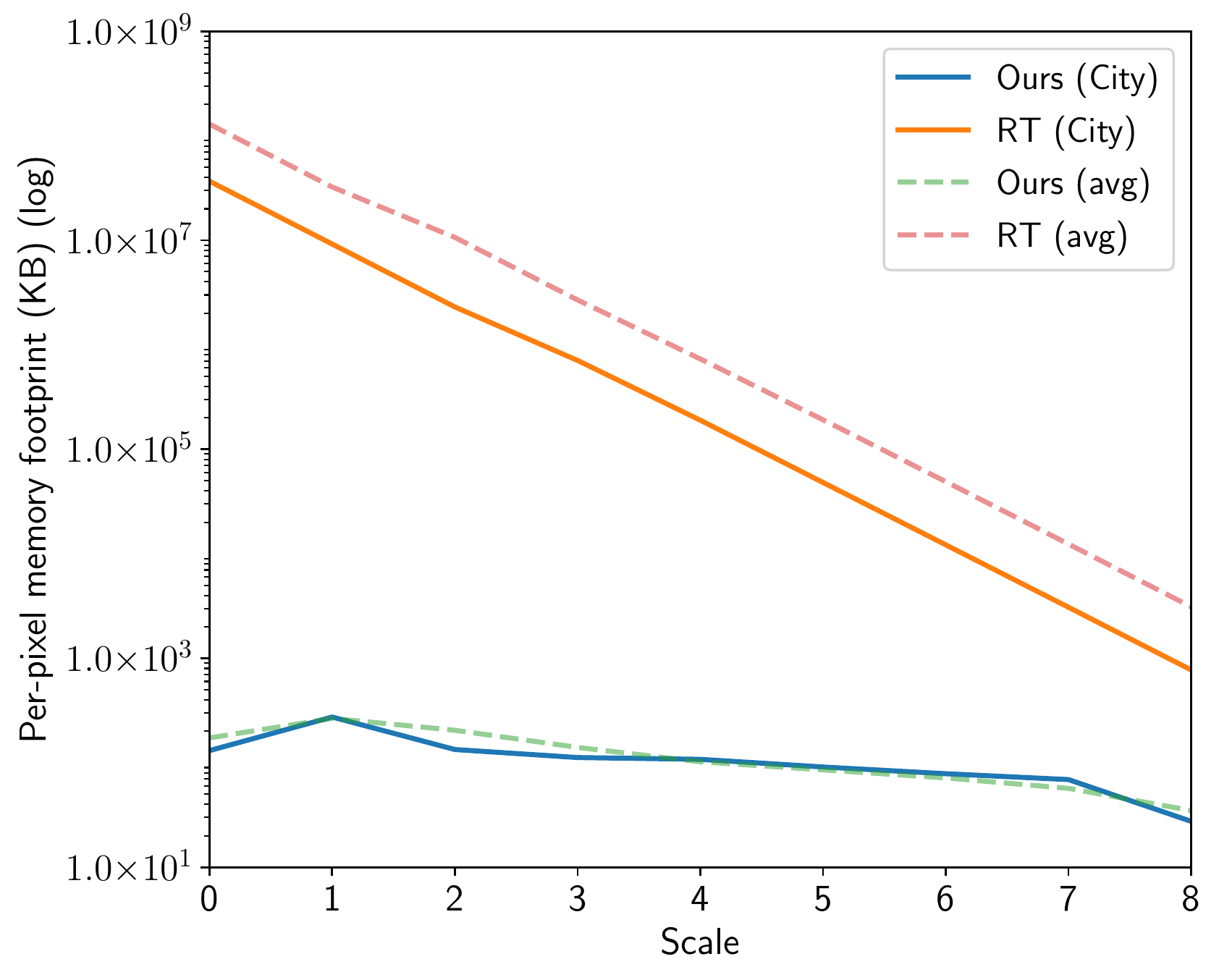} &
    \includegraphics[width=\plotW\linewidth]{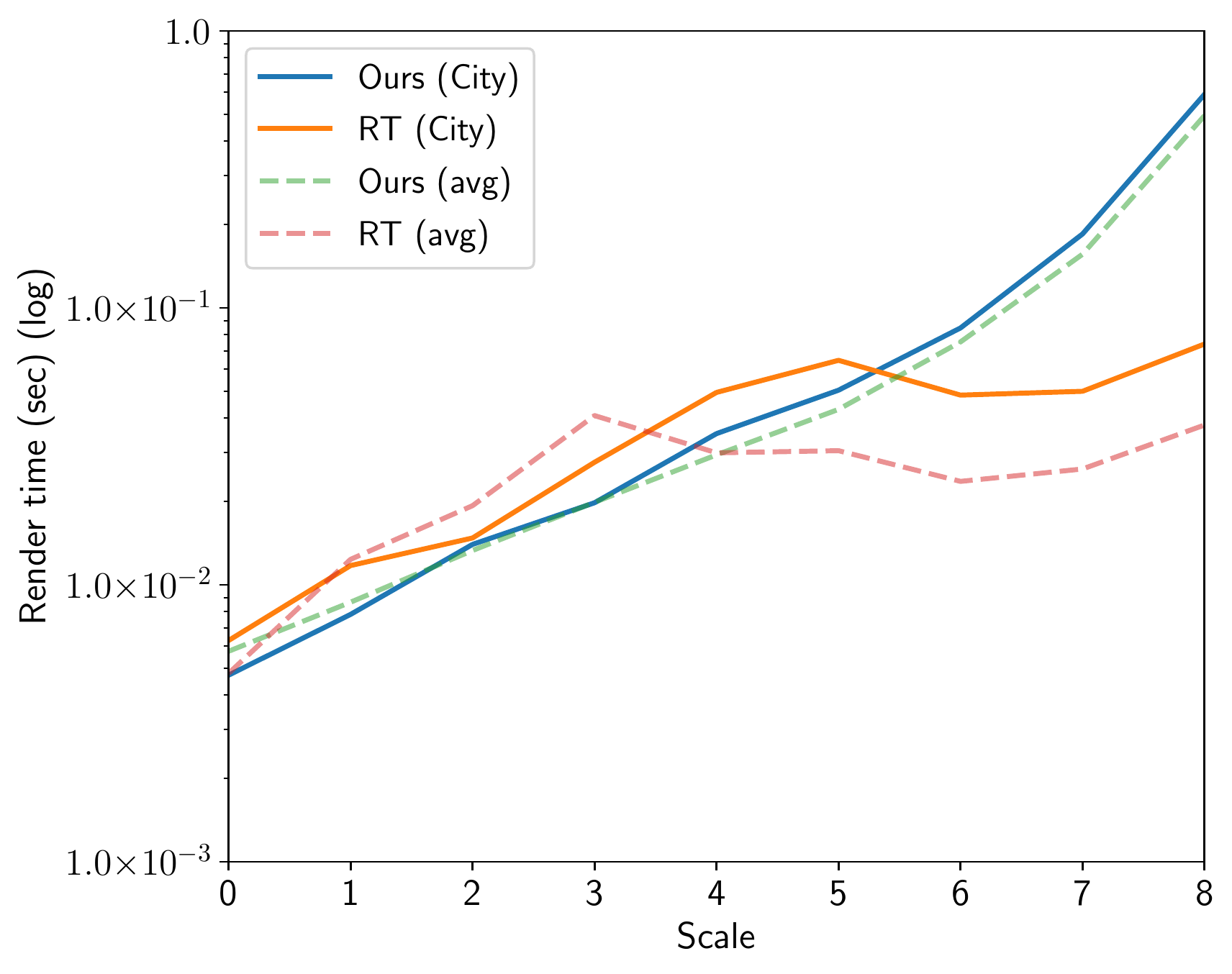}
    \\
  \end{tabular}
  \vspace{-3mm}
  \caption{The plot on the left shows the average MSE across scales for the scenes and methods from Fig.~\ref{fig:SotaComparisons}. Our approach performs favorably relative to previous state-of-the-art approaches at all scales. Note, we only plot results for scale 4 and above due to degenerative simplified meshes from these prefiltering approaches at coarser scales. The log plot in the middle shows the per-pixel memory footprint (in KB) of our approach compared to standard ray tracing for all scales of the \scene{City} scene and the average across all 7 scenes from Figs.~\ref{fig:SotaComparisons} and ~\ref{fig:ComplexScenes}. At the coarsest resolution (scale 0), the entire scene falls under a pixel footprint and we require 5 orders of magnitude less memory to render it. As the image resolution increases, ray tracing memory costs are amortized across the pixels and hence decrease. On the other hand, our footprint remains more-or-less constant across scales, but still yields an order of magnitude improvement at our finest resolution (scale 8), which corresponds to a $256\times256$ image. The rightmost plot shows our corresponding render times for these scenes as compared to an equal-quality ray traced baseline. At lower scales, our unoptimized prototype tends to be faster than ray tracing up until an inflection point at approximately scale 5, where the incurred cost of network inferencing over multiple batches exceeds the baseline.}
  \label{fig:Convergence}
\end{figure*}

\begin{figure}[!t]
  \centering
  \setlength{\tabcolsep}{1pt}
  \footnotesize%
  \begin{tabular}{lclc}
    \begin{sideways} \hspace{12mm}\textbf{City (scale 4)} \end{sideways} &
    \frame{\begin{overpic}[width=.45\linewidth,origin=u]{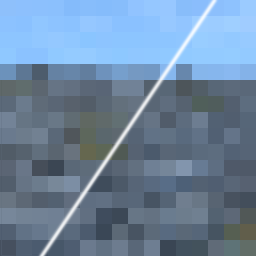}%
     \put(20,925){\color{white}\textbf{Ours}}\put(880,925){\color{white}\textbf{RT}}
    \end{overpic}}
    \begin{sideways} \hspace{12mm} \end{sideways} &
    
    \frame{\begin{overpic}[width=.45\linewidth,origin=u]{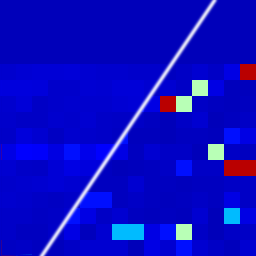}
     \put(20,925){\color{white}\textbf{Ours}}\put(880,925){\color{white}\textbf{RT}}\end{overpic}}
    \\
    \begin{sideways} \hspace{12mm}\textbf{City (scale 8)} \end{sideways} &
    \frame{\begin{overpic}[width=.45\linewidth,origin=u]{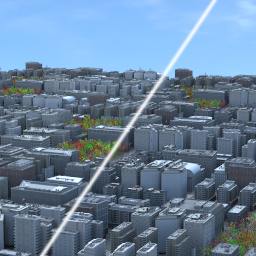}%
     \put(20,925){\color{white}\textbf{Ours}}\put(880,925){\color{white}\textbf{RT}}
    \end{overpic}}
    \begin{sideways} \hspace{12mm} \end{sideways} &
    
    \frame{\begin{overpic}[width=.45\linewidth]{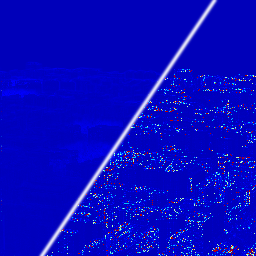}%
     \put(20,925){\color{white}\textbf{Ours}}\put(880,925){\color{white}\textbf{RT}}\end{overpic}}
    \\
  \end{tabular}
  \vspace{-1mm}
  \caption{The two images show complexity as a heatmap of per-pixel timings (blue to red with increasing cost) for our approach and equal-quality ray tracing across scales on the \scene{City} scene. Unlike ray tracing, our method accesses roughly a constant number of voxels at each pixel independently of scene complexity and scale.}
  \label{fig:Complexity}
   \vspace{-5mm} 
\end{figure}

		

We demonstrate how our method, called Deep Appearance Prefiltering (or \ours), can preserve appearance for level of detail (LoD) of previously unsupported materials, while also robustly handling heterogeneous cases of surfaces and volumes. Our pipeline also scales well to assets with more complex geometric and material properties. 

In our comparisons, we report mean squared error (MSE) to quantitatively assess the image quality with respect to the reference, and direct readers to the interactive viewer in the supplemental to see other metrics including the perceptually-based Structural Similarity Index (SSIM)~\cite{Wang04}. In addition, the supplemental has full color images and error heat maps across all scales.

Our OptiX~\cite{Parker10} \old{path}\new{ray} tracer is used with a single bounce to generate training data and results for the ``Reference'' method. The results corresponding to ``Ours'' are generated with a custom GPU beam tracer that traverses our SVOs and uses TensorRT to inference the trained decoder networks. The beam tracer does not have access to any geometry or materials of the original scene and generates the image simply by evaluating the networks for an ordered set of voxels at each pixel and combining the results using our transmittance model described in Sec.~\ref{sec:FrameworkOverview}. For all scenes, we use the Disney BRDF, which we ported to Mitsuba~\cite{Jakob10} as it is not supported out of the box. We verified that the high-sample-count, \old{path-traced} \new{ray-traced} images generate the same results across renderers. For the state-of-the-art comparisons in Mitsuba, we rendered 256 samples per pixel to ensure there is no visible noise since the approaches operate with rays. Note our approach still only has a single beam per pixel. Our intent is to instead demonstrate the fundamental limitations of the appearance models that our method overcomes.

Where applicable, the methods in this section sample according to the Nyquist frequency and utilize voxels from the next finer scale relative to the one determined by the pixel filtering kernel to avoid aliasing and overblurring, as is done in Loubet and Neyret~\shortcite{Loubet17}. We also display difference images with respect to the reference that are color coded from blue to red with increasing error. Finally, for all results and methods, we include a background layer that undergoes standard ray tracing without prefiltering, and which corresponds to our boundary term from Sec.~\ref{sec:TheoreticalOverview}. The images in this section are rendered at a resolution of $1024\times1024$. The complexity in terms of number of triangles for a single instance in each of our scenes can be found in Table~\ref{table:TriangleCount}.

\subsection{Appearance models}


We begin by comparing against state-of-the-art appearance models including the symmetric GGX microflake distribution (\sggx)~\cite{Heitz15}, the microfacet model with an ellipsoid normal distribution function (\eggx)~\cite{Dong15}, and the recent microflake model with self-shadowing (\selfshadow) from Loubet and Neyret~\shortcite{Loubet18}. For \selfshadow, we use the author implementation, which also included comparisons to \eggx ~and \sggx. Although \selfshadow~ was demonstrated for the application of appearance-preserving volumetric downsampling, we compare to it here for its novel microflake model which captures the self shadowing/occlusions that occur within a voxel for improved accuracy relative to \sggx. Since this method only supports a specular lobe, we also modified the original code to include an additional ellipsoid diffuse lobe. 

In general, due to their relatively simple phase functions, these approximate models are unable capture a wide range of effects that we highlight in Fig.~\ref{fig:SotaComparisons}. Note, \eggx ~still has access to the scene's mesh, while our approach, like \sggx, does not use the original geometry and materials yet is still the most faithful to the reference. Furthermore, since we include a diffuse EGGX lobe for \selfshadow, it is also not fully volumetric. The comparisons in Fig.~\ref{fig:SotaComparisons} are done using a directional light source, so we use our phase network, which outputs the throughput at a single point query.

The \scene{Cutlery} scene shows how the previous approaches struggle with specularities. In particular, they fail to preserve the anisotropic highlights along the spoon which are overblurred. This scenario has a watertight surface, so volumetric representations (\sggx ~and \selfshadow) do not perform as well as \eggx, which uses simplified geometric representations. Our approach does not try to fit an approximate model to this difficult scenario. Instead, it learns a significantly more precise phase function that was captured directly from the ray tracer, preserving the highlights. Meanwhile, in \scene{Oak}, \sggx~ and \selfshadow~ fail to capture the glints on the leaves, since sharp specularities are difficult to model with microflakes.

\subsection{Hybrid approaches}


We compare against the author-provided implementation of the state-of-the-art, hybrid mesh-volume approach (\hybrid) from Loubet and Neyret~\shortcite{Loubet17} in Fig.~\ref{fig:SotaComparisons}. This method uses heuristics in a mesh-based analysis to label regions of the asset as either macrosurfaces amenable for geometric simplification and an \eggx~ microfacet model or sub-resolution microgeometry better represented through volumetric rendering with an \sggx~ microflake distribution. The results of this method for the previous scenes are included in Fig.~\ref{fig:SotaComparisons} to demonstrate that, since \hybrid~ still uses established microflake models, it will also have the same issues as described earlier with preserving a wide range of complex effects and appearances, due to the limitations of using an approximate model. 

The \scene{Mossy Rock} scene shows moss growing on a rock, creating a rough, detailed surface on top of a large watertight mesh. The hybrid approach misclassifies this as sub-resolution geometry that can be rendered volumetrically, resulting in an overblurred appearance. On the other hand, our approach avoids classification altogether and will render the appearance more accurately. Furthermore, for the \scene{Oak} scene, the supplemental shows that for coarser scales (e.g., scale 7 and below), \hybrid~ tends to overblur the leaves of the tree as it relies increasingly on a volumetric representation.


In Fig.~\ref{fig:Convergence}, the plot on the left shows the average MSE loss for all methods for the scenes in Fig.~\ref{fig:SotaComparisons} across scales from coarse (scale 4) to fine (scale 8) corresponding to images of resolution $16\times16$ and $256\times256$, respectively. Note, we do not plot results for coarser scales to ensure fairness, as previous approaches can output degenerative meshes from their geometric simplification pipeline, which resulted in all black images. Our approach does not rely on explicit meshes and thus always works at all scales. Still, at the scales shown, our method has the lowest average error relative to the other state-of-the-art methods independently of scale.

 


\subsection{Complex scenes}

Since our framework can capture a wide range of appearances without having to rely on simplification of explicit geometry or volumetric microflake representations, we are able to demonstrate results on significantly more complex scenes (see Fig.~\ref{fig:ComplexScenes}). In their current implementations, both the \hybrid~ and \selfshadow~ code base operates on a single asset (e.g., as a texture-mapped obj) to perform the prefiltering and the environments shown here are extremely difficult to model as a single asset and texture due to their complexity. On the other hand, it is not a trivial extension to extend the algorithms to work on a collection of assets. Thus, we only compare against the reference \old{path-traced}\new{ray-traced} result, which we closely resemble in terms of final image quality, but with a significantly reduced per-pixel memory footprint. Note, the comparisons in Fig.~\ref{fig:ComplexScenes} use an environment map, so we use our phase-slice network, which outputs the throughput at a 2D slice for the particular view direction and can be used in a dot product with the environment map to determine the incoming radiance at all solid angles simultaneously.

Fig.~\ref{fig:ComplexScenes} shows our results for four scenes at a coarse and fine scale along with the reference renders. As shown in the color-coded difference images (closer to red is larger error), we accurately preserve the appearance of the ray traced result. However, we do so with only a fraction of the memory cost. Below each result, we report the per-pixel memory footprint. For the reference, this is computed by taking the memory size of a scene and dividing it by the number of pixels that the scene maps to (i.e., the projection of the asset to the image plane) when the image resolution matches the corresponding LoD scale. To illustrate, for an axis-aligned view, the bounding box of a prefiltered asset at LoD scale 4 would cover $16\times16$ pixels, while such a view at scale 8 would be $256\times256$. Setting a certain scale means that the voxels would be no finer than those found at that scale. For our method, the per-pixel memory footprint is computed by taking the number of touched voxels, multiplying by the memory size of the voxel data, and then dividing by the number of pixels for the scale. Note, we show our results here with instancing to highlight the compelling cases our method could be used. In a production setting, the tiles would correspond to different assets that are prefiltered to make up a scene, but the savings in our per-pixel memory footprint would persist since our approach would evaluate roughly the same number of voxels regardless of scene complexity.

As with all LoD methods of this nature, there is an inflection point where the benefit of prefiltering over standard ray tracing no longer exceeds the relative cost. We can see this in the center plot of Fig.~\ref{fig:Convergence}, which shows the per-pixel footprint for the \scene{City} scene across scales. If the \scene{City} were to fall entirely within one pixel, as in scale 0, a typical ray tracer would need to load the full scene (over 8GB) and send many samples to properly integrate over the large pixel footprint. This cost is amortized as the footprint covering the \scene{City} grows to 256 pixels in scale 8. However, in our approach each voxel requires only 3KB of memory as mentioned in Sec.~\ref{sec:Implementation}, so evaluating all the voxels along a pixel's footprint requires only about 100 KB of memory on average and is more-or-less constant across scales. This culminates in a savings of $\mathtt{\sim}\num[group-separator={,}]{280000}\times$ for scale 0 and $28\times$ at scale 8 relative to ray tracing. Eventually as the scales get finer, it is more advantageous to switch to ray tracing as one can extrapolate from the plot, but such scales are less common in production. For the typical LoD scales shown here, there is a clear memory savings and this behavior holds true when considering the average memory footprint across all 7 scenes from Figs.~\ref{fig:SotaComparisons} and ~\ref{fig:ComplexScenes}.

\subsubsection{Timings and complexity}
As expected for LoD methods, there is an inflection point where ray tracing becomes faster. For example, at the finest image scale (scale 8), an equal-quality Monte Carlo render (EQMC) requires 128 spp, which takes 74ms on the standard GPU ray tracer, while our GPU beam tracer needs 591ms. However, when we start minifying details (e.g., at scale 5 and lower), our method starts to outperform ray tracing, as illustrated in the rightmost plot of Fig.~\ref{fig:Convergence}. For a single instance of \scene{City} at scale 4, EQMC would require 128 spp, which takes 50ms on the ray tracer and 35ms on our beam tracer using our framework. The inflection point depends on the implementation efficiency of both algorithms and, with an unoptimized implementation of our algorithm, we observed the inflection point to be around scale 5 across all 7 scenes. One implementation inefficiency occurs at higher image resolutions, where the total number of pixels and inference calls increases and we are technically limited to running a maximum batch of 4096 elements through our pipeline at a time, resulting in low GPU utilization. 

The heatmaps in Fig.~\ref{fig:Complexity} correspond to the per-pixel complexity as a function of time for our method and EQMC across scales. Specifically, each pixel in ours represents the number of voxels evaluated at that pixel multiplied by the cost of evaluating each voxel. Meanwhile, for EQMC, each pixel shows the number of samples required to reach the same level of MSE error as our approach times the cost of each sample. Note, the per-pixel timings shown here do not capture the overhead of swapping network batches, as discussed in the previous paragraph, since the image could simply be evaluated as tiles in parallel across different GPUs to avoid this cost. Since our framework needs to process roughly the same number of voxels everywhere and with the same evaluation cost, we demonstrate constant timings across pixels and scales relative to EQMC, which needs to evaluate more samples based on the complexity of the appearance falling within a pixel's footprint. 

\subsection{Temporal stability}
In the supplemental, we include video fly-throughs for the scenes in  Fig.~\ref{fig:SotaComparisons} and Fig.~\ref{fig:ComplexScenes}. Our results are temporally coherent when switching across different voxel scales without any noticeable flickering or popping even though such consistency is not enforced in the training loss for the networks. Moreover, in scenes such as \scene{Stormtrooper Army}, the network produces anti-aliased results for distant stormtroopers along the horizon despite using only a single beam per pixel, while the ray-traced result requires significant samples in this region to reduce flickering. The network is also able to smoothly interpolate across both view and light directions, as a result of the stochastic data generation that continuously updated the voxel data and allowed arbitrary queries.

Note, we did not retrain or use separate networks across scales. By having our network train on voxels from all scales simultaneously, it is able to smoothly interpolate across scales with voxels of varying degrees of geometric and material complexity. In the future, it would be interesting to investigate how interpolation in the latent space of voxels could produce non-linear appearance changes for regions in between LoD scales. This could replace the traditional linear combinations used by previous LoD methods and in our approach.

\section{Discussion, limitations, and future work}
\label{sec:Discussion}

\begin{figure}[tp]
	\setlength{\fboxrule}{10pt}%
	\setlength{\insetvsep}{20pt}%
	\setlength{\tabcolsep}{1pt}
	\renewcommand{\arraystretch}{1}%
	\footnotesize%
	\begin{tabular}{cccc}
		
		\setInset{A}{red}{384}{0}{128}{128}
		\setInset{B}{OliveGreen}{768}{768}{128}{128}
		& Ours & Ours & Reference (256K spp) \\
		\begin{sideways} \hspace{-27mm} \textbf{Cutlery (phase)} \end{sideways} & \addBeautyCrop{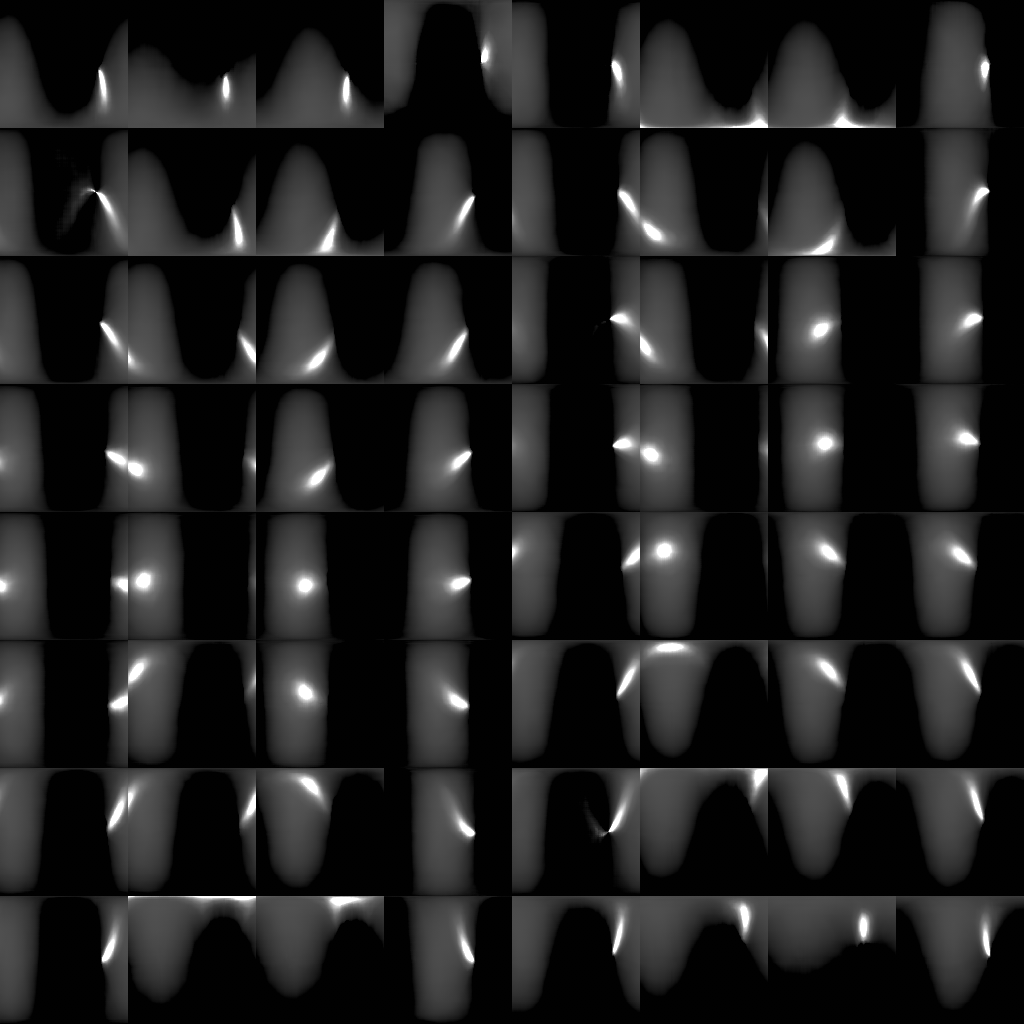}{0.22}{1024}{1024}{0}{0}{1024}{1024} &
		\addInsets{Figures/Discussion/netRecon/phase/denoiseNet.png} &
		\addInsets{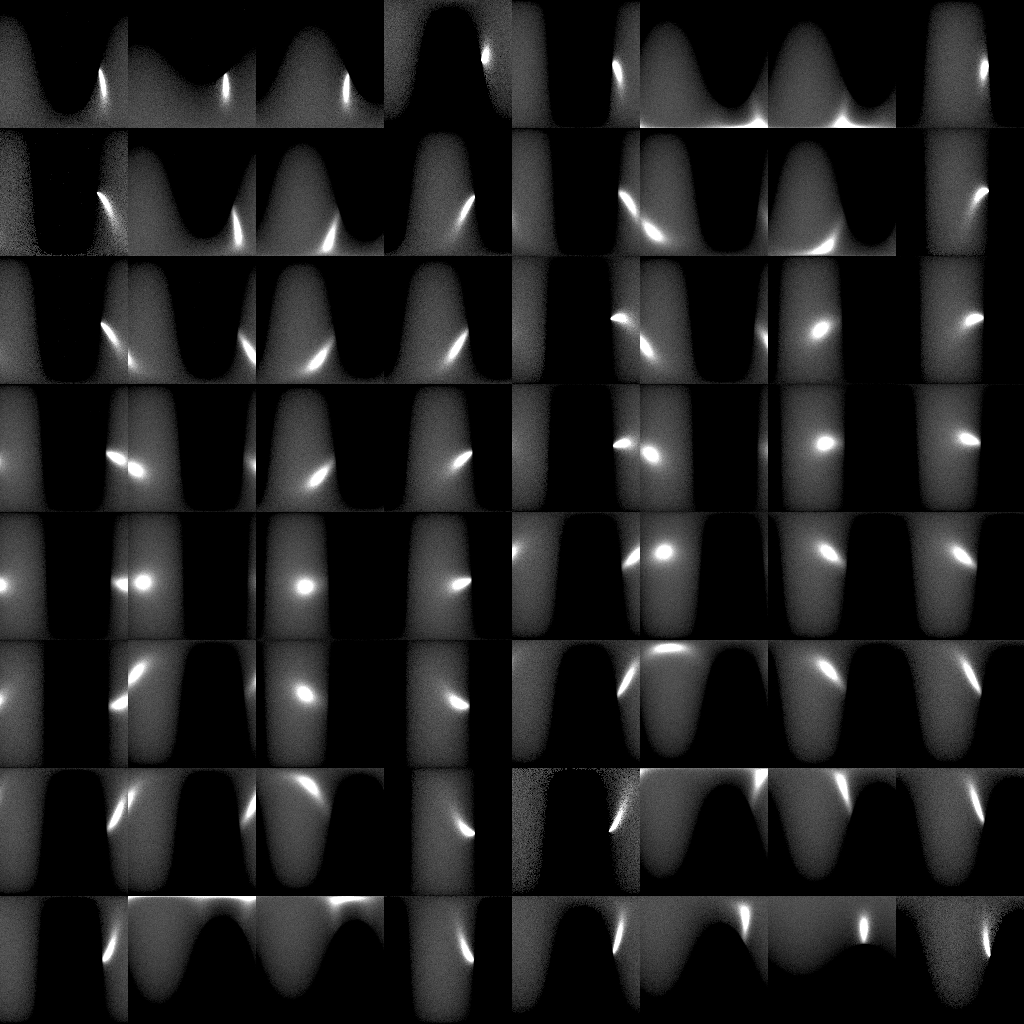} \\


	\end{tabular}
	\caption{The full image on the left is a visualization of different view directions along a uniform stride reconstructed using our phase point query network. Each tile is a slice showing the incident throughput (single channel) corresponding to a given outgoing direction. The insets demonstrate how our network is able to fit to the underlying data despite the noise in the reference. Across the two insets, the network faithfully denoises the phase along various supports, while preserving anisotropic specularities.}
	\label{fig:Denoising}
	\vspace{-2mm}
\end{figure}

\begin{figure}[!t]
  \centering
  \setlength{\tabcolsep}{1pt}
  \footnotesize%
  \begin{tabular}{cc}
    
    \includegraphics[width=0.5\linewidth]{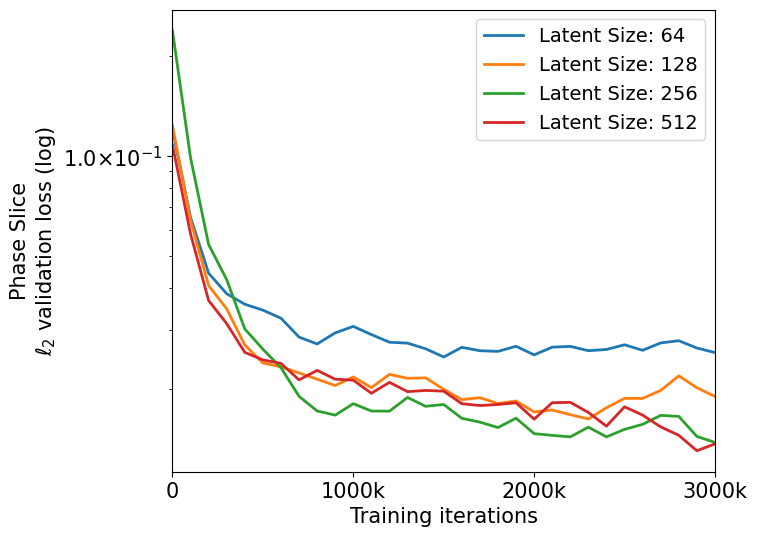} &
    \includegraphics[width=0.5\linewidth]{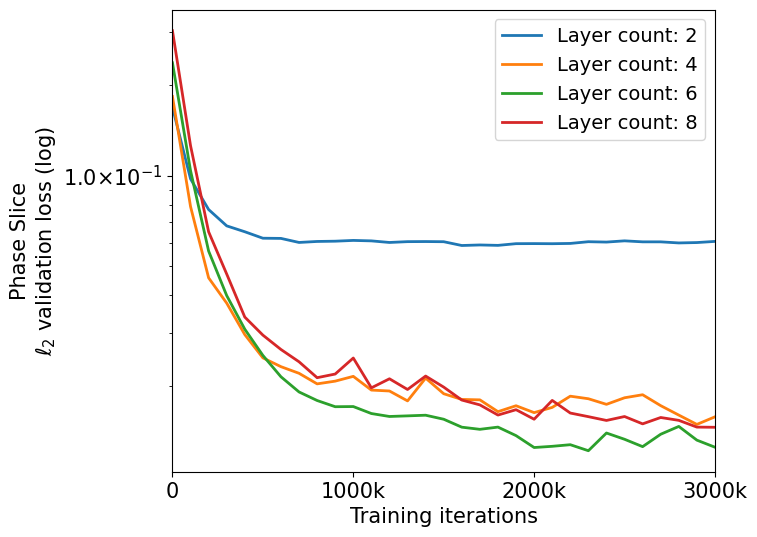} \\
  \end{tabular}
  \vspace{-4mm}
  \caption{Ablation studies plotting validation errors ($\ell_2$) for the phase-slice network across training iterations using various configurations in the \scene{Cutlery} scene. Increasing the latent size results in better performance, yet gains become marginal for larger sizes (left plot). Increasing the number of network layers also improves performance up to a point (right plot). In our implementation, we use 256 floats for the latent vector and 6 layers in the decoder. See additional ablation analysis in Appendix~\ref{appendix:AppendixA}.}
  \label{fig:Ablation}
\end{figure}

\begin{figure}[tp]
	\setlength{\fboxrule}{10pt}%
	\setlength{\insetvsep}{20pt}%
	\setlength{\tabcolsep}{1pt}
	\renewcommand{\arraystretch}{1}%
	\footnotesize%
	\begin{tabular}{ccc}
		
		& Ours & Coverage \\
		\begin{sideways} \hspace{9mm} \textbf{Mossy Rock} \end{sideways} &
		\hspace{0.5mm}\frame{\includegraphics[width=0.35\linewidth]{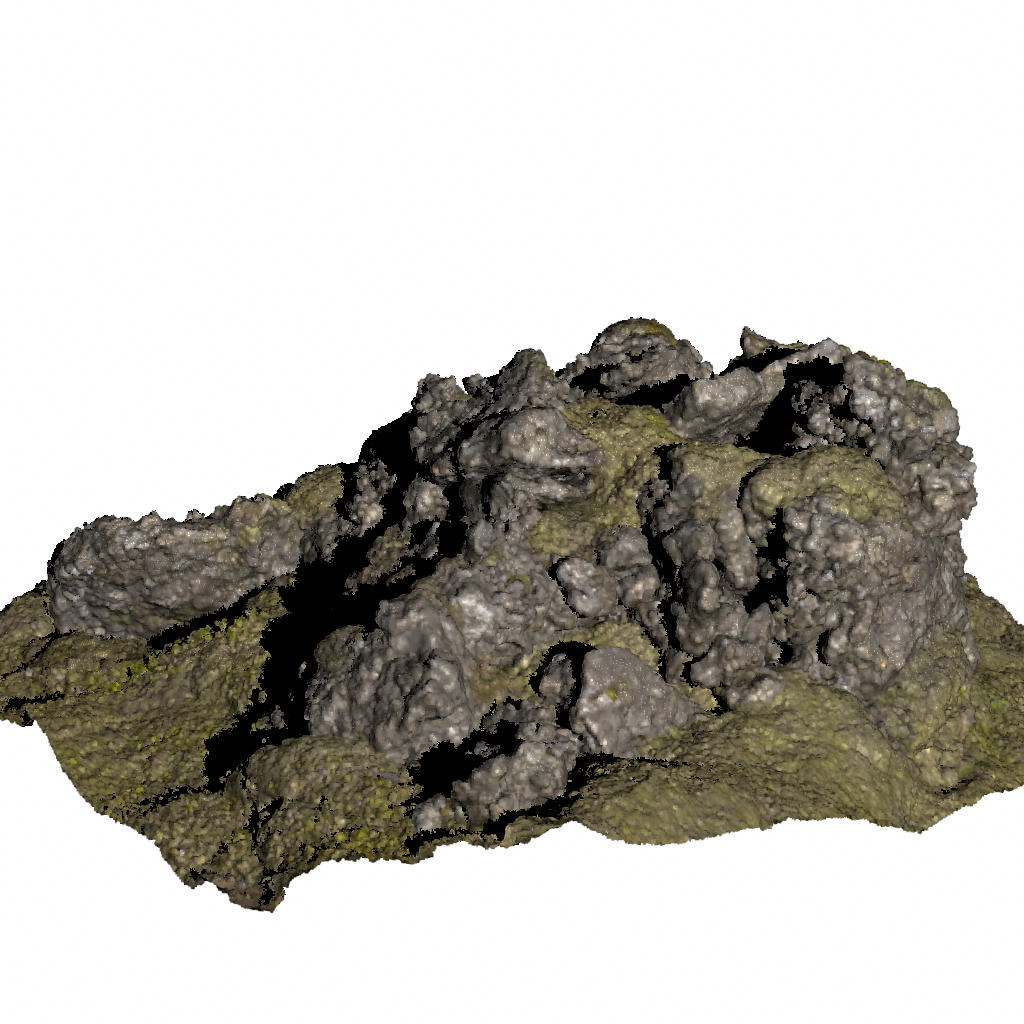}} & 
		\hspace{1mm} \includegraphics[width=0.35\linewidth]{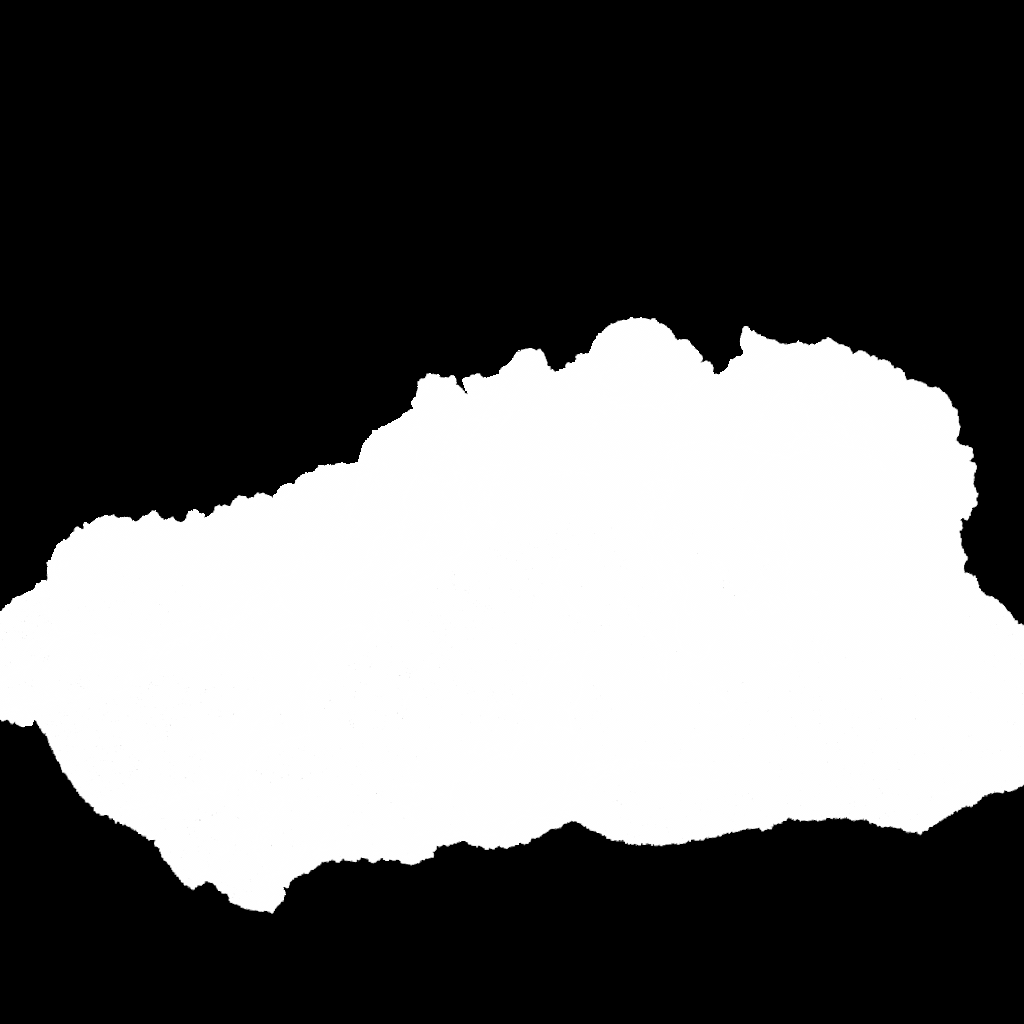} \\
		\begin{sideways} \hspace{13mm} \textbf{Oak} \end{sideways} &
		\hspace{0.5mm}\frame{\includegraphics[width=0.35\linewidth]{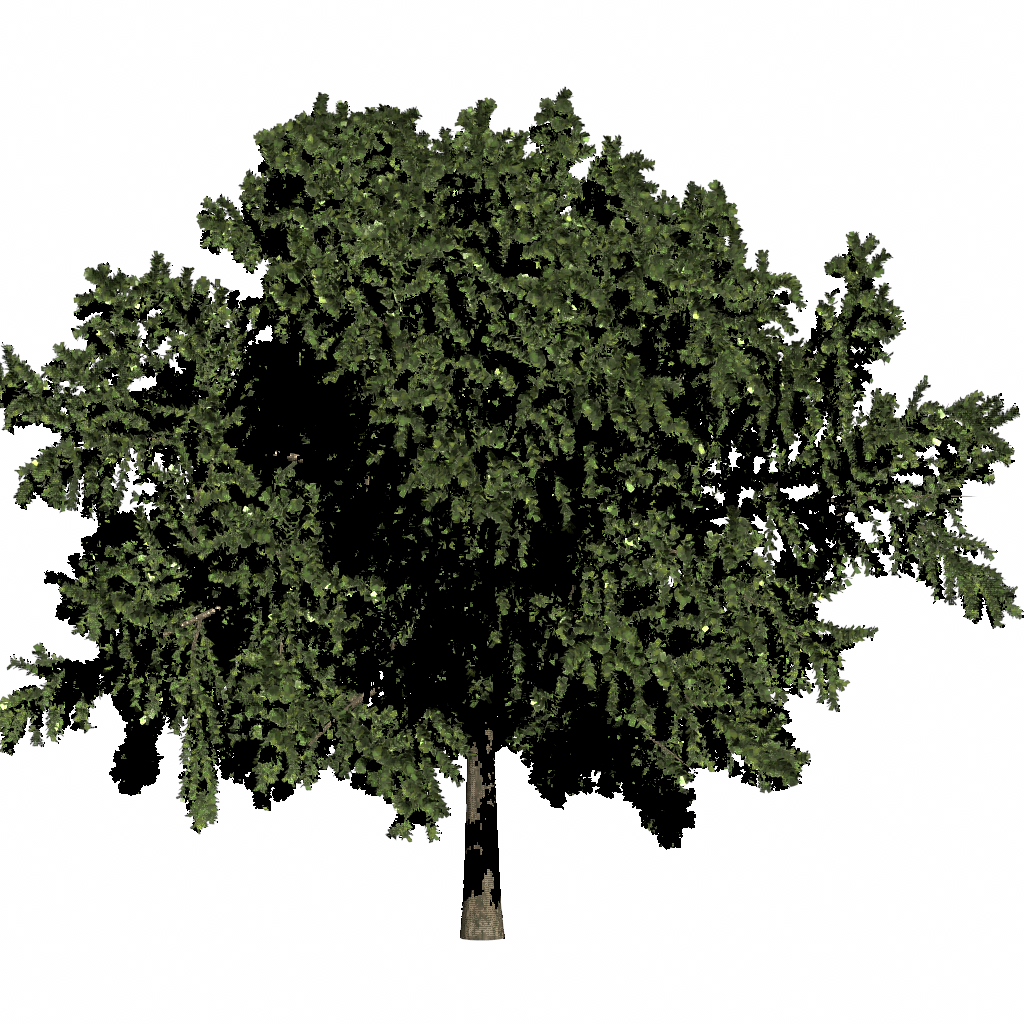}} & 
		\hspace{1mm} \includegraphics[width=0.35\linewidth]{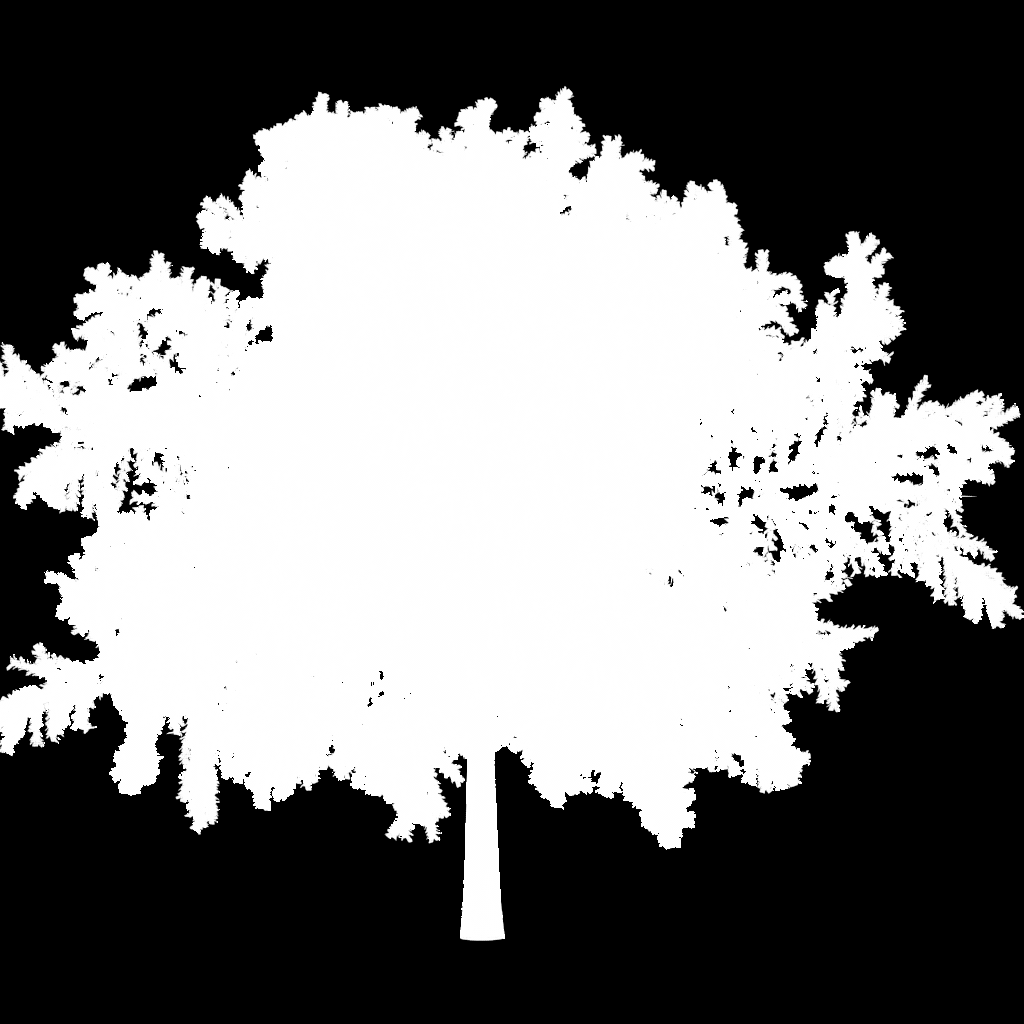} \\

	\end{tabular}
	\vspace{-2mm}
	\caption{Total coverage for the \scene{Mossy Rock} scene (top), where transmittance behaves as a watertight surface, and the \scene{Oak} scene (bottom), which behaves volumetrically. Note, the white background does not bleed through the asset.}
	\label{fig:Coverage}
\end{figure}

\begin{figure}[!t]
	\setlength{\fboxrule}{10pt}%
	\setlength{\insetvsep}{20pt}%
	\setlength{\tabcolsep}{1pt}
	\renewcommand{\arraystretch}{1}%
	\footnotesize%
	\begin{tabular}{ccc}
		
		& Ours & Ours w/ modified albedo \\
		\begin{sideways} \hspace{3mm} \textbf{Stormtrooper Army} \end{sideways} &
		\hspace{0.5mm}\frame{\includegraphics[width=0.35\linewidth]{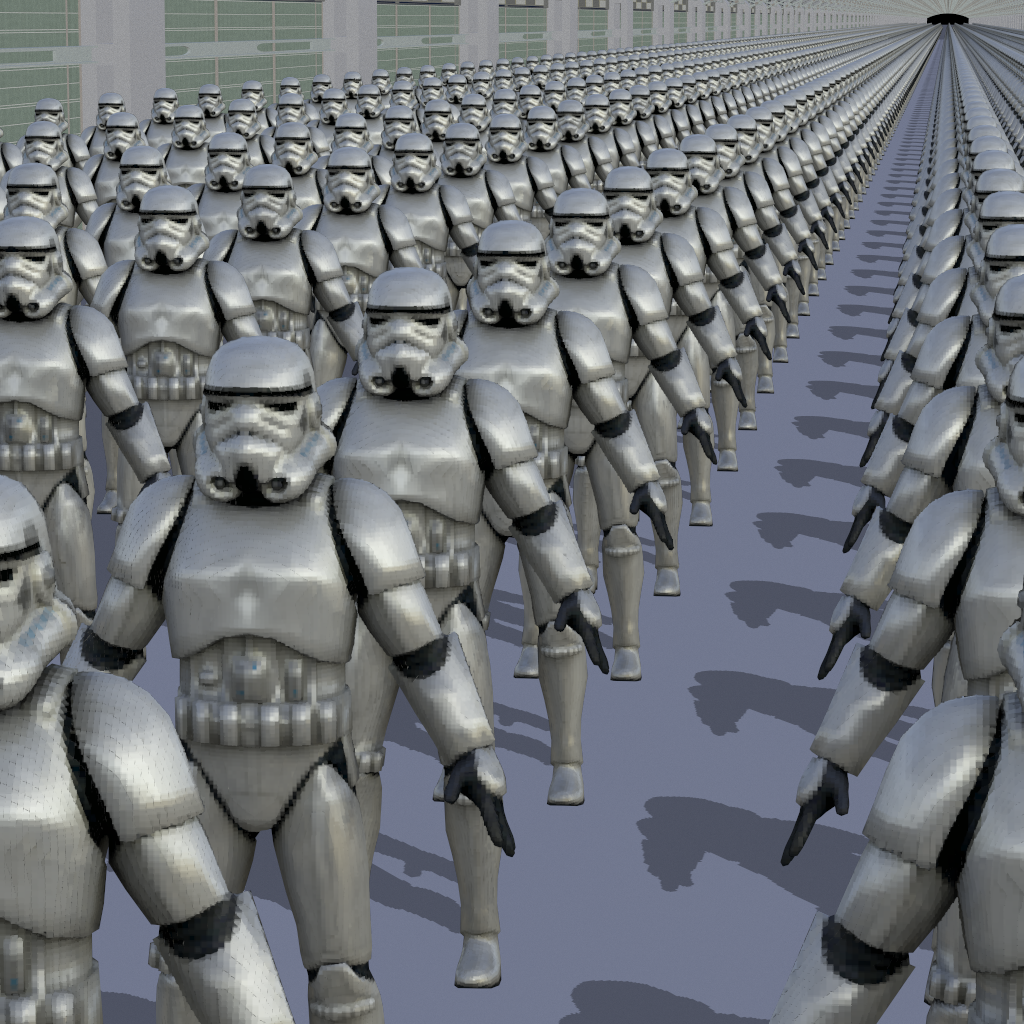}} & 
		\hspace{1mm} \includegraphics[width=0.35\linewidth]{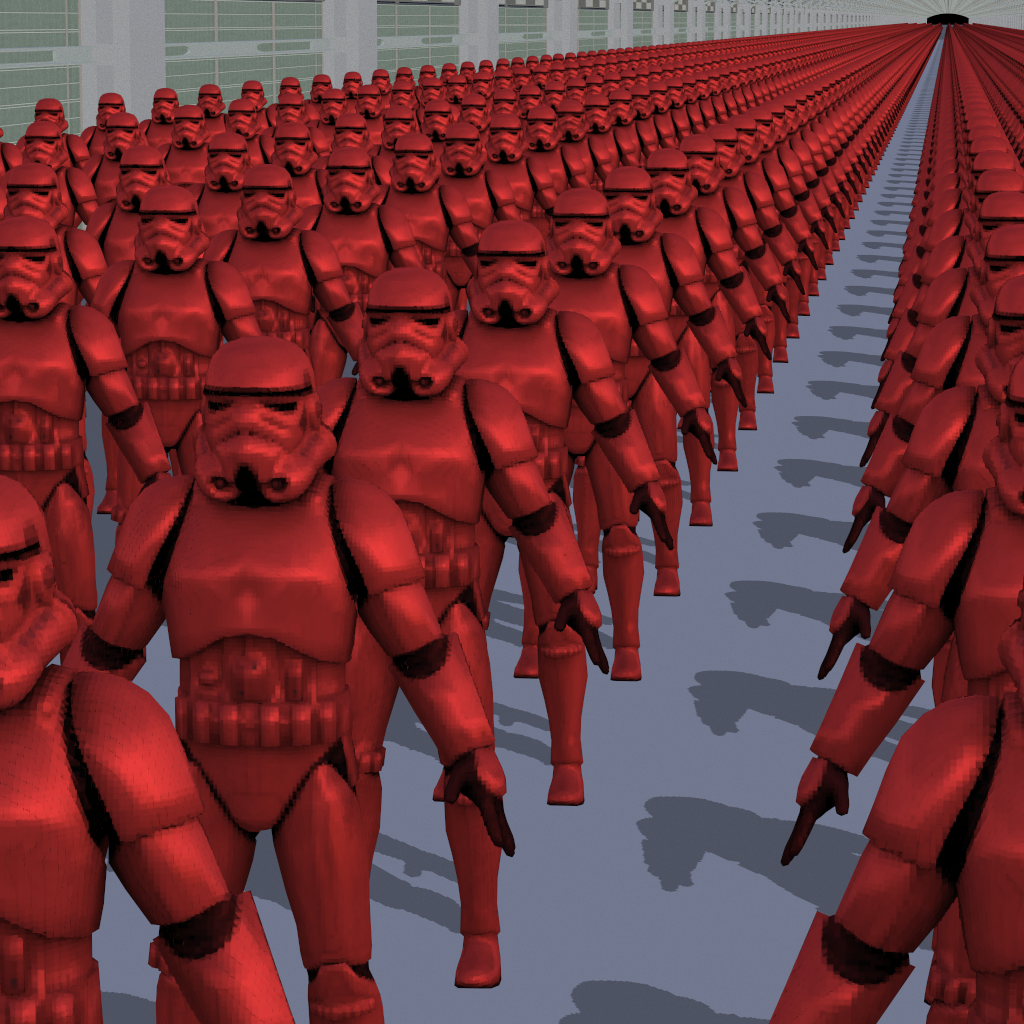} 	
	\end{tabular}
	\vspace{-2mm}
	\caption{Our framework's design offers flexibility with its ability to expose certain parameters such as the diffuse albedo to easily modify the scene appearance. Without any additional training or data generation, we can easily change the stormtroopers from the classic white to a modern red.}
	\label{fig:ChangingAlbedo}
	\vspace{-2mm}
\end{figure}

\begin{figure}[!t]
	\setlength{\fboxrule}{10pt}%
	\setlength{\insetvsep}{20pt}%
	\setlength{\tabcolsep}{1pt}
	\renewcommand{\arraystretch}{1}%
	\footnotesize%
	\begin{tabular}{ccc}
		
		LUT & Ours & Reference (16K spp) \\
		\hspace{0.5mm}\frame{\includegraphics[width=0.3\linewidth]{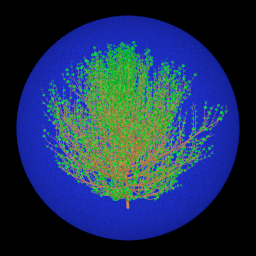}} & 
		\hspace{1mm} \includegraphics[width=0.3\linewidth]{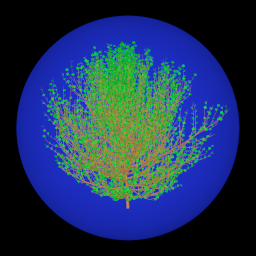} &
		\hspace{1mm} \includegraphics[width=0.3\linewidth]{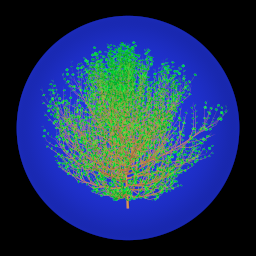}
	\end{tabular}
	\vspace{-4mm}
	\caption{Results on a toy example of a scene with a bush in front of a blue sphere to demonstrate how well geometric and material correlations are tracked. On the left, we use a lookup table (LUT) to generate the image directly. In the center, we use our network-based compression of the tabular data. Both LUT and ours closely match the ray-traced reference (right).}
	\label{fig:ToyExample}
	\vspace{-3mm}
\end{figure}

\begin{figure}[!t]
	\setlength{\fboxrule}{10pt}%
	\setlength{\insetvsep}{20pt}%
	\setlength{\tabcolsep}{1pt}
	\renewcommand{\arraystretch}{1}%
	\footnotesize%
	\begin{tabular}{ccccc}
		
		\setInset{A}{red}{500}{500}{150}{150}
		\unsetInset{B}
		& Ours (hybrid) & Ours & Ours (hybrid) & Ref. (16K spp) \\
		\begin{sideways} \hspace{-14mm} \textbf{Parking Lot} \end{sideways} &
		\addBeauty{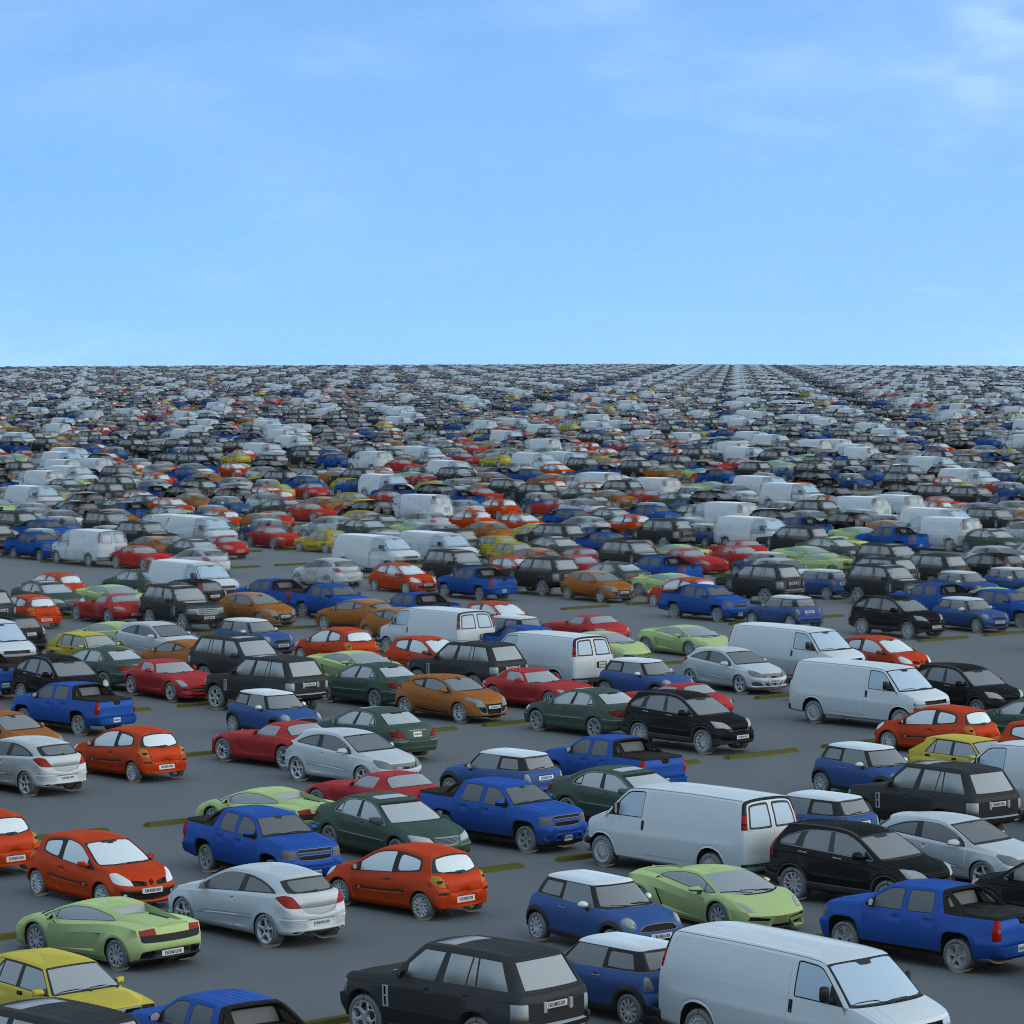}{0.1}{1024}{1024} &
		\addInsets{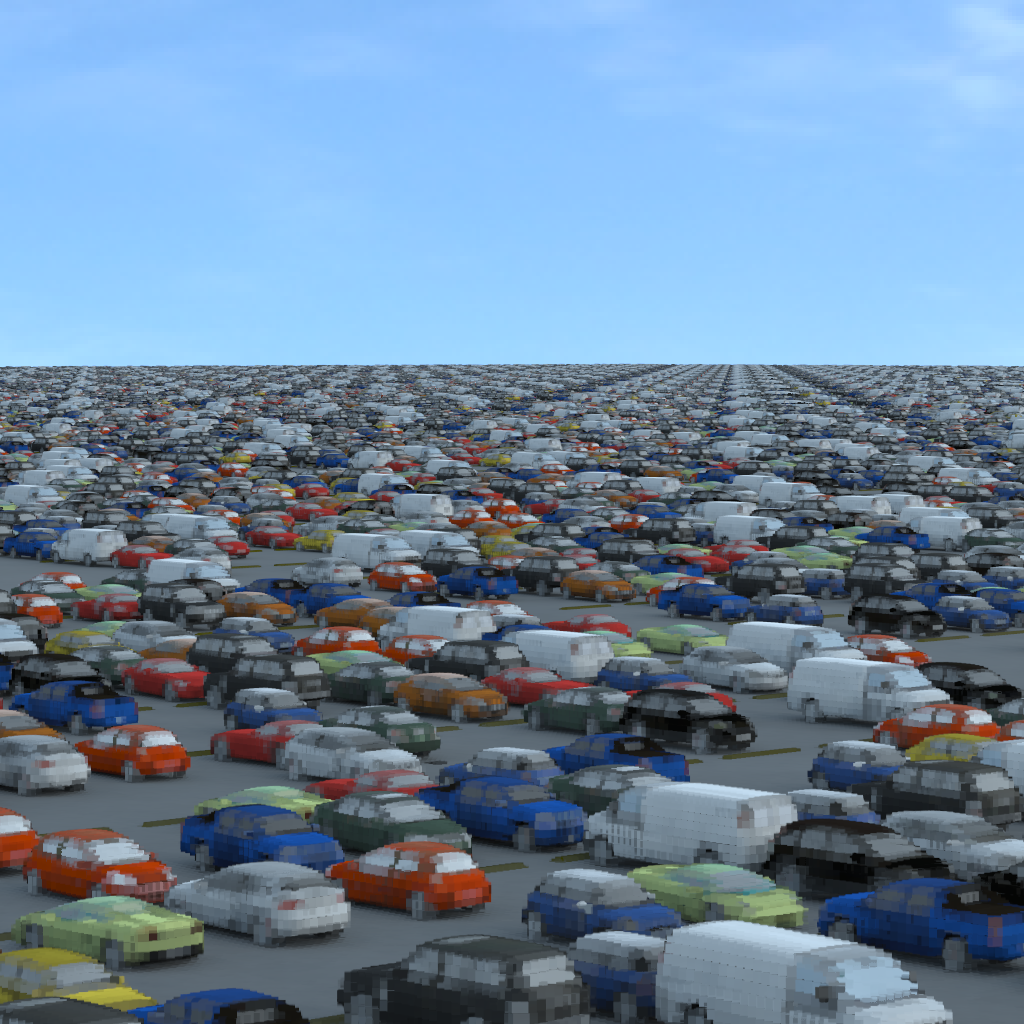} &
		\addInsets{Figures/Discussion/Hybrid/Ours-Hybrid-img.png} &
		\addInsets{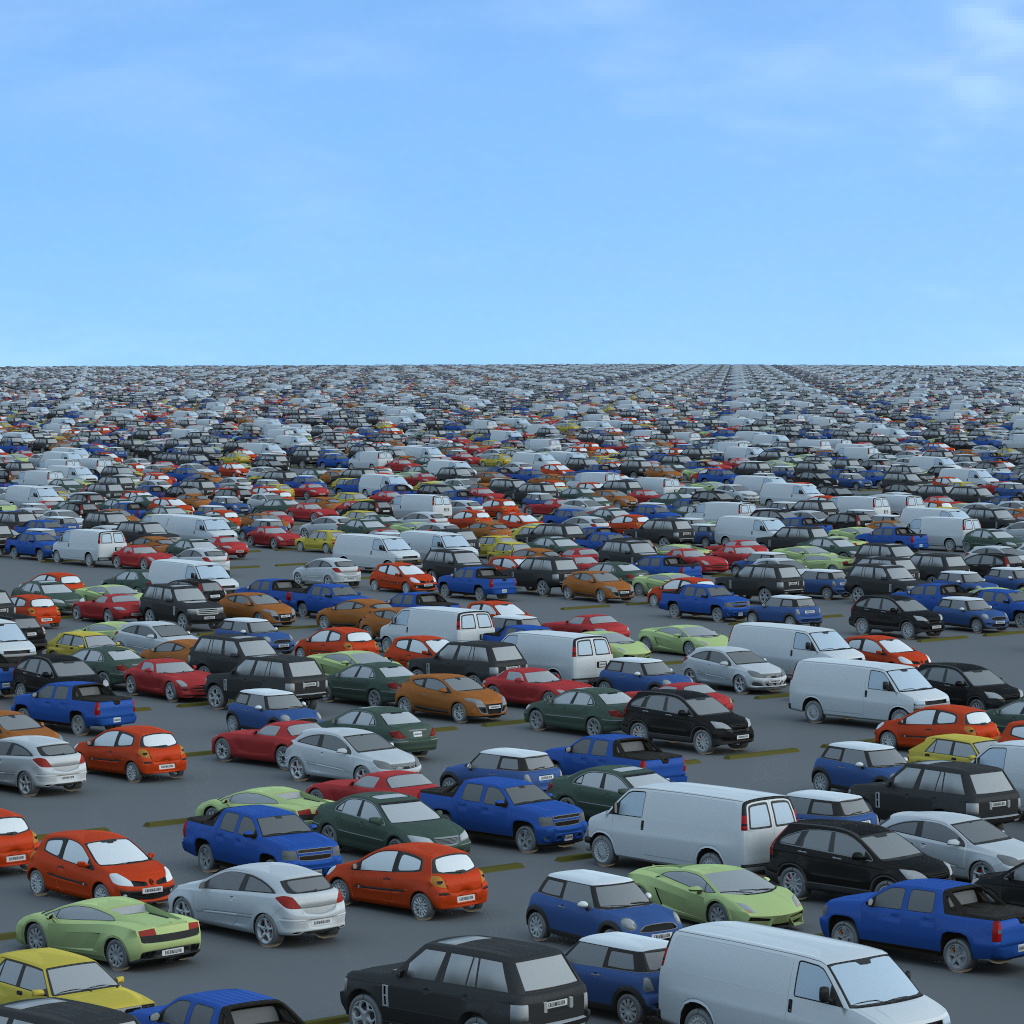}
		
	\end{tabular}
	\vspace{-3mm}
	\caption{Results using a hybrid system with our approach for the minified regions and standard ray tracing for the foreground. Our original approach has blocky artifacts in the foreground since the size of the voxels at the maximum LoD scale precomputed by our framework is significantly larger than the beam at each pixel. The inset straddles a region where the foreground and background layers meet and are blended in a hybrid system. }
	\label{fig:Hybrid}
	\vspace{-3mm}
\end{figure}

\begin{figure}[t]
	\setlength{\fboxrule}{10pt}%
	\setlength{\insetvsep}{20pt}%
	\setlength{\tabcolsep}{1pt}
	\renewcommand{\arraystretch}{1}%
	\footnotesize%
	\begin{tabular}{ccccc}
		
		& Ours & Ours & Ref. (16K spp) & \\
		\setInset{A}{red}{560}{840}{100}{100}
		\unsetInset{B}
		\begin{sideways} \hspace{-10mm} \textbf{City} \end{sideways} & \addBeauty{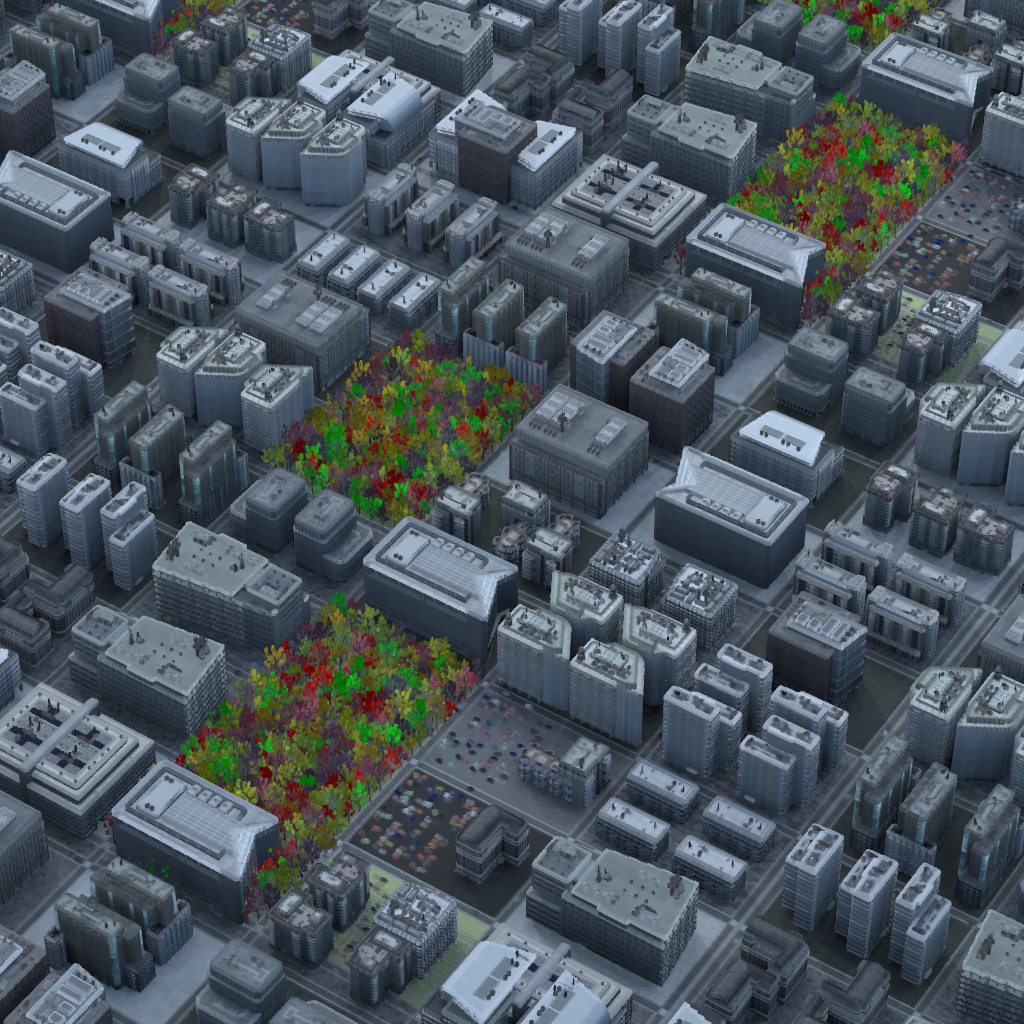}{0.1}{1024}{1024} &
		\addInsets{Figures/Discussion/Limitations/Limitations_city_net.png} &
		\addInsets{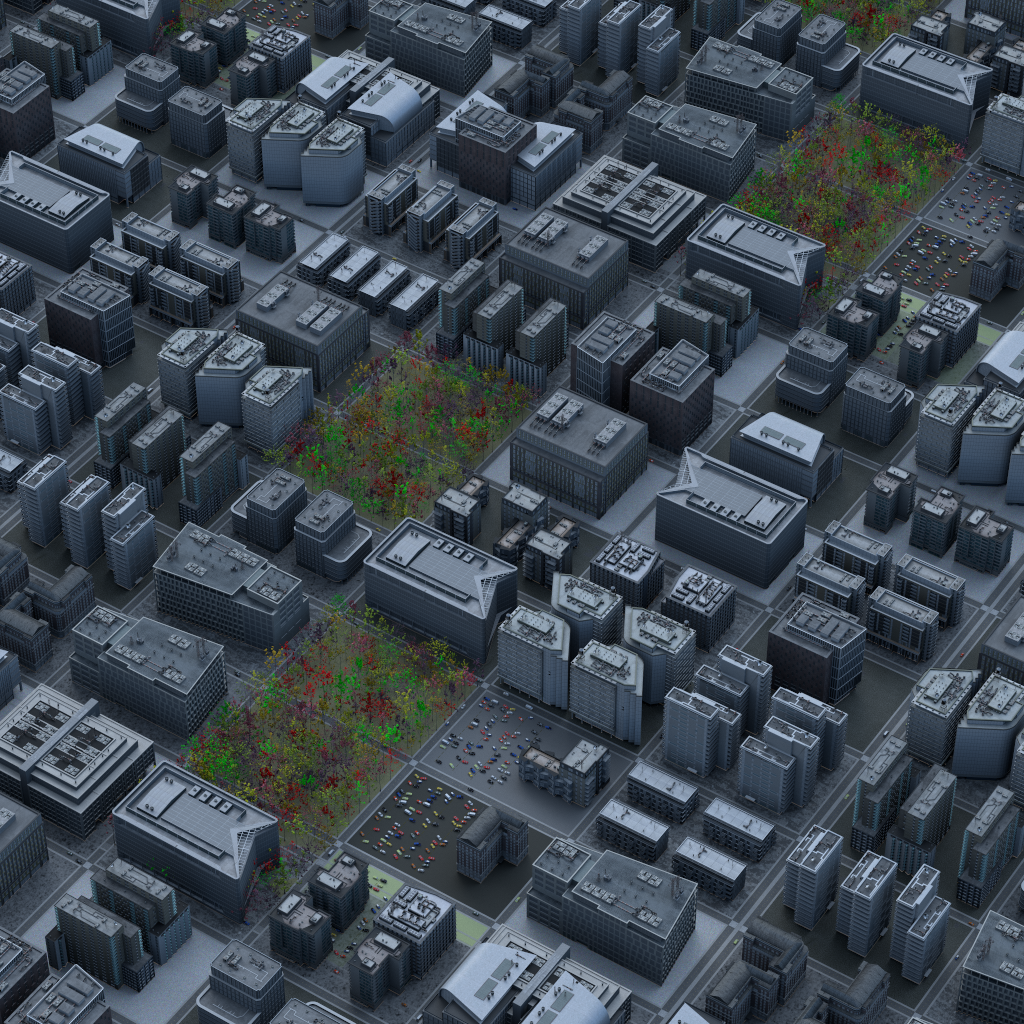} & 
		\multirow{2}{*}{\includegraphics[width=.21\linewidth]{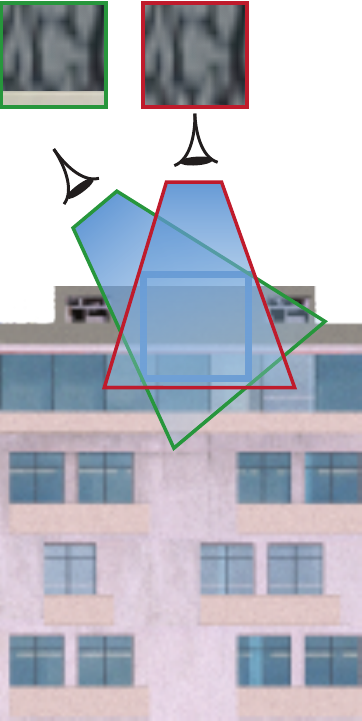}} \\
        \vspace{-2mm} \\ 
		\setInset{A}{red}{425}{575}{75}{75}
		\begin{sideways} \hspace{-13.5mm}\textbf{Mossy Rock} \end{sideways} & \addBeauty{Figures/\resultsDirSota/MossyRock/data/\oursResult\suffixComplex-img.png}{0.1}{1024}{1024} &
		\addInsets{Figures/\resultsDirSota/MossyRock/data/\oursResult\suffixComplex-img.png} &
		\addInsets{Figures/\resultsDirSota/MossyRock/groundtruth\suffixComplex.png} & \\
	\end{tabular}
	\vspace{-3mm}
	\caption{Limitations of our approach. The first row shows a roof in the \scene{City} scene for which our results are not as dark as the reference. This artifact stems when a voxel straddles occluded geometry with a different base color than the unoccluded region. Specifically, the roof has a room directly below it that has a light colored floor and which is included in the voxels containing the roof. When calculating the average RGB albedo for the directly overhead view (red beam) the albedo contains only the roof texture, however grazing views (green beam) can include the floor's albedo causing the lighter color we see. The albedo of the projected cross-section of the voxel is shown for these corresponding views. Another limitation can be found in the bottom row for the \scene{Mossy Rock} scene. Sharp delta highlights that only occur over a very small solid angle are not always preserved perfectly across all voxels and can get slightly overblurred.}
	\label{fig:Limitations}
	\vspace{-4mm}
\end{figure}

It is interesting to note that in our system the networks actually do more than just compression; they actually perform both denoising and interpolation. To keep the computation in our data generation step tractable, we limit the number of samples sent out for each outgoing direction which resulted in some noise in the 4D tabular data (see Fig.~\ref{fig:Denoising}). Here each 2D tile corresponds to a different outgoing direction and each pixel within the tile corresponds to a different incident direction, and each tile can be thought of as a slice of the 4D light field. However, as was observed in the recent Noise2Noise work~\cite{Lehtinen18}, using reference data with slight noise did not cause optimization issues and instead allowed the network to denoise the data and smoothly fit the underlying function. 


In Fig.~\ref{fig:Ablation}, we show ablation plots for the \scene{Cutlery} scene. The first plot shows the $\ell_2$ error of the phase-slice validation data (a subset of the voxels in the training data since we are overfitting to a given scene) with varying sizes of latent vectors. We see that as the size increases from 64 to 512 floats, the converged error decreases due to the network's ability to store more information within these latent features and its additional coefficients. However, the data suggests there are only marginal returns once the vector is sufficiently large (e.g., increasing from 256 to 512 floats has comparable performance despite requiring double the memory). This tradeoff was not beneficial, so we kept the vector size at 256 floats. 

The second plot of Fig.~\ref{fig:Ablation} evaluates the impact of decoder size on validation error. As before, we observed better performance as the number of layers increased from 2 to 6 layers. At 8 layers, we found that performance slightly decreased relative to 6 layers (used in our final implementation), in particular for the phase-slice network. This could be due to difficulty optimizing the larger network. In the other decoders, 6 and 8 layers performed comparably, so we decided to use the more lightweight 6-layer architecture consistently across decoders. Appendix~\ref{appendix:AppendixA} has additional analysis on plots corresponding to the remaining three decoders for the \scene{Cutlery} scene.

Fig.~\ref{fig:Coverage} serves as a sanity check of our transmittance model by placing the assets in front of a white background. The top row shows the \scene{Mossy Rock} asset which is a large macrogeometry with a rough, yet watertight, surface. Meanwhile, the bottom row shows the \scene{Oak} asset which contains many small leaves in random locations, where this sub-resolution microgeometry behaves relatively more like a volume. In both cases, the white background does not bleed through showing that our beam has achieved full coverage and our transmittance model was sufficiently accurate.

Fig.~\ref{fig:ChangingAlbedo} demonstrates another application of our approach. Once an asset is prefiltered in our framework, it can still be modified with trivial changes without any computation of additional data or further training. For example, in the \scene{Stormtrooper Army} scene, we can modify each voxel's diffuse albedo by scaling up the red channel of the albedo information in their latent vectors to switch from white stormtroopers to red ones. Moreover, a simple extension to our current approach would be to additionally save out the specular albedo to have control over that as well. Although we show modification in the diffuse albedo across all the voxels in this scene, selecting only a subset of voxels to modify, perhaps through a GUI, is readily realizable. Finally, it would be interesting to explore how to expose additional material parameters, such as roughness, for manipulation within our latent data to control the glossiness of the full asset or portions of it. 

In Fig.~\ref{fig:ToyExample}, we set up a toy example with a bush in front of a blue sphere in order to demonstrate how our framework tracks material and geometric correlations and synthesizes an image. If we use the saved out tabular data as a look up table (LUT) we can generate an image that closely resembles the high-sample-count reference rendered with a ray tracer showing that our assumptions and approximations hold up sufficiently. Moreover, our full network-based approach successfully compresses the data without a noticeable loss in quality. In fact, our full approach is able to remove the small noise found in the LUT result since the network denoises the tabular data as shown in Fig.~\ref{fig:Denoising}.

Some of the scenes in Fig.~\ref{fig:ComplexScenes} contain blocky artifacts in the foreground regions of scale 8. This is from the pixel footprint being significantly smaller than the size of the discrete voxels found at our finest scale. For example, in the \scene{Parking Lot} scene shown in Fig.~\ref{fig:Hybrid} we see that our precomputed voxels are too coarse to accurately represent the foreground objects. However, we can use a hybrid system, a more pragmatic solution for a production pipeline, where we blend between the foreground and the simplified background based on the size of the pixel footprint. Specifically, for pixels where the voxels are too coarse, we simply use standard rendering and use our approach on pixels where the voxels are a pixel or smaller. With this strategy, we can accurately represent larger objects that are close to the viewer, while still saving on costs by using our prefiltering approach on the background layer. Note, with additional computation finer scales can be precomputed in order to render a larger region of the foreground with our framework. 

However, our method has some limitations that are the subject of future work. \new{As noted previously, our approach requires precomputing the voxel data and training on a per-scene basis, which can be a significant computational cost for some pipelines. Constant improvements in real-time and GPU rendering enable faster data generation to alleviate the former issue, while leveraging faster training methods~\cite{Mueller22} could help with the latter.}

\new{Another limitation is shown in} the top row of Fig.~\ref{fig:Limitations} depicting the roof of a building in the \scene{City} scene that is lighter than the ray traced reference. The voxels on the roof straddle a room whose interior is occluded normally. Thus, the albedo of the room's floor (which is lighter than the roof) is included in the average base color we store. To properly handle such a case, we could instead use an RGBA mask (instead of just the alpha channel as with the coverage mask), a $4\times$ increase in correlation-related data. By having the albedo spatially represented, we could determine which regions are occluded and omit them from the average base color.

Extending the phase function to have a spatial component instead of only a scalar can be similarly justified. However, this would increase the data by $M\times M$, where $M$ is the desired spatial resolution of a single slice (e.g., the element of a fixed outgoing direction would be a 4D LUT of size $N\times N\times M\times M$, where $N = 128$ in our implementation). Note, these estimates of increased data do not take into account the likely necessity of increasing the size of the latent vectors for each voxel as well as the networks in order to more accurately capture the additional data. Thus, in this case, there is a high potential for both the memory footprint and the runtime during rendering to increase. Moreover, the single RGB base color per view direction and single throughput per incoming and outgoing direction was sufficient for the scenes shown here. Finally, the current framework demonstrated results with a diffuse base color, but it would be interesting to extend the framework to additionally handle a specular albedo, perhaps in a similar fashion with an additional latent vector and sub-network or to combine the latent vectors with a single albedo decoder.

We used a relatively simplistic transmission model for combining adjacent voxels. In particular, we chose a coverage mask to determine occlusions while tracing beams through our SVO. This approach is limited by the resolution of our masks and there could be inaccuracies from subpixel occlusions. \old{\st{Moreover, we only calculate the coverage mask in the outgoing direction. Ideally, for improved accuracy, we would calculate a 2D spatial coverage mask for every 4D pair of incoming/outgoing directions, resulting in a 6D table and a $16\times16$ increase in correlation-related data for our implementation. Thus, where the coverage in the incoming and outgoing directions are not the same, there is currently a potential for errors. One could also imagine assuming separability of each of the 2D coverage masks for the outgoing and incoming direction and multiply them together to determine the net transparency contribution at the cost of having to evaluate an additional coverage mask per voxel (for the incoming direction). We leave investigation of how effective this strategy is for future work.}} Properly accounting for correlation in volumetric rendering is active research~\cite{Bitterli18,Jarabo18,Kettunen21,Vicini21} and more sophisticated formulations could be applied in our framework.


In the second row of Fig.~\ref{fig:Limitations}, we show another limitation present in the \scene{Mossy Rock} scene. If the highlights of a voxel are present in an extremely small solid angle of only a few bins, it poses a difficult scenario for our networks to capture exactly, especially when training on the order of a million voxels.  Thus, such sharp highlights in some cases can be slightly overblurred by our approach.



Here, we focused on properly handling primary beams, which tend to be among the most difficult aspects since their contribution, and any artifacts, would be directly visible. It would be interesting to apply our framework to render effects such as global illumination by casting multiple beams. This could be done by spawning new beams at every intersected voxel and accumulating the radiance contribution of that sub-beam. Nothing changes from a theoretical perspective, as our algorithm is presented as a general framework with an arbitrary number of bounces (see Sec.~\ref{sec:TheoreticalOverview}). Instead, the main challenge would be to efficiently evaluate multiple sub-beams without accumulating costs beyond that of \old{path}\new{ray} tracing. 
Furthermore, to avoid inefficient, brute-force sampling for generating additional beams or to enable common sampling strategies such as multiple importance sampling (MIS) in our beam tracer, we would need a mechanism for importance sampling the phase function of each voxel. The most straightforward way to do this in our current framework would be to normalize our decoder's phase-slice estimate to convert it to a PDF and then create an inverse CDF for importance sampling. Ideally, however, a learned solution would be more lightweight and flexible. One prospect would be to train a network to take in a set of random numbers and a latent representation of the phase (or its PDF) and output the next sample direction and its probability. Ensuring unbiasedness with such a strategy would be a compelling research direction. 

\vspace{-1mm}
\section{Conclusion}
We present the first deep learning framework for prefiltering complex 3D environments into multi-scale LoDs. Unlike some previous work, we do not rely on surface-only or volumetric-only models to simplify assets. Instead, we voxelize the scene and convert it to an SVO representation. The full appearance of each voxel is then captured through a ray-traced, data-generation step that saves the phase function, albedo, and coverage information in tabular form. By saving the true appearance in this manner, we avoid heuristics for classifying regions of assets, as in hybrid approaches, and can capture effects including sharp specularities that volumetric approaches cannot. Our novel, learning-based compression scheme compresses the rendered data into small latent vectors that can be easily stored and used by a beamtracer to render the final image by utilizing lightweight decoder networks, and without requiring any access to the original geometry or materials. To facilitate community involvement, \new{our code and supplementary material are available online.\footnote{\href{https://doi.org/10.7919/F4NK3C21}{https://doi.org/10.7919/F4NK3C21}}}\old{we will release our code, trained weights, and training set upon publication} Finally, we compare favorably to state-of-the-art prefiltering approaches and demonstrate significant memory savings relative to ray tracing on a variety of complex scenes.

\vspace{-1mm}
\new{\section{Acknowledgments}
We thank Delio Vicini, Anjul Patney, Zhao Dong, and Warren Hunt for helpful discussions. Special thanks to Anton Sochenov for his tremendous help including prototyping a real-time demo. We are very appreciative to Guillaume Loubet for releasing code for comparisons of both of his papers. We are grateful for Matt Chapman's help in generating many of the scenes in the paper. We thank the following artists/sources for the models we used in our scenes: GetWreckedDJ (stormtrooper), avi9526 (hangar), Quixel (mossy rock asset and textures), and Ndakasha (cutlery cloth). We thank Facebook Reality Labs for their unwavering support. This work was partially funded by National Science Foundation grants \#IIS-1619376 and \#IIS-1911230.}
\vspace{-3mm}

\bibliographystyle{ACM-Reference-Format}
\bibliography{Bibliography}

\appendix
\begin{figure}[!h]
	\centering
	\setlength{\tabcolsep}{1pt}
	\footnotesize%
	\begin{tabular}{cc}
		
		\includegraphics[width=0.5\linewidth]{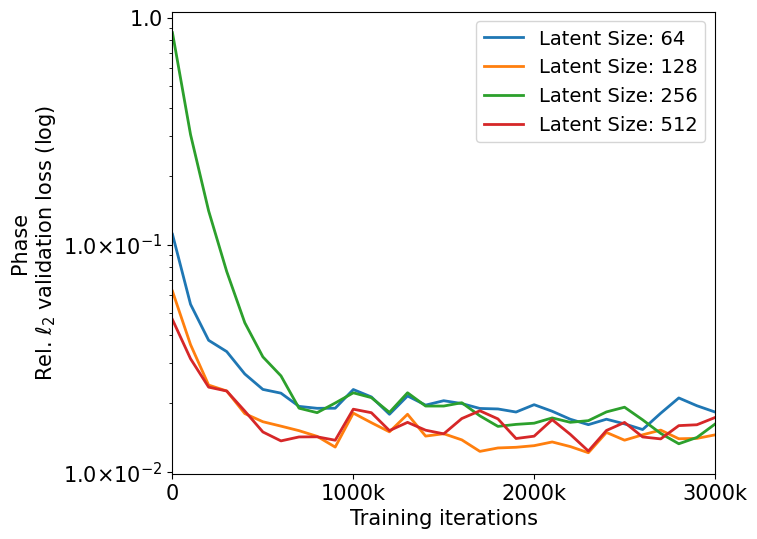} &
		\includegraphics[width=0.5\linewidth]{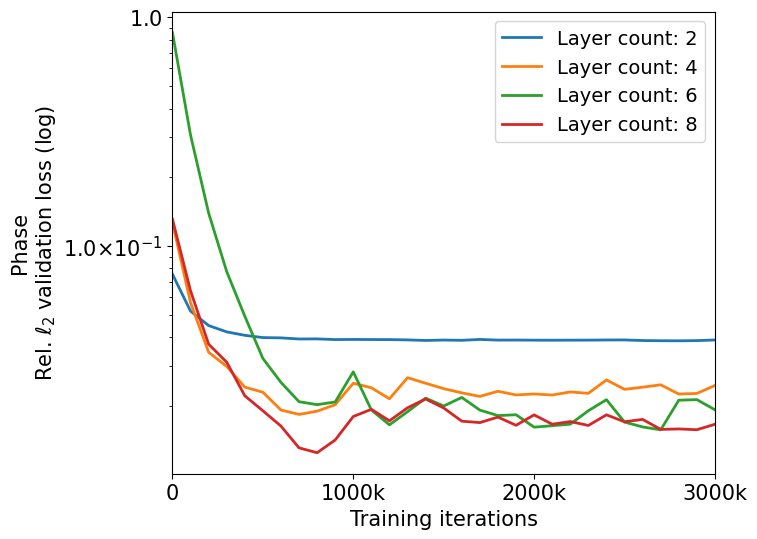} \\
		\includegraphics[width=0.5\linewidth]{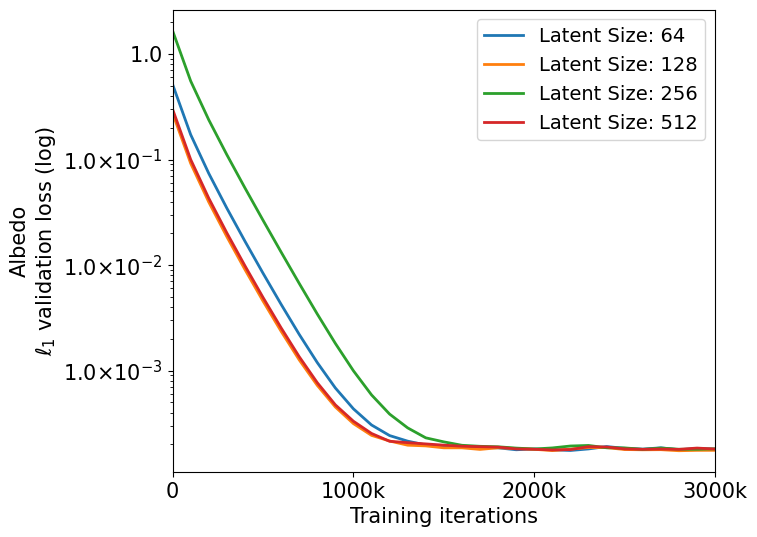} &
		\includegraphics[width=0.5\linewidth]{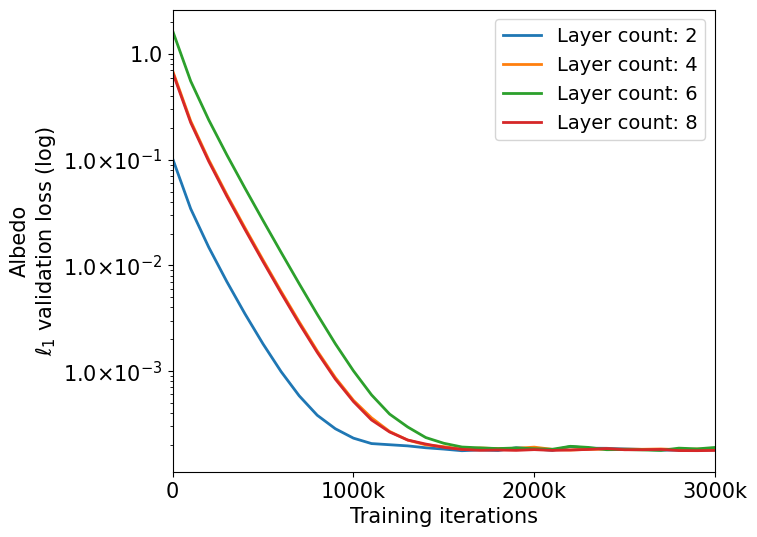} \\
		\includegraphics[width=0.5\linewidth]{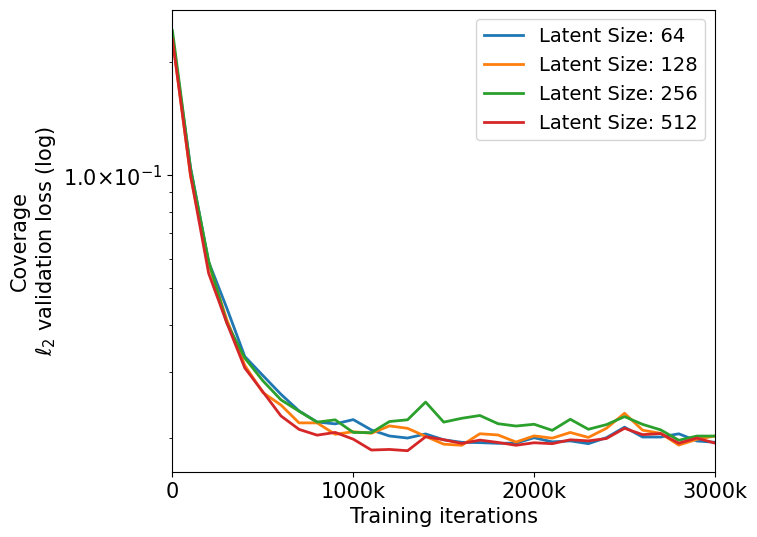} &
		\includegraphics[width=0.5\linewidth]{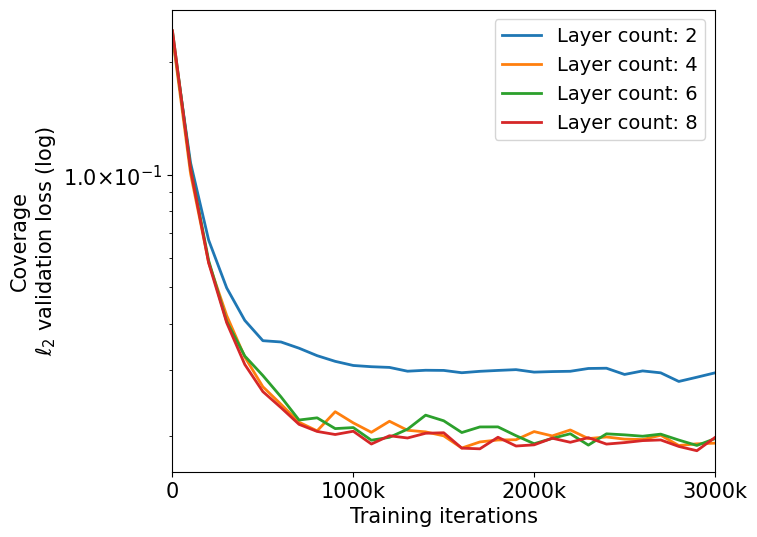} \\

	\end{tabular}
	\vspace{-4mm}
	\caption{Additional ablation results plotting validation errors for the other decoders not discussed in the main text. The behavior of the phase network (first row) is similar to the phase-slice network (see Fig.~\ref{fig:Ablation}) where increasing the latent size results in slightly better performance, yet gains become marginal for larger sizes (left plot). Moreover, increasing the number of network layers also improves performance up to a point (right plot). Meanwhile, the coverage and albedo sub-networks have a relatively lower threshold for latent sizes and layer counts as they do not improve their performance after reaching a certain size. To ensure all cases are handled robustly with a consistent architecture, we use 256 floats for the latent vector and 6 layers in the decoder, as this performs well in all cases and for all decoders.}
	\label{fig:AblationAdditional}
	\vspace{-4mm}
\end{figure}

\vspace{-4mm}
\section{Additional Ablation Analysis}
\label{appendix:AppendixA}
Fig.~\ref{fig:AblationAdditional} shows additional results from the two ablation experiments for the three remaining decoders (the phase-slice network plots were shown in Fig.~\ref{fig:Ablation}): phase, albedo, and coverage. The first row shows the results of these two ablations for the phase network. We see a similar behavior to that of the phase-slice results in that there are only marginal gains in performance for the converged networks with latent size of 256 floats and 6 layers, respectively. Next, the coverage network (second row) has roughly the same converged result for all latent sizes and for networks with 4 or more layers. Finally, the albedo network in the last row, performs well at all latent and network sizes. For simplicity and a consistent architecture for all sub-networks, we use 256 floats for each latent component and 6 layers for each decoder. Note, these experiments suggest there is potential to further minimize memory footprint by using albedo and coverage latent sizes of 64 or 128 floats in future implementations. 

\vspace{-1mm}
\section{Pseudocode}
\label{appendix:AppendixB}
\makeatletter
\newcommand{\removelatexerror}{\let\@latex@error\@gobble}
\makeatother
\new{In Algorithm~\ref{alg:pseudocode},} we present pseudocode for various components of our system to facilitate comprehension. \new{Our full code is available at \href{https://doi.org/10.7919/F4NK3C21}{our paper page} at to allow for comparisons and future work.}

\removelatexerror
\newcommand\mycommfont[1]{\itshape\sffamily\textcolor{black}{#1}}
\SetCommentSty{mycommfont}
\vspace{2mm}
\begin{algorithm}[ht!]
\DontPrintSemicolon
\SetAlgoLined
 \KwIn{Scene to prefilter, level of detail (LoD) scale \pseudoScale, scene for final rendering (e.g., camera and lighting parameters)}
 \KwOut{Rendered image, trained network weights \pseudoWeights, sparse voxel octree \pseudoSVO, latent voxel encodings \pseudoLatent}\;
 \SetKwBlock{DoParallel}{do in parallel}{end}
 \SetNoFillComment
	\tcc{SVO creation (Sec.~\ref{sec:FrameworkOverview})}
	\pseudoSVO ~= Voxelize(\pseudoScale)\;\;
	\tcc{Data generation (Sec.~\ref{subsec:DataGeneration}) and training (Sec.~\ref{subsec:Training})}
	\ForAll{training iterations}{
		\DoParallel{
			\pseudoData ~= GenerateVoxelData(\pseudoSVO)~\tcp{Includes brute-force rendering}
			\pseudoWeights ~= TrainNetwork(\pseudoData, \pseudoWeights)\;
		}
	}\;
	\tcc{Prerendering (Sec.~\ref{subsec:Runtime})}
	\pseudoLatent ~= GenerateLatentVectors(\pseudoData, \pseudoWeights)~\tcp{Save each voxel's latent vector}
	\pseudoShadow ~= GenerateShadowMaps(\pseudoSVO, \pseudoLatent, \pseudoWeights)\;
	LoadFinalSceneParameters(~)\;\;

	\tcc{Runtime (Sec.~\ref{subsec:Runtime})}
	\ForAll{pixels \pseudoPixel ~in final image \pseudoImage}{
	    \;
	    \tcc{Return list of voxels the beam touches in front to back order}
		\pseudoVoxelList ~= TraceBeam(\pseudoPixel, \pseudoSVO)\;
		\;
		\pseudoBeamCoverage ~= InitializeBeamCoverage(~)\;
		\pseudoColor ~= 0\;
		\pseudoTransmittance ~= ComputeTransmittance(\pseudoBeamCoverage)\;
		\While{\pseudoTransmittance < 1 \textbf{and} Count(\pseudoVoxelList) > 0}{
			\pseudoOut, \pseudoInc ~= GetOutgoingAndIncomingDirections(~)~\tcp*{Sec.~\ref{sec:FrameworkOverview}}
			\pseudoLatentVec ~= EncodeQueryAndConcatenate(\pseudoVox, \pseudoOut, \pseudoInc)\;
			\pseudoPhase, \pseudoAlbedo, \pseudoCoverage ~= EvaluateDecoders(\pseudoLatentVec)\;
			\pseudoPhaseFull ~= ComputePhase(\pseudoPhase, \pseudoAlbedo)~\tcp*{Eq.~\ref{eq:FullPhase}}
			\pseudoBeamCoverage ~= UpdateWavefrontAndCoverage(\pseudoCoverage)~\tcp*{Eq.~\ref{eq:TrackedCoverage}}
			\pseudoTransmittance ~= ComputeTransmittance(\pseudoBeamCoverage)~\tcp*{Eq.~\ref{eq:Transmittance}}
			\pseudoScattering ~= ComputeScatteringTerm(\pseudoPhaseFull, \pseudoTransmittance, \pseudoShadow)~\tcp*{Eq.~\ref{eq:ScatteringTermPrefilter}}
			\pseudoEmissionPrefilter ~= ComputePrefilteredEmission(~)~\tcp*{Eq.~\ref{eq:EmissionTermPrefilterExplicit}}
			\pseudoEmission ~= ComputeEmissionTerm(\pseudoEmissionPrefilter, \pseudoTransmittance)~\tcp*{Eq.~\ref{eq:EmissionTermPrefilter}}			
			\pseudoBoundary ~= ComputeBoundaryTerm(~)~\tcp*{Eq.~\ref{eq:BoundaryTerm}\hspace{1.5mm}}			
			\pseudoColor ~= UpdateColor(\pseudoScattering, \pseudoEmission, \pseudoBoundary)~\tcp*{Eq.~\ref{eq:PixelFlux}\hspace{1.5mm}}			
		}
	}
	\Return \pseudoImage
 \caption{Deep Appearance Prefiltering (DAP) Algorithm}
 \label{alg:pseudocode}
\end{algorithm}

\end{document}